\documentclass[aps,prd,twocolumn,superscriptaddress,amsmath,amssymb,nofootinbib,longbibliography,preprintnumbers,floatfix]{revtex4-2}
\usepackage[dvipsnames]{xcolor}
\usepackage[colorlinks]{hyperref}
\usepackage{xspace}

\usepackage{tikz}
\usetikzlibrary{tikzmark}
\usetikzlibrary{positioning}
\usepackage[normalem]{ulem}
\usepackage{tablefootnote}
\usepackage{aas_macros}
\usepackage{multirow}

\hypersetup{linkcolor=BrickRed,citecolor=Green,
filecolor=Mulberry,
urlcolor=NavyBlue,
menucolor=BrickRed,
runcolor=Mulberry
}

\definecolor{lime}{HTML}{A6CE39}
\DeclareRobustCommand{\orcidicon}{\hspace{-1mm}
	\begin{tikzpicture}
	\draw[lime, fill=lime] (0,0) 
	circle [radius=0.16] 
	node[white] {{\fontfamily{qag}\selectfont \tiny \,ID}};
	\draw[white, fill=white] (-0.0525,0.095) 
	circle [radius=0.007];
	\end{tikzpicture}
	\hspace{-3mm}
}

\foreach \x in {A, ..., Z}{\expandafter\xdef\csname orcid\x\endcsname{\noexpand\href{https://orcid.org/\csname orcidauthor\x\endcsname}
			{\noexpand\orcidicon}}
}

\usepackage{graphicx}
\usepackage{bbm}

\makeatletter
\newcommand{\abs}{\@ifstar\abssmall\absbig}
\newcommand{\absbig}[1]{\left \lvert #1 \right \rvert}
\newcommand{\abssmall}[1]{\lvert #1 \rvert}
\makeatother

\newcommand{\x}{\ensuremath{\mathrm{x}}\xspace}
\newcommand{\y}{\ensuremath{\mathrm{y}}\xspace}
\newcommand{\z}{\ensuremath{\mathrm{z}}\xspace}

\newcommand{\Tr}{\mathrm{Tr}}

\newcommand{\dd}{\mathrm{d}}

\newcommand{\bF}{\overline{F}}
\newcommand{\bP}{\overline{P}}

\newcommand{\bE}{\overline{E}}

\newcommand{\half}{\mathrm{\frac{1}{2}}}

\newcommand{\N}{\mathcal{N}}
\newcommand{\F}{\mathcal{F}}
\newcommand{\bcN}{\overline{\mathcal{N}}}
\newcommand{\bcF}{\overline{\mathcal{F}}}

\newcommand{\beq}{\begin{equation}}
\newcommand{\eeq}{\end{equation}}
\newcommand{\ben}{\begin{enumerate}}
\newcommand{\een}{\end{enumerate}}

\newcommand{\flash}{\texttt{FLASH}\xspace}
\newcommand{\flashri}{\texttt{FLASH} $(ri)$\xspace}
\newcommand{\emu}{\texttt{Emu}\xspace}
\newcommand{\imo}{\ensuremath{{\rm Im}(\Omega)_{\rm max}}\xspace}
\newcommand{\kmax}{\ensuremath{k_{\rm max}}\xspace}
\newcommand{\tsat}{\ensuremath{t_{\rm sat}}\xspace}

\newcommand{\smod}{\ensuremath{{\rm smod}}\xspace}

\newcommand{\cnumber}{$c$-number\xspace}
\newcommand{\cnumbers}{$c$-numbers\xspace}

\newcommand{\Mcal}{\ensuremath{\mathcal{M}}\xspace}
\newcommand{\avgnex}{\ensuremath{\langle|N_{ex}|\rangle}\xspace}
\newcommand{\avgnee}{\ensuremath{\langle N_{ee}\rangle}\xspace}

\renewcommand{\vec}{\mathbf}

\newcommand{\edit}[1]{\textnormal{\color{black}{#1}}}

\begin{document}
\preprint{N3AS-25-001}

\title{Advection Algorithms for Quantum Neutrino Moment Transport}

\author{Evan Grohs\orcidD{}}
\email{ebgrohs@ncsu.edu}
\affiliation{Department of Physics, North Carolina State University, Raleigh, NC 27695, USA}

\author{Sherwood Richers\orcidC{}}
\affiliation{Department of Physics, University of Tennessee Knoxville, Knoxville, TN 37996, USA}

\author{Julien Froustey\orcidA{}}
\affiliation{Department of Physics, North Carolina State University, Raleigh, NC 27695, USA}
\affiliation{Department of Physics, University of California Berkeley, Berkeley, CA 94720, USA}
\affiliation{Department of Physics, University of California San Diego, La Jolla, CA 92093, USA}

\author{Francois Foucart\orcidF{}}
\affiliation{Department of Physics \& Astronomy, University of New Hampshire, 9 Library Way, Durham, NH 03824, USA}

\author{James P. Kneller\orcidG{}}
\affiliation{Department of Physics, North Carolina State University, Raleigh, NC 27695, USA}

\author{Gail C. McLaughlin\orcidH{}}
\affiliation{Department of Physics, North Carolina State University, Raleigh, NC 27695, USA}

\begin{abstract}
Neutrino transport in compact objects is an inherently challenging multi-dimensional problem. This difficulty is compounded if one includes flavor transformation -- an intrinsically quantum phenomenon requiring one to follow the coherence between flavors and thus necessitating the introduction of complex numbers. To reduce the computational burden, simulations of compact objects that include neutrino transport often make use of momentum-angle-integrated moments (the lowest order ones being commonly referred to as the energy density and flux) and these quantities can be generalized to include neutrino flavor, i.e., they become quantum moments. Numerous finite-volume approaches to solving the moment evolution equations for classical neutrino transport have been developed based on solving a Riemann problem at cell interfaces. In this paper we describe our generalization of a Riemann solver for quantum moments, specifically decomposing complex numbers in terms of a (signed) magnitude and phase instead of real and imaginary parts. We then test our new algorithm in numerous cases showing a neutrino fast flavor instability, varying from toy models with analytic solutions to snapshots from neutron star merger simulations. Compared to previous algorithms for neutrino transport with flavor mixing, we find uniformly smaller growth rates of the flavor transformation along with concomitantly larger length-scales, and that the results are a better match with the growth rates seen from multi-angle codes. 
\end{abstract}

\maketitle

\section{Introduction}
\label{sec:introduction}

Multi-messenger observations of binary neutron star mergers (NSMs) \cite{2020ARNPS..70...95R} and core-collapse supernovae (CCSNe) \cite{1989ARA&A..27..629A,muller2020hydrodynamics,burrows2021core} present a tremendous opportunity for breakthroughs not only in our understanding of the astrophysical mechanisms that drive these phenomena but also of the origin of the elements and fundamental nuclear physics.  
To model gravitational wave, electromagnetic signals, and neutrino emission from NSMs and CCSNe, global 3D simulations of these systems are required. 
These simulations would ideally evolve at least Einstein's equation for the space-time metric, the relativistic equations of (magneto)hydrodynamics, and neutrino radiation transport with flavor mixing. 
The prohibitive cost of including and resolving all the physics in CCSNe and NSMs that is known to affect the outcome remains an important problem (see, e.g.~\cite{Burrows:2020qrp,Kyutoku:2021icp,Foucart:2022bth,Mezzacappa:2020pkk,Kiuchi:2024lpx} for recent reviews). 
Focusing just on the neutrino transport, global simulations of CCSNe and NSMs currently face three important difficulties: the inclusion of all relevant reactions involving neutrinos with sufficient physical realism, the implementation of accurate yet cost-effective radiation transport algorithms, and the inclusion of flavor mixing. Even when neutrino flavor mixing is ignored, approximate transport algorithms are often used to make global simulations of CCSNe and NSMs feasible on available computing resources, although recently, direct discretization of Boltzmann's equations has been possible using Monte-Carlo transport~\cite{Abdikamalov:2012zi,richers2015monte,Miller:2019gig,Foucart:2021mcb} and discrete ordinate transport~\cite{Sumiyoshi:2012za}. 
Adding neutrino flavor mixing introduces a wealth of new phenomenology~\cite{Tamborra:2020cul,Volpe:2023met}. The presence of electrons and positrons in NSMs and CCSNe will induce a flavor-dependent mixing potential \cite{1995ARA&A..33..459H,2023ecnp.book..367F}, and nonlinear mixing potentials arise from the presence of (anti)neutrinos themselves \cite{1992PhLB..287..128P,1993PhRvD..48.1462S}. 
The two potentials beget phenomena such as bipolar oscillations \cite{2006PhRvD..74l3004D,2006PhRvD..74j5010H}, energy spectral swaps \cite{2007PhRvD..76h1301R,2007PhRvL..99x1802D,2007PhRvD..76l5008R}, matter-neutrino resonances \cite{Malkus:2012ts,Malkus:2014iqa,Wu:2015fga,Frensel:2016fge,Tian:2017xbr,2018PhRvD..97h3011V}, slow collective modes \cite{2006PhRvL..97x1101D,2007PhRvD..75l5005D,2010ARNPS..60..569D}, collisional instabilities \cite{Johns:2021qby,Johns:2022yqy,Xiong:2022vsy,Xiong:2022zqz,Liu:2023vtz,Akaho:2023brj}, and fast flavor instabilities (FFIs) \cite{Sawyer:2005jk,Dasgupta:2016dbv,Izaguirre:2016gsx,Wu:2017qpc,Abbar:2018shq, Nagakura:2021hyb,Richers:2022zug,Nagakura:2023mhr}.

Although the effects of neutrino flavor transformation upon CCSNe and NSMs would be most satisfyingly provided by first-principle calculations of the quantum kinetic equations (QKEs), directly doing so is currently cost prohibitive. 
An alternative approach to full neutrino transport is instead to consider angular moments of the distribution. What could be described as “classical" moments continue to be used extensively to describe the transport of neutrinos in many supernova and compact object merger simulations -- see Refs.\ \cite{ISRAEL1979341,1981MNRAS.194..439T,2018ApJ...854...63O} for examples. The benefit of this approach is that one can truncate the tower of moment evolution equations and thus the computational expense.
Typically the evolution equations are truncated at the first one or two moments (the number/energy density and the number/energy flux). However, the truncation means one needs to supply a “closure" that allows for the computation of the higher-order moments which appear in the evolution equations but are not evolved. We refer the reader to Refs.\ \cite{minerbo_maximum_1978,Smit_closure,Shibata:2011kx,Murchikova:2017zsy,2020PhRvD.102h3017R,2023ApJ...943...78W} for detailed discussions and studies of classical closures. The generalization of this approach for neutrino transport including flavor transformation --- what could be called “quantum" moments --- was introduced by Ref.\ \cite{strack:2005} and further developed in Refs.\ \cite{Zhang:2013lka,johns2020neutrino,2022PhRvD.105l3036M}. Note that the introduction of quantum moments also means that we need to generalize the closure. The important point for the reader to remember about the quantum moments is that they are matrices containing complex numbers (\cnumbers in brief) whereas the classical moments are real scalars. In classical transport, the moments are often evolved using flux-conservative shock capturing methods developed in the context of fluid dynamics. In particular, Riemann solvers are frequently used at cell interfaces, and we shall do the same by adopting the Riemann solver initially developed by Harten, Lax, van Leer, and Einfeldt \cite{HLL_1983,1988SJNA...25..294E} -- the HLLE Riemann solver. 
We note that O'Connor and Couch \cite{2018ApJ...854...63O} implemented the HLLE Riemann solver with a few modifications, which we will refer to later.  The HLLE solver is well-tested for the evolution of real-valued moments but needs to be adapted for quantum moments. The purpose of this paper is to describe and test these adaptations.

The outline of the remainder of this paper is as follows.
Section \ref{sec:methods} contains the theory behind the method. We establish the equations of motion that we use to model neutrino radiation transport, 
the signed moduli we use for the flux moments, the elements of the HLLE solver and the cell interface reconstruction algorithm we adopt. 
From there, we move on to Sec.~\ref{sec:implementation}, which introduces our particular implementation of a new advection algorithm with the code \flash, which we will call in the following the “modulus phase implementation.” We then test our solver on a range of FFI situations in Sec.~\ref{sec:results}, with several configurations obtained in a classical NSM moment simulation along with relevant simplified-model cases, further outlined in Appendices~\ref{sec:beam}, \ref{app:fid_90d} and \ref{app:NSM4}. These various examples show that our new approach better predicts the characteristics of FFI than the previous implementation of quantum moment transport (which we will call “real-imaginary implementation”). We conclude in Sec.\ \ref{sec:conclusion} by summarizing our results, discussing future improvements.

Before we begin, we comment on our notation and the conventions used in this work.  First, we will write our equations with natural units where $c=\hbar=1$, and give the numerical results of our simulations in cgs units.  Repeated spatial indices $j,k,...$ imply summations over directions $\{\x,\y,\z\}$ unless otherwise stated explicitly. Much of this work details the difference between advection of real-valued versus \cnumber moments.  To delineate between the two schemes, which are used for different elements of the flavor matrix representation of the quantum moments, we will call the set of real-valued moments “flavor diagonal elements”, or diagonal elements in brief. Conversely, we refer to the \cnumber moments as “flavor off-diagonal elements”, or off-diagonal elements in brief.  
We will compare the results of the modulus phase implementation with those of the real-imaginary implementation from Ref.\ \cite{flashri}.  When showing results in tables and plots, we will label the previous real-imaginary implementation \flashri, and the new modulus phase implementation simply as \flash.

\section{A Riemann Solver for Quantum Moments}
\label{sec:methods}

\subsection{Moment Transport}
The neutrino flavor eigenbasis is not coincident with the mass eigenbasis, leading to the quantum phenomenon of flavor oscillations. We need to adopt a formalism which can follow quantum flavor oscillations and still capture the statistical properties of the system in the aggregate.  Our approach is to use a mean field approximation to treat the neutrino distributions as generalized one-body reduced density matrices, which are functions of time, position, and momentum.
We term these matrices $\varrho(t,\vec{x},\vec{p})$ for neutrinos and $\overline{\varrho}(t,\vec{x},\vec{p})$ for antineutrinos. These matrices are Hermitian and arranged by the two flavor indices: electron-type ($e$) and heavy-lepton type ($x$).  They can be combined and expanded to encompass more flavor and chiral states but we only present results for the two flavor, left-handed neutrino (and right-handed antineutrino) oscillating system in this work.  The diagonal entries of $\varrho$ give the differential number density of neutrinos for a given flavor in phase space, e.g., $\varrho_{ee}$ for an electron neutrino. Those diagonal entries are analogous to the real classical distributions used in non-oscillatory simulations.  When oscillations are allowed to occur, the off-diagonal entries of the density matrices encode a flavor coherence, i.e., a measure of the entanglement between two flavor states. These components, $\varrho_{ex}$ and $\overline{\varrho}_{ex}$, are necessarily \cnumbers.

The QKEs are the dynamical equations of motion for the density matrices. Introducing a generic emission/absorption/scattering term $\mathcal{C}$, we write the QKEs as
\begin{equation}
    \frac{\partial\varrho}{\partial t} + \dot{\mathbf{x}}\cdot\frac{\partial\varrho}{\partial\mathbf{x}} + \dot{\mathbf{p}}\cdot\frac{\partial\varrho}{\partial\mathbf{p}}
    = -\imath\,[H,\varrho] + \mathcal{C} \, ,
    \label{eq:qke_general}
\end{equation}
where the dot over the vector indicates a time derivative and $H$ is the Hamiltonian-like operator. A similar expression exists for $\overline{\varrho}$. There is extensive literature on solving the QKEs with various geometries and source terms (see in particular Refs.\ \cite{1991NuPhB.349..743B,Sigl:1993ctk,strack:2005,2007JPhG...34...47B,Volpe:2013uxl,2013PrPNP..71..162B,2014PhRvD..89j5004V,2014PhRvD..90l5040S,2015PhLB..747...27C,2015IJMPE..2441009V,Blaschke:2016xxt,Richers:2019grc,Froustey:2020mcq,Volpe:2023met}). In what follows, we shall ignore the momentum derivative (which contains higher-order refractive and relativistic effects) and delay an exposition of the Hamiltonian until later sections when we present simulation results for specific cases. Our focus in this paper is neutrino advection, i.e., the spatial derivatives of the lhs of Eq.~\eqref{eq:qke_general}.

From the one-body density matrix, we define the differential energy, flux, and pressure moments as:
\begin{align}
  E(t,\mathbf{x},p) &\equiv \frac{p^3}{(2\pi)^3}\int d\Omega_p\, \varrho(t,\mathbf{x},\mathbf{p}),\label{eq:mom_0}\\
  F^j(t,\mathbf{x},p) &\equiv \frac{p^3}{(2\pi)^3}\int d\Omega_p\frac{p^j}{p}\, \varrho(t,\mathbf{x},\mathbf{p}),\label{eq:mom_1}\\
  P^{jk}(t,\mathbf{x},p) &\equiv \frac{p^3}{(2\pi)^3}\int d\Omega_p\frac{p^j p^k}{p^2}\, \varrho(t,\mathbf{x},\mathbf{p}),
\label{eq:mom_2}
\end{align}
where $j,k \in \{\x,\y,\z\}$ are spatial indices. 
In addition, because it will appear later in our test problems, we define the differential number density moment as
\begin{equation}
    N(t,\vec{x},p) \equiv \frac{1}{p}\, E(t,\vec{x},p) \, .
\end{equation}
Applying these same integrals to Eq.~\eqref{eq:qke_general} yields the QKEs for the first two angular moments~\cite{Zhang:2013lka,Richers:2019grc}:
\begin{subequations}
\label{eq:moment_QKEs}
\begin{align}
    \frac{\partial E}{\partial t} + \frac{\partial F^j}{\partial \mathrm{x}^j} &= \mathcal{S}_E \, , \\
    \frac{\partial F^j}{\partial t} + \frac{\partial P^{jk}}{\partial \mathrm{x}^k} &= \mathcal{S}_{F}^{j} \, ,
\end{align}
\end{subequations}
where the source terms $\mathcal{S}_E$ and $\mathcal{S}_{F}^{j}$ are 
\begin{align}
  \mathcal{S}_E(t,\mathbf{x},p) &= \frac{p^3}{(2\pi)^3}\int d\Omega_p\,\left( -\imath\,[H,\varrho] + \mathcal{C} \right),\label{eq:SE}\\
  \mathcal{S}_{F}^{j}(t,\mathbf{x},p) &= \frac{p^3}{(2\pi)^3}\int d\Omega_p\frac{p^j}{p}\,\left( -\imath\, [H,\varrho] + \mathcal{C}\right),\label{eq:SF}
\end{align}
and similarly for antineutrinos. 
The equations for quantum moment transport look very similar to their classical counterparts, but the change from scalars to matrices introduces a whole new set of challenges, in particular regarding the advection algorithm.

\subsection{The Signed Modulus Representation}

Foreshadowing the issues we will encounter later, we digress to describe how the moments, particularly the flux, are represented. For classical moments the sign of the flux indicates the direction of the flow of the energy density. For quantum moments, we run into a conceptual difficulty with how to assign a direction for the off-diagonal elements of the flux. 
To solve this problem we make use of the signed modulus for the elements of the flux. To show the difference between the more familiar modulus and the signed modulus, consider the generic $ab,\; a\neq b$ flavor-space elements of the energy density moment $E_{ab}$ and of the flux density moment $F_{ab}$ (omitting the spatial index $j$ for clarity). We define
\begin{equation}
    \label{eq:smod_def}
\begin{aligned}
  E_{ab} &= |E_{ab}| e^{\imath \phi_{E,ab}} &&& F_{ab} &= \smod(F_{ab})\,e^{\imath \phi_{F,ab}},\\
  0&\le |E_{ab}| <\infty &&& -\infty&<\smod(F_{ab})<\infty,\\
  -\pi&\leq \phi_{E,ab}<\pi &&& -\pi&\leq \phi_{F,ab}<\pi
\end{aligned}
\end{equation}
We will use the sign of the signed modulus for the flux to indicate the net direction of flow. The use of the signed modulus for the flux appears to introduce degenerate representations of the flux: the same complex number can be obtained by inverting the sign of the signed modulus and changing the phase by $\pi$. In practice we solve this ambiguity at the steps in the code where we increment the flux by using the protocol that the signed modulus of the sum of two \cnumbers is the same as the summand with the largest absolute modulus. Once the sign of the signed modulus is determined, the phase is unambiguous. 

\subsection{The Quantum Riemann Solver}
\label{subsec:Riemann_solver}

As in many approaches to classical transport, we discretize the computational volume into a grid of finite-volume cells. For simplicity, we consider in this section a case where the moments only vary in the $\rm{x}$ direction with the understanding that our equations can be easily generalized to apply to three dimensions. Nonetheless, we shall continue to indicate spatial indices on the flux and pressure. The cell centers are denoted as ${\rm x}_i$ and we make the spatial grid uniform so that each cell has the same size $\Delta \x$. The moments are also discretized and the value of the moment within a given cell will be indicated with an index $i$. The values of the energy and flux that we update as we step forward in time are said to “live" at the center of the grid cells. Finally, since we are focusing upon the advection of the moments which does not mix the flavors in the absence of “sources" in Eq.~\eqref{eq:moment_QKEs}, we shall generally not indicate the flavor indices of the moments nor on any other quantity that could possess them, though the relations apply to each component separately. 

After discretizing the space, we treat neutrino advection using familiar finite-volume methods with some modifications to deal with the complex numbers. 
The algorithm we employ is based on the HLLE Riemann solver, see~\cite{HLL_1983,1988SJNA...25..294E,2018ApJ...854...63O}. Our quantum version of the HLLE Riemann solver starts in the usual fashion. The numerical spatial derivative for the flux is
\begin{align}
  \frac{\partial F^j}{\partial \x}(\x_i) &= \dfrac{\mathbb{F}^{j}(\x_i+\frac{\Delta \x}{2}) - \mathbb{F}^{j}(\x_i-\frac{\Delta \x}{2})}{\Delta\x} \,,
  \label{eq:num_edf}
\intertext{while for the pressure}
  \frac{\partial P^{j k}}{\partial \x}(\x_i) &= \dfrac{\mathbb{P}^{j k}(\x_i+\frac{\Delta \x}{2}) - \mathbb{P}^{j k}(\x_i-\frac{\Delta \x}{2})}{\Delta\x} \, , 
  \label{eq:num_mdf}
\end{align}
where $\mathbb{F}$ and $\mathbb{P}$ are the solutions to a Riemann problem at each interface of cell $i$. For further ease of notation we introduce the symbols $\mathbb{F}^j_{i\pm\half}$ defined to be
\begin{equation}
    \mathbb{F}^j_{i\pm\half} = \mathbb{F}^j\left(\x_i\pm\frac{\Delta \x}{2}\right) \, ,
\end{equation}
\edit{which represent the solutions for $\mathbb{F}$ from the Riemann problem at the interface} between grid cells $i$ and $i-1$ (denoted $i-\half$), and cells $i$ and $i+1$ (denoted $i+\half$),  and similarly for $\mathbb{P}^{jk}_{i\pm \half}$. 
The interface fluxes in Eq.~\eqref{eq:num_edf} or pressures in Eq.~\eqref{eq:num_mdf} are evaluated in the HLLE Riemann solver in two steps. First, the value of the moments on each side of a cell interface are reconstructed; second, those pairs of values are used to determine the Riemann solution. 
Note that since there is no spatial derivative of $E$ in Eq.~\eqref{eq:moment_QKEs} we only require the Riemann solutions for $\mathbb{F}^j$ and $\mathbb{P}^{jk}$.

\paragraph*{Cell interface reconstruction —} The first step of the HLLE solver is to manufacture the value of the moments at the cell interfaces. If we use the energy density as an example and consider the interface between cell $i$ and $i+1$, we construct at the interface between the cells the quantities $E^{(R)}\left(\x_i+\frac{\Delta \x}{2}\right)$ and $E^{(L)}\left(\x_i+\frac{\Delta \x}{2}\right)$ which we denote as $E_{i+\half}^{(R)}$ and $E_{i+\half}^{(L)}$ respectively. The superscripts $(R)$ and $(L)$ indicate the “right" side of cell $i$ and the “left" side of cell $i+1$.  We can reconstruct the equivalent quantities for the flux moment which we denote in a similar way as 
$F_{i+\half}^{j,(R)}$ and $F_{i+\half}^{j,(L)}$. 
We shall refer to the algorithm to compute these values as the “reconstruction,” which we shall discuss in detail in Section~\ref{sec:reconstruction}. 

\paragraph*{Solution of the Riemann problem at the interface —} Whatever the details of the reconstruction algorithm, once the values of $E_{i+\half}^{(L/R)}$ and $F_{i+\half}^{j,(L/R)}$ are determined, we compute $P_{i+\half}^{jk,(L/R)}$ using the closure. The Riemann solutions at the interface between cells $i$ and $i+1$ are then given by:
\begin{align}  
\mathbb{F}^j_{i+\half} =& \frac{1}{\lambda^j_{\rm max} - \lambda^j_{\rm min}} \Big[ \lambda^j_{\max}F^{j,(L)}_{i+\half} - \lambda^j_{\min}F^{j,(R)}_{i+\half}  \nonumber \\
   & \qquad\qquad  +  \lambda^j_{\min}\lambda^j_{\max}\big(E^{(R)}_{i+\half} - E^{(L)}_{i+\half}\big) \Big] \, \label{eq:FRiemann},\\   
\mathbb{P}^{jk}_{i+\half} =& \frac{1}{\lambda^j_{\rm max} - \lambda^j_{\rm min}} \Big[\lambda^j_{\max}P^{jk,(L)}_{i+\half} - \lambda^j_{\min}P^{jk,(R)}_{i+\half} \nonumber \\
  & \qquad\qquad + \lambda^j_{\min}\lambda^j_{\max}\big(F^{k,(R)}_{i+\half} - F^{k,(L)}_{i+\half}\big)\Big] \, ,
  \label{eq:PRiemann}
\end{align}
where the the $\lambda$'s are the characteristic velocities for the interface $i+\half$. In order to be physical, these velocities are real.
Note that there is no Einstein summation over $j$ in these equations. 
The HLLE Riemann solver provides an algorithm for determining the $\lambda$ velocities depending on whether the characteristic curves are all positive, all negative, or a mixture. The extremal characteristic velocities are
\begin{align}
  \lambda^{j}_{\rm min} &= \min\left[\lambda^{j,(L)}_{\rm min}, \lambda^{j,(R)}_{\rm min}, 0\right] \, ,\\
  \lambda^{j}_{\rm max} &= \max\left[\lambda^{j,(L)}_{\rm max}, \lambda^{j,(R)}_{\rm max}, 0\right] \, .
\end{align}
Rather than evaluating the characteristic velocities directly, we follow \cite{2018ApJ...854...63O} in approximating the characteristic velocities by interpolating between the values in the optically thick and thin limits: 
\begin{align}
  \lambda^{j,(S)}_{\rm min} &= \frac{3\,\chi^{(S)}-1}{2}\lambda^{j,(S)}_{\rm thin,min} + \frac{3\,(1-\chi^{(S)})}{2}\lambda_{\rm thick,min} \, , \\
\lambda^{j,(S)}_{\rm max} &= \frac{3\,\chi^{(S)}-1}{2}\lambda^{j,(S)}_{\rm thin,max} + \frac{3\,(1-\chi^{(S)})}{2}\lambda_{\rm thick,max}  \, ,
\end{align}
where $S \in \{L,R\}$ and $\chi^{(S)}$ is the Eddington factor — provided by the closure — using the flux and energy density moments reconstructed on either side of the interface.\footnote{Note that we are assuming that $\chi^{(L/R)}$ is a real number. In cases where the closure would give a complex Eddington factor for the off-diagonal elements of the moments, this equation would need to be modified.} 

Up to this point, there appears to be no difference with a classical Riemann solver: compare with \cite{2015ApJS..219...24O,2018ApJ...854...63O}.
However, we remind the reader that we have suppressed the flavor indices for the sake of clarity and the algorithm we are describing has to be repeated for each flavor space element of the moments. It is at this point that we specifically point out that for the quantum case, the Eddington factor could be different for the flavor space diagonal and off-diagonal elements. 
In addition, we also pause to reiterate that our motivation for using the signed arithmetic for complex numbers in our paper is to ensure the limiting characteristic velocities have the correct signs for the flavor off-diagonal elements of the moments. 
In the thin limit, the maximal characteristic velocities read
\begin{align}
  \lambda^{j,(S)}_{\rm thin,min} &= \min\left\{-\frac{|F_{i+\half}^{j,(S)}|}{|\vec{F}_{i+\half}^{(S)}|},\frac{|E_{i+\half}^{(S)}|\times\smod\left[F_{i+\half}^{j,(S)}\right]}{|\vec{F}_{i+\half}^{(S)}|^2}\right\} \, ,\label{eq:lambda_thin_min_smod}\\
  \lambda^{j,(S)}_{\rm thin,max} &= \max\left\{+\frac{|F_{i+\half}^{j,(S)}|}{|\vec{F}_{i+\half}^{(S)}|},\frac{|E_{i+\half}^{(S)}|\times\smod\left[F_{i+\half}^{j,(S)}\right]}{|\vec{F}_{i+\half}^{(S)}|^2}\right\}\label{eq:lambda_thin_max_smod} \, .
\end{align}
where $\smod(F)=s|F|$ with $s=\pm1$ such that $s$ indicates the direction of flow.
\edit{In Eqs.\ \eqref{eq:lambda_thin_min_smod} and \eqref{eq:lambda_thin_max_smod}, the flavor off-diagonal fluxes have well-defined signed moduli and the value of $s$ is known at the beginning of the time step.  Determining the value of $s$ for the cell-interface interpolated fluxes will be detailed in Sec.\ \ref{sec:reconstruction}.}
These expressions are to be contrasted with the classical ones~\cite{2015ApJS..219...24O,2018ApJ...854...63O}, valid for the flavor on-diagonal elements of the moments,\footnote{Note that we can generally write Eqs.~\eqref{eq:lambda_thin_min_smod}--\eqref{eq:lambda_thin_max_smod} for all flavor components. Indeed, they reduce to \eqref{eq:lambda_thin_min}--\eqref{eq:lambda_thin_max} for the on-diagonal elements, since the energy density is then a positive quantity and $\smod[F]=F$ if $F \in \mathbb{R}$.}
\begin{align}
  \lambda^{j,(S)}_{\rm thin,min} &= \min\left\{-\frac{|F_{i+\half}^{j,(S)}|}{|\vec{F}_{i+\half}^{(S)}|},\frac{E^{(S)}_{i + \half} \times F^{j,(S)}_{i+\half}}{|\vec{F}^{(S)}_{i+\half}|^2}\right\},\label{eq:lambda_thin_min}\\
  \lambda^{j,(S)}_{\rm thin,max} &= \max\left\{+\frac{|F_{i+\half}^{j,(S)}|}{|\vec{F}_{i+\half}^{(S)}|},\frac{E^{(S)}_{i+\half} \times F^{j,(S)}_{i+\half}}{|\vec{F}^{(S)}_{i+\half}|^2}\right\}.\label{eq:lambda_thin_max}
\end{align}

In the thick limit, the expressions are the same in the classical and quantum cases,
\begin{align}
  \lambda_{\rm thick,min} &= -\frac{1}{\sqrt{3}} \, , \\
  \lambda_{\rm thick,max} &= +\frac{1}{\sqrt{3}} \, .
\end{align}
Finally, we mention that Appendix B of Ref.\ \cite{2018ApJ...854...63O} introduces Peclet numbers and limiting flux terms to correct for deviation from hyperbolicity in high opacity regions. We follow the same prescription and have implemented \cnumber limiting flux terms to correct the HLLE solution in Eqs.~\eqref{eq:FRiemann} and \eqref{eq:PRiemann}.

\subsection{Cell Interface Reconstruction}
\label{sec:reconstruction}

To compute the Riemann solutions of the flux and pressure moments we first need to “reconstruct" the moments at either side of each interface between the grid cells. Denoting a generic moment in cell $i$ as $\Mcal_i$, we need to compute $\Mcal_{i-\half}^{(R)}$ and $\Mcal_{i+\half}^{(L)}$. Note that for $n$ cells along an axis $j$, we have $n+1$ interfaces, implying we must use ghost cells/boundary conditions on each side for $\Mcal^{(L)}_{1-\half}$ and $\Mcal^{(R)}_{n+\half}$. 
The flavor indices of the moment have been suppressed for the sake of clarity and we remind the reader that the reconstruction procedure needs to be applied to each flavor space element of the moments. For the flavor space diagonal elements, reconstructing $E^{(L/R)}$, $F^{j(L/R)}$ and $P^{jk,(L/R)}$ involves only real numbers (because the moments are Hermitian) and so we adopt the same Minmod procedure as in the classical transport case, which we recall below.

\paragraph*{Classical minmod interpolation –} In classical transport, we calculate moment values at the interface using a special first-order interpolator
\begin{equation}
\label{eq:M_minmod_class}
\begin{aligned}
  \Mcal_{i-\half}^{(R)} &= \Mcal_i - \frac{1}{2}{\rm minmod}\left(\Mcal_{i+1} - \Mcal_i, \Mcal_i - \Mcal_{i-1}\right) \, ,\\
  \Mcal_{i+\half}^{(L)} &= \Mcal_i + \frac{1}{2}{\rm minmod}\left(\Mcal_{i+1} - \Mcal_i, \Mcal_i - \Mcal_{i-1}\right) \, ,
\end{aligned}
\end{equation}
where the minmod function is defined to be
\begin{equation}
\label{eq:minmod}
  {\rm minmod}(g,h) \equiv \frac{1}{2}[{\rm sgn}(g) + {\rm sgn}(h)]{\rm min}(|g|,|h|) \, ,
\end{equation}
for real numbers $g,h$ and where ${\rm sgn}(g)$ is the sign function. As mentioned above, this algorithm is still used for the on-diagonal components $E_{aa}, \, \vec{F}_{aa}$.

The reconstruction of the \emph{off-diagonal} elements of the moments, which cannot be achieved directly with the same procedure as we have to deal with \cnumbers, is the focus of the remainder of this section. 

\paragraph*{Interpolation for off-diagonal components —} In previous studies, e.g.~\cite{Grohs:2022fyq}, we applied a standard linear reconstruction with a Minmod limiter to the real and imaginary parts of the flavor space off-diagonal elements of the moments separately, and found it gives reasonable results. In this work we change the reconstruction scheme to one based on interpolating along circular arcs in the complex plane but retain the spirit of the standard Minmod reconstruction algorithm~\eqref{eq:M_minmod_class}. Specifically, for cell $i$ we choose to build the interpolant using $\Mcal_i$ and $\Mcal_n$, where $\Mcal_n$ is \emph{either} $\Mcal_{i+1}$ or $\Mcal_{i-1}$ based on a “minimal difference” criterion. The minimal difference criterion we adopt is the comparison of the quantities $|\delta \Mcal_{i-\half}| = |\Mcal_i-\Mcal_{i-1}|$ and $|\delta \Mcal_{i+\half}| = |\Mcal_{i+1}-\Mcal_{i}|$ (with complex magnitudes). $\Mcal_{i-1}$ is used for $\Mcal_n$ if $|\delta \Mcal_{i-\half}| <|\delta \Mcal_{i+\half}|$, and $\Mcal_{i+1}$ is used otherwise.

Even once we decide which pair of values to interpolate between, we can imagine many different interpolation schemes between them. Before we describe the interpolation scheme we chose, we will first describe a few seemingly intuitive approaches and describe why they do not suit our needs. First, one could create an interpolating scheme between $\Mcal_i$ and $\Mcal_{n}$ with a linear function for the real and imaginary parts separately. This would have the advantage that cell-centered quantities would represent both the values of the moments at the centers and the true cell-averaged quantities. However, this scheme does not in general preserve the monotonicity of the \emph{magnitudes} of the complex numbers; for example, in the case that $\Mcal_i$ and $\Mcal_n$ have the same magnitude but lie in opposite quadrants of the complex plane, the midpoint between them will have a smaller magnitude. Second, one could consider the interpolating function to be $\Mcal_i^t\,\Mcal_n^{1-t}$ with $0 \leq t \leq 1$. The choice $t=1/2$ bisects the phase angle but yields a \cnumber with a magnitude closer to the smaller of the two magnitudes $|\Mcal_i|$ and $|\Mcal_n|$: a value for $t$ can be found, which yields a \cnumber with a magnitude halfway between $|\Mcal_i|$ and $|\Mcal_n|$, but the phase will be closer to whichever complex number has the largest magnitude. Another idea would be to create a linear function for the magnitude and phase separately, which also generally yields points which are not equally distant from the two points we interpolate between. Third, one could create a circle in the complex plane that intersects all three of the points $\Mcal_i$ and $\Mcal_{i\pm1}$ however, this would break from the spirit of the minmod approach we use to maximize the stability of the system. Finally, one could use a standard high-order reconstruction method like ENO or WENO algorithms. This could be advantageous in the long run, since these methods have well-defined stability and convergence properties, but are generally less stable than minmod reconstruction schemes, and thus would require significant additional development (in particular regarding the generalization to complex quantities). We choose to maintain the minmod-like approach to maximize the stability of the code at the cost of accuracy for a given grid resolution.

The interpolation scheme we adopt uses two different methods of interpolation 
in the complex plane: a line and circular arc.  We choose the particular interpolation method based on the three values of the arguments $\phi_{i-1}, \phi_{i}$ and $\phi_{i+1}$ of $\mathcal{M}_{i-1}, \mathcal{M}_i$, and $\mathcal{M}_{i+1}$ respectively, such that the resulting interpolated point is equidistant from the original points. 
Let $\delta\phi_{i+\half}=\phi_{i+1}-\phi_i$ and $\delta\phi_{i-\half}=\phi_i-\phi_{i-1}$.  If the following conditions hold
\begin{align}
\delta\phi_{i+\half} &= 0 \ {\rm modulo}\,\pi,\quad {\rm and}\label{eq:line_cond1}\\
\delta\phi_{i-\half} &= 0 \ {\rm modulo}\,\pi,\label{eq:line_cond2}
\end{align}
then $\mathcal{M}_{i-1}, \mathcal{M}_i, \mathcal{M}_{i+1}$ all lie on a straight line in the complex plane passing through the origin i.e. they all have the same phase (modulo $\pi$), and so we can attempt to interpolate on that line.  If the conditions in Eqs.\ \eqref{eq:line_cond1} and \eqref{eq:line_cond2} do not hold, then we proceed to the arc interpolation algorithm.

\subsubsection{Interpolating on a line}
Even if the three complex numbers fall on a line in the complex plane, interpolation is not as straightforward as one might expect. To interpolate we first need to test whether the numbers are “ordered" along the line. We do this by comparing the magnitudes of $\delta \Mcal_{i-\half}$, $\delta \Mcal_{i+\half}$ and $\Mcal_{i+1}-\Mcal_{i-1}$. If 
\beq\label{eq:line_safety}
|\delta \Mcal_{i-\half}|,|\delta \Mcal_{i+\half}|<|\Mcal_{i+1} - \Mcal_{i-1}|
\eeq
then the three moments are “ordered" and one can regard $\mathcal{M}_i$ as being “between" $\mathcal{M}_{i-1}$ and $\mathcal{M}_{i+1}$. If the above test fails, then we set the moment values at the left and right sides of the cell equal to the value in the cell center, i.e., $\Mcal^{(R)}_{i-\half}=\Mcal^{(L)}_{i+\half} =\Mcal_i$. If the test in Eq.\ \eqref{eq:line_safety} passes, then we adopt the minmod interpolation as usual.  
We first find the minimum of the moduli $|\delta \Mcal_{i-\half}|$ and $|\delta \Mcal_{i+\half}|$, i.e.
\begin{equation}
  \Delta \equiv {\rm min}\left(|\delta \Mcal_{i-\half}|,|\delta \Mcal_{i+\half}|\right) \, ,
\end{equation}
and then assign the signed modulus of the moment at the left and right sides of the cell as 
\begin{subequations}
\label{eq:line_smod}
\begin{align}
  \smod[\Mcal_{i+\half}^{(L)}] &= \smod[\Mcal_i] + \frac{\Delta}{2} \, , \\
  \smod[\Mcal_{i-\half}^{(R)}] &= \smod[\Mcal_i] - \frac{\Delta}{2} \, , \\  
  \arg[\Mcal_{i+\half}^{(L)}] &= \arg[\Mcal_{i-\half}^{(R)}] = \arg[\Mcal_i] \, .
\end{align}
\end{subequations}
For the energy density, we can write the same equation with the convention $\smod[E] = |E| \geq 0$ [recall Eq.~\eqref{eq:smod_def}]. However, we must then ensure that the interpolated values we obtain using the equations above are non-negative. For this reason we institute an additional check to ensure the energy density modulus remains positive. If, at any interface between cells, Eq.~\eqref{eq:line_smod} leads to $\smod[E^{(S)}]<0$, where $S \in \{L,R\}$, then we invert the sign of the modulus add $\pi$ to the phase, that is
\begin{multline}
    E^{(S)} = - |\smod[E^{(S)}] | e^{\imath \arg[E^{(S)}]} \\ \text{i.e.} \quad \left\{
    \begin{aligned}  |E^{(S)}|& \leftarrow |\smod[E^{(S)}]| \, , \\ \arg[E^{(S)}]& \leftarrow \arg[E^{(S)}] + \pi \, . \end{aligned} \right.
\end{multline}

\begin{figure}[!ht]
    \centering      
    \includegraphics[width=\linewidth]{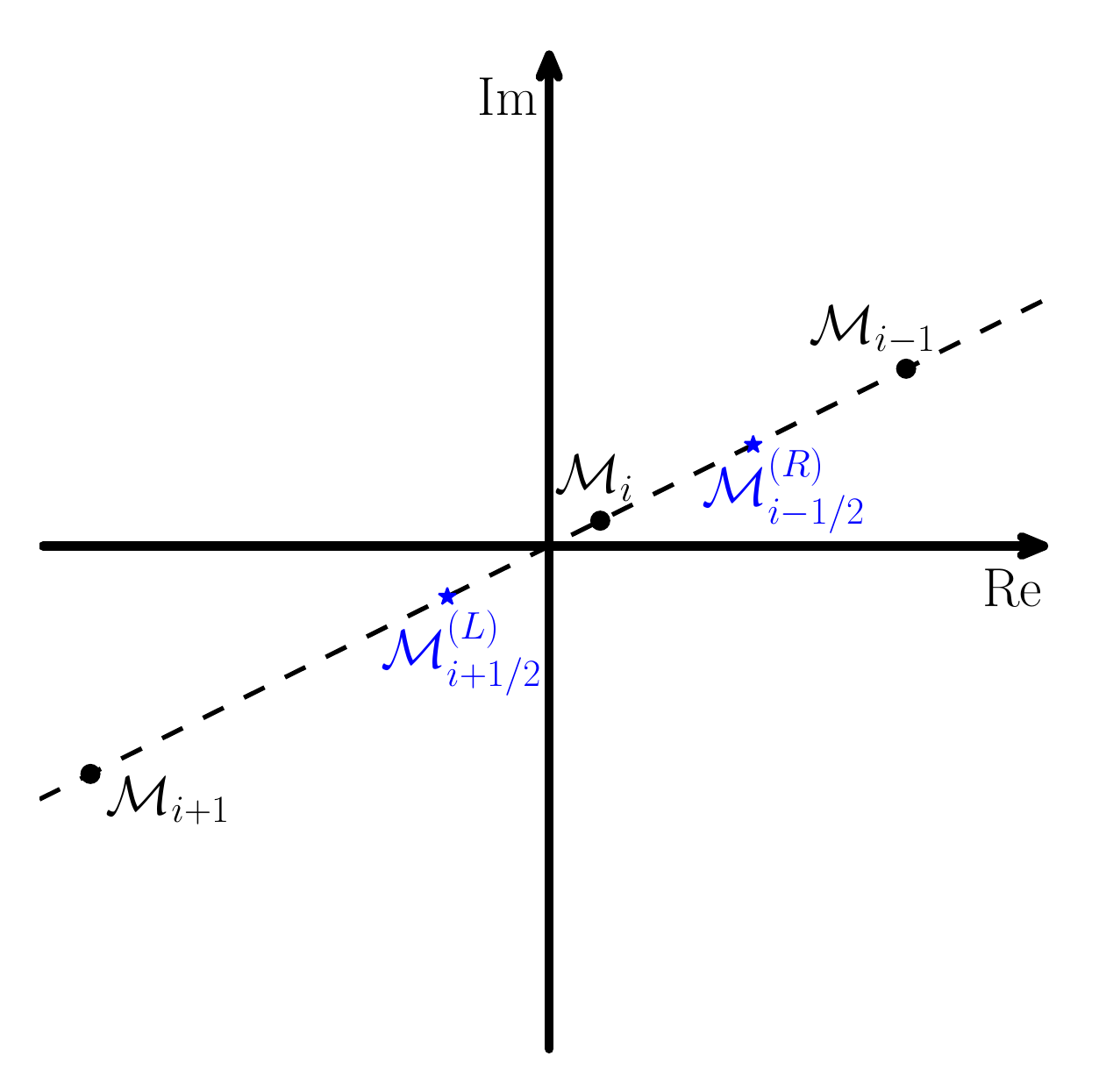}
    \caption{Example of line interpolation procedure (see text for explanation). 
    \label{fig:line}
    }
\end{figure}

Figure \ref{fig:line} shows a diagram of the line interpolation procedure. We use black dots to locate three points in the complex plane representing the moment values $\Mcal_{i-1},\Mcal_i,\Mcal_{i+1}$. Blue stars in Fig.\ \ref{fig:line} give the interpolated values $\Mcal_{i+\half}^{(L)}$ and $\Mcal_{i-\half}^{(R)}$. For the example drawn here, the “nearest" neighbor to $\Mcal_i$ is $\Mcal_{i-1}$ and so the actual interpolated  value used is the one between this pair. In this example $\smod[\Mcal^{(R)}_{i-\half}]$ will have the same sign as $\smod[\Mcal_i]$ and the phases will be identical. However $\Mcal^{(L)}_{i+\half}$ is on the “other side" of the origin in the complex plane from $\Mcal_i$. If $\Mcal_i$ is a flux, then $\smod[\Mcal^{(L)}_{i+\half}]$ will have the opposite sign as compared to $\smod[\Mcal_{i}]$, but the \emph{same} phase.  On the other hand, if $\Mcal_i$ is an energy density, then $|\Mcal^{(L)}_{i+\half}|=|\Mcal_{i}|$ but $\arg[\Mcal^{(L)}_{i+\half}]=\arg[\Mcal_i]+\pi$.

\subsubsection{Interpolating on a circular arc}

Interpolation on the line in the previous section is a 2D problem rotated into 1D where we must take care to assign the correct modulus and corresponding phase.  If the differences of the phases (modulo $\pi$) of the three moments are not zero then we consider interpolation along a circular arc (even if the three moments $\mathcal{M}_{i-1}, \mathcal{M}_i$, and $\mathcal{M}_{i+1}$ are colinear on a line which does not pass through the origin). Since interpolation along a circular arc is a more complicated algorithm, it is useful to begin with a figure so that the reader can visualize the quantities that we describe in the algorithm.
\begin{figure}[!ht]
    \centering      
    \includegraphics[width=0.8\linewidth]{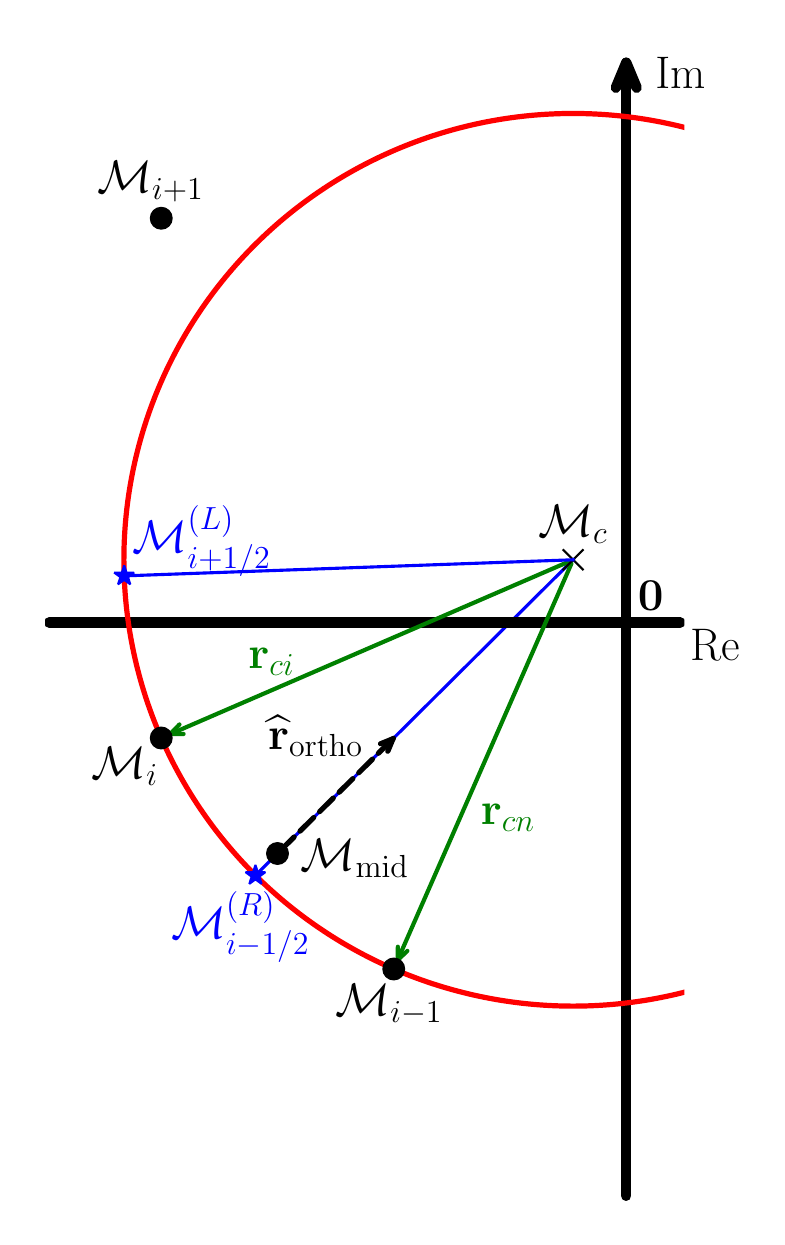}
    \caption{Example diagram of circular arc interpolation in the complex plane. See manuscript text for explanation.
    \label{fig:circle}
    }
\end{figure}

\begin{figure}
    \centering      
    \includegraphics[width=0.8\linewidth]{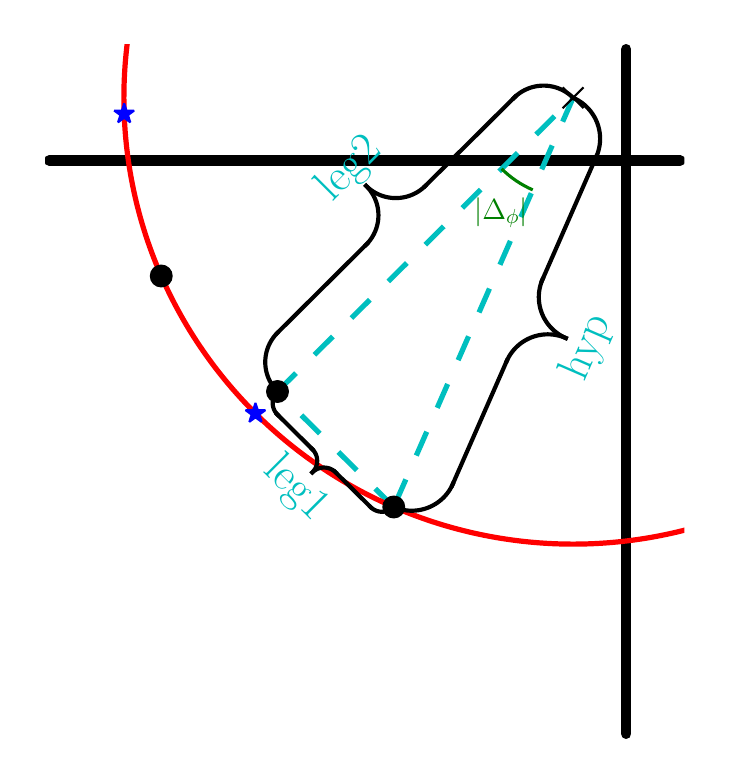}
    \caption{Definition of some quantities in the same example as Fig.~\ref{fig:circle}.
    \label{fig:triangle}
    }
\end{figure}
Figure \ref{fig:circle} shows an example of the circular arc interpolation procedure. We use black dots to locate three points $\Mcal_{i-1}, \, \Mcal_i, \, \Mcal_{i+1}$. 
Let us outline the steps of the algorithm before providing the details. 
We first consider four initial tests to ensure monotonicity and avoid null values.  If those four tests pass, we construct a circular arc through $\Mcal_{i}$ and $\Mcal_n$ (we recall that $\Mcal_n$ is either $\Mcal_{i+1}$ or $\Mcal_{i-1}$ such that $\lvert \Mcal_n - \Mcal_i\rvert$ is minimal). For the example drawn on Fig.~\ref{fig:circle}, $\Mcal_n = \Mcal_{i-1}$. Using those two points we then find the point $\Mcal_c$ which defines the center of the circle arc. Next we find the vectors $\vec{r}_{ci}$ and $\vec{r}_{cn}$ which point from $\Mcal_c$ to $\Mcal_i$ and $\Mcal_n$ respectively, and then we rotate the vectors around $\Mcal_c$ to obtain candidate values $\Mcal^{(R)}_{i-\half}$
and $\Mcal^{(L)}_{i+\half}$, see Eq.~\eqref{eq:candidate_values}. We call these “candidate” values because we consider four more tests to ensure our interpolated moment values maintain monotonicity and are not too extreme. Finally, if those additional tests are passed, we assign a sign to the flux moduli, see Eq.~\eqref{eq:signed_mod_rule}.

The four initial tests determine whether we are able to use this interpolation procedure on the circular arc.  First, we check the differences in the signed moduli. The sequence $\{\smod[\Mcal_{i-1}], \, \smod[\Mcal_i], \, \smod[\Mcal_{i+1}]\}$ must be monotonically increasing or decreasing.  We do a similar test for the phases, with the important qualification that we add/subtract $2\pi$ if the phases cross the dividing angle. The third check is to ensure that all three phases lie in the same half plane of the cartesian axis. Finally, we do a check to ensure there are no null values in the differences of the signed moduli and the phases.  If any of these tests fail, we set 
$\Mcal^{(R)}_{i-\half}=\Mcal^{(L)}_{i+\half} =\Mcal_i$, similar to the case when we interpolate along the line.

If all four tests pass, then we can proceed to drawing an arc between $\Mcal_i$ and $\Mcal_n$. 
\begin{enumerate}
  \item Find the midpoint between $\Mcal_i$ and $\Mcal_n$, calling it $\Mcal_{\rm mid}$
  \beq
    \Mcal_{\rm mid} \equiv \half(\Mcal_i + \Mcal_n) \, .
  \eeq  
  \item Find the distance between this midpoint and $\Mcal_i$
  \beq
    {\rm leg}_1=|\Mcal_i-\Mcal_{\rm mid}| \, .
  \eeq
  ${\rm leg}_1$ comprises the first leg of a right triangle.
  
  \item Set the hypotenuse of the right triangle to be the average of the moment magnitudes
  \beq
    {\rm hyp} = \half(|\Mcal_i| + |\Mcal_n|)
  \eeq
  The hypotenuse will serve as a radius for the circle.
  
  \item Using ${\rm hyp}$ and ${\rm leg}_1$, calculate the other leg of the right triangle
  \beq
    {\rm leg}_2 = \sqrt{{\rm hyp}^2 - {\rm leg}_1^2}.
  \eeq

  \item Find the unit modulus complex number $\widehat{\vec{r}}_{\rm ortho}$ orthogonal to $\Mcal_i - \Mcal_n$, labeled as\footnote{Note that we use the notation $\Mcal_{\dots}$ for the moments and interpolated quantities, which are complex numbers represented in the cartesian $\mathbb{R}^2$ plane (see Fig.~\ref{fig:circle}). We use the equivalent vector notation $\mathbf{r}_{\dots}$ for \emph{differences} between these complex numbers, which only serve as intermediaries to go from one physical quantity $\Mcal_{\dots}$ to another.} $\widehat{\vec{r}}_{\rm ortho}$
  \beq
    \widehat{\vec{r}}_{\rm ortho} = \frac{{\rm Im}(\Mcal_i - \Mcal_n) - \imath{\rm Re}(\Mcal_i-\Mcal_n)}{|\Mcal_i-\Mcal_n|}
  \eeq

  \item Orient the unit modulus complex number $\widehat{\vec{r}}_{\rm ortho}$ to be pointing in the half-plane where the cartesian axis origin is located
  \begin{align}
    \widehat{\vec{r}}_{\rm ortho} \rightarrow -{\rm sgn}[\hphantom{+}&{\rm Re}(\widehat{\vec{r}}_{\rm ortho}){\rm Re}(\Mcal_{\rm mid})\nonumber\\
    +& {\rm Im}(\widehat{\vec{r}}_{\rm ortho}){\rm Im}(\Mcal_{\rm mid})] \, \widehat{\vec{r}}_{\rm ortho}
  \end{align}

  \item Find the center of the circle with radius equal to the hypotenuse which passes through $\Mcal_i$ and $\Mcal_n$  
  \beq
    \Mcal_c =\Mcal_{\rm mid} + {\rm leg}_2\,\widehat{\vec{r}}_{\rm ortho}.
  \eeq

  \item Find the displacements from $\Mcal_c$ to $\Mcal_i$ and $\Mcal_n$
  \begin{equation}
  \begin{aligned}
    \vec{r}_{ci} &= \Mcal_i - \Mcal_c,\\
    \vec{r}_{cn} &= \Mcal_n - \Mcal_c.
  \end{aligned}
  \end{equation}

  \item Find the polar angles of $\vec{r}_{ci}$ and $\vec{r}_{cn}$
  \begin{equation}
  \begin{aligned}
    \phi_{ci} &= \arg[\Mcal_i-\Mcal_c],\\
    \phi_{cn} &= \arg[\Mcal_n-\Mcal_c].    
  \end{aligned}
  \end{equation}

  \item Find the angle between $\vec{r}_{ci}$ and $\vec{r}_{cn}$ and divide it by two
  \begin{equation}
    \Delta_{\phi} \equiv \half\begin{cases}
    \phi_{ci} - \phi_{cn} \ \,{\rm if}\,\,n=i-1,\\
    \phi_{cn} - \phi_{ci} \ \,{\rm if}\,\,n=i+1\\
    \end{cases}.
  \end{equation}
  It is possible that $\Delta_\phi$ could be negative, depending on the location of $\Mcal_n$ relative to $\Mcal_i$.

  \item Rotate $\vec{r}_{ci}$ by $\pm\Delta_\phi$ to obtain candidate values for the right and left sides of the cell in reference to the center of the circle
  \begin{equation}
  \begin{aligned}
    \vec{r}_{c,i+\half} & = e^{+\imath\Delta_\phi}\vec{r}_{ci} \, , \\
    \vec{r}_{c,i-\half} & = e^{-\imath\Delta_\phi}\vec{r}_{ci}.    
  \end{aligned}
  \end{equation}

  \item Add $\Mcal_c$ to $\vec{r}_{c,i\pm\half}$ to obtain the unsigned-modulus test candidate values for the moment at the left/right sides of the cell $i$ as 
  \begin{equation}
  \label{eq:candidate_values}
  \begin{aligned}
    \Mcal^{(R)}_{i-\half} &= \Mcal_c + \vec{r}_{c,i-\half},\\
    \Mcal^{(L)}_{i+\half} &= \Mcal_c + \vec{r}_{c,i+\half}.
  \end{aligned}
  \end{equation}

  \item Finally, we calculate the phases and unsigned moduli of $\Mcal^{(R)}_{i-\half}$ and $\Mcal^{(L)}_{i+\half}$.
 
\end{enumerate}
As we stated in the outline for the algorithm, the goal of the above procedure is to find \emph{candidate} values for the reconstructed moments on the left and right sides of the $i$'th cell and put those into the form of an absolute modulus and polar angle.  Like the minmod procedure in 1D, the above procedure uses the moment $\Mcal_i$ and either $\Mcal_{i-1}$ or $\Mcal_{i+1}$ in the interpolation. Unlike minmod, however, there is no guarantee that the above procedure preserves monotonicity. Therefore, after obtaining candidate values for the particular moment under consideration, we do four tests on both $\Mcal^{(R)}_{i-\half}$ and $\Mcal^{(L)}_{i+\half}$ to determine whether we will accept the candidate reconstructed moments 
or whether we will choose other values.

Let us first consider $\Mcal^{(L)}_{i+\half}$.
The first test examines whether the candidate modulus is too extreme compared to the signed moduli.  If the candidate modulus is too different from $|\Mcal_i|$ compared to the difference of signed moduli between $\Mcal_{i}$ and $\Mcal_{i+1}$, we set the unsigned modulus to the value in cell $i+1$. In other words,
\begin{multline}
\left\lvert |\Mcal^{(L)}_{i+\half}| - |\Mcal_i| \right\rvert > \Big| \smod[\mathcal{M}_{i+1}] - \smod[\mathcal{M}_{i}] \Big| \\
  \implies \quad |\Mcal^{(L)}_{i+\half}| \leftarrow |\Mcal_{i+1}| \, .\label{eq:sc_1}
\end{multline}
We use the difference in signed moduli in the above check as it is designed to identify values which are too extreme. If the test in Eq.\ \eqref{eq:sc_1} passes, then we check for monotonicity in the unsigned moduli. If this second test fails, the modulus of $\Mcal_{i+\half}^{(L)}$ is set to $|\Mcal_{i}|$. Namely,
\begin{multline} 
\mathrm{sgn}\left(|\Mcal^{(L)}_{i+\half}| - |\Mcal_i|\right) \neq   \mathrm{sgn}\left(|\Mcal_{i+1}| - |\Mcal^{(L)}_{i+\half}|\right) \\
  \implies \quad \left\lvert\Mcal^{(L)}_{i+\half}\right\rvert \leftarrow |\Mcal_{i}| \, .\label{eq:sc_2}
\end{multline}
If the first test were to fail, the logic in Eq.\ \eqref{eq:sc_1} acts to set the modulus value to that of cell $i+1$.  If the first test were to pass but the second test failed, \eqref{eq:sc_2} sets the modulus to that of cell $i$.  If both tests pass, then we accept the test candidate modulus.

The third and fourth checks are the phase analogs to the checks in Eq.\ \eqref{eq:sc_1} and \eqref{eq:sc_2}.  
We check for extremeness in the phase
\begin{multline}
\left\lvert \arg[\Mcal^{(L)}_{i+\half}] - \arg[\Mcal_i] \right\rvert > \Big\lvert \arg[\mathcal{M}_{i+1}] - \arg[\mathcal{M}_{i}]  \Big\rvert \\
   \implies \quad \arg[\Mcal^{(L)}_{i+\half}] \leftarrow \arg[\Mcal_{i+1}] \, .
\end{multline}
If this test passes, the fourth check tests for monotonicity
\begin{multline}
\mathrm{sgn}\left(\arg[\Mcal^{(L)}_{i+\half}] - \arg[\Mcal_i]\right) \\ \neq \mathrm{sgn}\left(\arg[\Mcal_{i+1}] - \arg[\Mcal^{(L)}_{i+\half}]\right)   \\
  \implies \quad \arg[\Mcal^{(L)}_{i+\half}] \leftarrow \arg[\Mcal_{i}] \, .
\end{multline}
If the third and fourth tests pass, then we accept the candidate value for the phase.
The above four checks correspond to the left side of the $i+\half$ interface, i.e., the right side of cell $i$.  We repeat the four checks for the left side of cell $i$ with the permutations $i+1\rightarrow i-1$ and $\Mcal^{(L)}_{i+\half}\rightarrow\Mcal^{(R)}_{i-\half}$. 

Once we have the moduli and phases of the reconstructed moments, the final step in the procedure is to assign signs to the \emph{signed moduli} if the moment in question is a flux.  Without loss of generality, let us assume that we interpolated on the $n=i-1$ side.
Then the right side inherits the sign from $\mathcal{M}_i$, i.e., $\smod[\mathcal{M}^{(L)}_{i+\half}]\leftarrow {\rm sgn}\left(\smod[\mathcal{M}_i]\right)|\Mcal^{(L)}_{i+\half}|$.
For the left side, $\Mcal^{(R)}_{i-\half}$ inherits the sign from the point with the \emph{closest} absolute modulus.  In other words,
\begin{widetext}
\begin{equation}
\label{eq:signed_mod_rule}
  \smod[\mathcal{M}^{(R)}_{i-\half}] \longleftarrow |\Mcal^{(R)}_{i-\half}|
  \times\begin{cases}
  {\rm sgn}\left(\smod[\mathcal{M}_{i-1}]\right) &{\rm if}\quad\Big||\Mcal_{i-1}| - |\Mcal^{(R)}_{i-\half}|\big|<\Big||\Mcal_i| - |\Mcal^{(R)}_{i-\half}|\Big|\\
  {\rm sgn}\left(\smod[\mathcal{M}_{i}]\right) &{\rm if}\quad\big||\Mcal_{i-1}| - |\Mcal^{(R)}_{i-\half}|\big|\geq \Big||\Mcal_i| - |\Mcal^{(R)}_{i-\half}|\Big|
  \end{cases}.
\end{equation}
\end{widetext}
If the signed moduli change signs, i.e., if we assign a minus sign to either $\Mcal^{(S)}_{i\pm\half}$ where $S \in \{L,R\}$, we add $\pi$ to the phase so as to preserve the complex number calculated in the arc-interpolation procedure.  This ends the arc-interpolation procedure, and the cell interface reconstruction algorithm.

Figures \ref{fig:circle} and \ref{fig:triangle} give an illustration of the algorithm.
As the nearest neighbor to $\Mcal_i$ is $\Mcal_{i-1}$, the interpolation is done using this pair.
$\Mcal_{\rm mid}$ is the midpoint between $\Mcal_i$ and $\Mcal_n = \Mcal_{i-1}$.
The dashed line connecting $\Mcal_{\rm mid}$ and $\Mcal_{i-1}$ on Fig.~\ref{fig:triangle} has a length ${\rm leg}1$.
We calculate the unit vector orthogonal to $\Mcal_i - \Mcal_n$ and locate it at $\Mcal_{\rm mid}$.
From $\Mcal_{\rm mid}$, we scale $\widehat{\vec{r}}_{\rm ortho}$ by ${\rm leg}2$ to find $\Mcal_c$, denoted by a black cross.
$\Mcal_c$ is the center of a circle, with a radius given by the hypotenuse of the right triangle with legs ${\rm leg}_1$ and ${\rm leg}_2$.
The angle subtended between ${\rm leg}_2$ and ${\rm hyp}$ on Fig.\ \ref{fig:triangle} is $|\Delta_\phi|$, where the sign of $\Delta_\phi$ indicates the direction of rotation.
From $\Mcal_c$, we draw the vectors to $\Mcal_i$ and $\Mcal_{i-1}$, indicated on Fig.~\ref{fig:circle} by the green vectors $\vec{r}_{ci}$ and $\vec{r}_{cn}$, respectively.
We rotate $\vec{r}_{ci}$ by $\pm\Delta_\phi$ to obtain the vectors $\vec{r}_{c,i\pm\half}$.
When we add $\Mcal_c$ to $\vec{r}_{c,i\pm\half}$, we obtain the blue stars $\Mcal^{(S)}_{i\pm\half}$.
For the three points in Fig.\ \ref{fig:circle}, the sequence of moduli $\{|\Mcal_{i-1}|, \, |\Mcal_{i-\half}^{(R)}|, \, |\Mcal_i|, \, |\Mcal_{i+\half}^{(L)}|, \, |\Mcal_{i+1}|\}$
is monotonic.  The same applies for the phase across the dividing angle $\pi$.  Therefore, both $\Mcal^{(L)}_{i+\half}$ and $\Mcal^{(R)}_{i-\half}$ pass all four checks and we accept the candidate values for unsigned modulus and phase. If $\Mcal$ is a flux factor, then $\smod[\Mcal^{(L)}_{i+\half}]$ will have the same sign as $\smod[\Mcal_{i}]$.  For the left side of the cell, $\smod[\Mcal^{(R)}_{i-\half}]$ would inherit the sign from $\smod[\Mcal_{i-1}]$, since $\lvert\Mcal^{(R)}_{i-\half}\rvert$ is closer to $\lvert \Mcal_{i-1} \rvert$ than to $\lvert \Mcal_i \rvert$.

\subsubsection{Difference between circular arc and minmod}

The cell interface reconstruction algorithm gives moment values on the left and right sides of the $i^{\rm th}$ cell. In this section we give a brief interlude on the difference between the circular-arc procedure for the modulus-phase implementation, and applying minmod in an implementation based on representing the moments with real-imaginary components — the algorithm we mentioned previously as being the one used in \cite{Grohs:2022fyq}.

\begin{figure}[!ht]
    \centering      
    \includegraphics[width=0.8\linewidth]{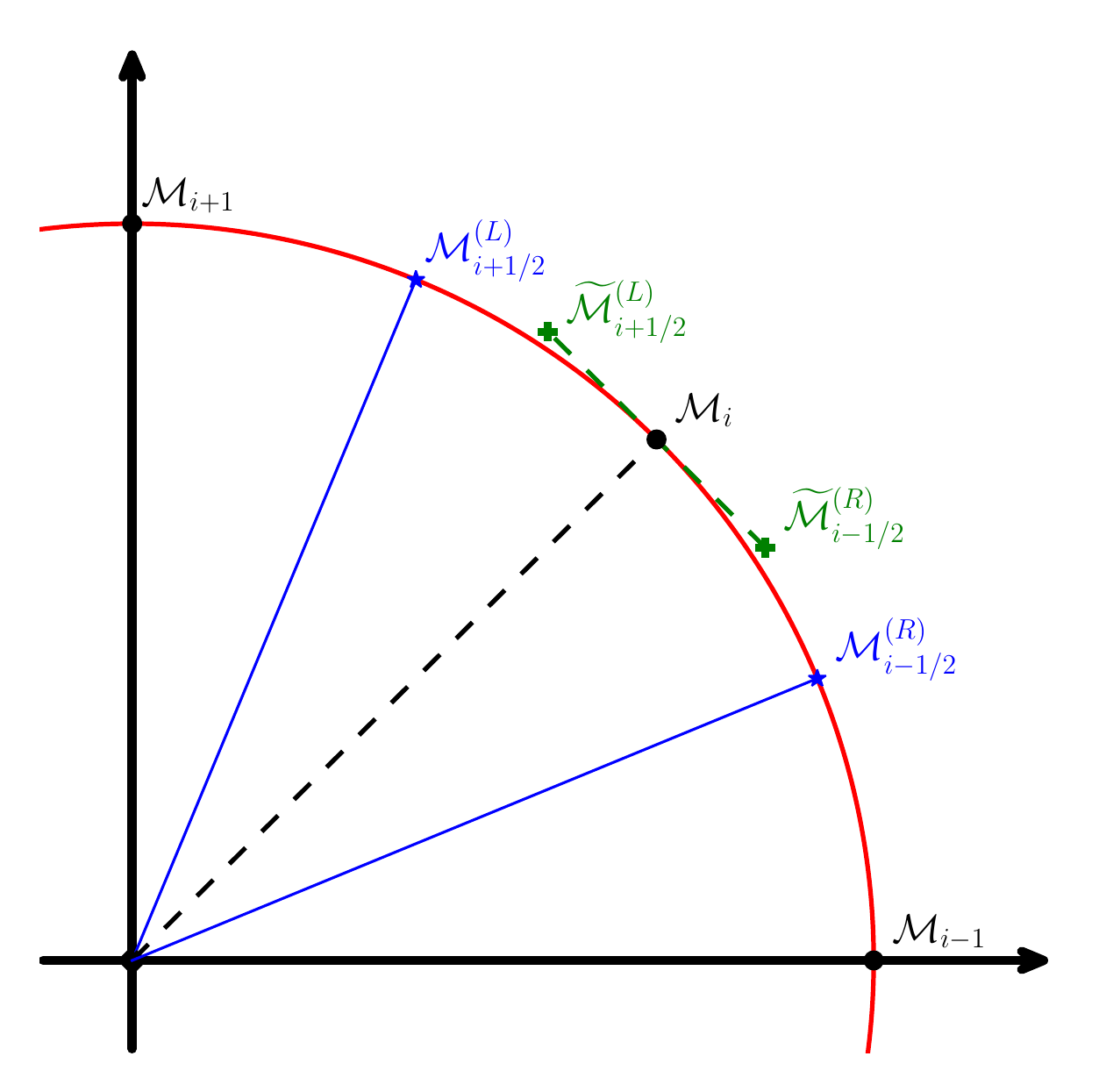}
    \caption{Diagram, with moments located on the unit circle, illustrating the difference between a component-wise interpolation using Minmod and our Modulus-Phase algorithm.
    \label{fig:mm_v_MP}
    }
\end{figure}

Figure \ref{fig:mm_v_MP} shows a diagram in the complex plane where we use the circular arc interpolation procedure and contrast that to a component-wise calculation using minmod.  We have picked the values $\Mcal_{i-1}, \, \Mcal_i, \, \Mcal_{i+1}$ to lie on the unit circle with increasing phase.  We have drawn $\Mcal_i$ with a polar angle of $\pi/4$
\begin{equation}
  \Mcal_i = \frac{1}{\sqrt{2}}(1,1) \, ,
\end{equation}
and set $\Mcal_{i-1}=(1,0)$ and $\Mcal_{i+1}=(0,1)$.
The circular arc interpolation gives candidate values also along the unit circle with phases $\pm\pi/8$ away from $\pi/4$, that is, 
\begin{equation}
\begin{aligned}
  \Mcal^{(R)}_{i-\half} & = \left(\cos\left(\frac{\pi}{4} + \frac{\pi}{8}\right),\sin\left(\frac{\pi}{4} - \frac{\pi}{8}\right)\right) \, , \\
  \Mcal^{(L)}_{i+\half} & = \left(\cos\left(\frac{\pi}{4} - \frac{\pi}{8}\right),\sin\left(\frac{\pi}{4} + \frac{\pi}{8}\right)\right)   \, .
\end{aligned}
\end{equation}
Conversely, we can separate the real and imaginary parts of $\Mcal_{i-1,i,i+1}$
and use minmod to find alternative reconstructed values for the moment, which we shall write as ${\widetilde{\Mcal}}^{(L/R)}$. To be explicit, transposing to $\mathrm{Re}(\Mcal)$ and $\mathrm{Im}(\Mcal)$, and using the interpolation described in Eq.~\eqref{eq:M_minmod_class}, we would get:
\begin{equation}
\begin{aligned}
  \mathrm{Re}(\widetilde{\Mcal}_{i-\half}^{(R)}) &= \frac{1}{\sqrt{2}} - \half{\rm minmod}\left(0-\frac{1}{\sqrt{2}}, \frac{1}{\sqrt{2}} - 1\right) \, , \\
  \mathrm{Im}(\widetilde{\Mcal}_{i-\half}^{(R)}) &= \frac{1}{\sqrt{2}} - \half{\rm minmod}\left(1-\frac{1}{\sqrt{2}}, \frac{1}{\sqrt{2}} - 0\right) \, ,
\end{aligned}
\end{equation}
and similarly for $\widetilde{\Mcal}_{i+\half}^{(L)}$ except the $-$ signs before minmod become $+$ signs. This would lead to the reconstructed moments:
\begin{equation}
\begin{aligned}
   \widetilde{\Mcal}_{i-\half}^{(R)} = \frac{1}{2\sqrt{2}}\left(1+\sqrt{2},
  3-\sqrt{2}\right) \, , \\
  \widetilde{\Mcal}_{i+\half}^{(L)} = \frac{1}{2\sqrt{2}}\left(3-\sqrt{2},
  1+\sqrt{2}\right)  \, .
\end{aligned}
\end{equation}
The green $+$ symbols on Fig.\ \ref{fig:mm_v_MP} show the values for both sides of the $i^{\rm th}$ cell if we use the minmod procedure on the real and imaginary parts of the moments. The reader can see they do not have unit magnitude and are closer to $\Mcal_i$ than $\Mcal_{i+1}$ and $\Mcal_{i-1}$. In contrast, the reconstructed moments using the circular arc are midway along the arcs.

We stress that neither interpolation procedure is correct nor incorrect.  We constructed the example in Fig.\ \ref{fig:mm_v_MP} to give incommensurate interpolated values for the two procedures.  However, this specific example does serve an important pedagogical purpose.  For the example drawn in Fig.\ \ref{fig:mm_v_MP}, the minmod-interpolated values produce moduli that are larger than the unit radius.  If we are reconstructing flux factors, then minmod is producing fluxes with larger magnitudes than the circular arc interpolator.  It may not matter directly since the interpolated values will be used to compute the spatial fluxes $\mathbb{F}$ and the pertinent quantity will be the difference of fluxes, not the stand-alone values themselves.  Nevertheless, the point remains that the component-wise minmod procedure does not guarantee conservation of the magnitude of a \cnumber in a case where we expect it.  We conjecture that this lack of conservation has led to larger advective derivatives of the moments between cells in previous works and hence faster rates of building/destroying coherence in the system, as observed in~\cite{flashri}.  The circular arc interpolation, on the other hand, can preserve the magnitude of a \cnumber.  Therefore, we hypothesize that the circular arc interpolation procedure will lead to smaller rates of creating/destroying coherence.  Section \ref{sec:results} will bear out this hypothesis.

\section{Implementing the Quantum Riemann Solver}
\label{sec:implementation}

We now proceed to implement the quantum Riemann algorithm we have developed. 
The previous implementation of moment neutrino transport with the \flash code, used in \cite{Grohs:2022fyq,flashri}, was a straightforward generalization of the classical transport algorithm. 
The flavor-space off-diagonal elements of the energy density and flux were represented as real-imaginary pairs and the advection routines were applied separately to the real and imaginary components. We will refer from now on to this former implementation as \flashri. 
This approach, albeit promising, was not fully satisfying. For example, $\mathrm{Re}(E_{ex})$ is not a nonnegative quantity, and the “flux factor” $\mathrm{Re}(F_{ex}^j)/\mathrm{Re}(E_{ex})$ is not limited to the range $[-1,1]$. To incorporate the use of signed modulus and phase as an alternative to real and imaginary components for \cnumbers we have rewritten a large amount of the \flash code, but the approach to evolving the equations remains the same as that described in~\cite{2015ApJS..219...24O,2018ApJ...854...63O}.

\paragraph*{Mitigating unphysical solutions —}

Our moment calculations advect real energy densities $E_{ee}$ and $E_{xx}$, and complex energy density coherence $E_{ex}$ (and analogous antineutrino densities). Since the energy density $E$ is positive definite these quantities are constrained by the requirement that $E_{ee}\,E_{xx} - |E_{ex}|^2 \geq 0$.
\edit{However, our advection scheme treats these three quantities separately without any consideration to the previous constraint. The possibility arises that the energy density moments could enter an unphysical solution space.  Implementing the constraint into the advection scheme a priori would require a species-coupled Riemann solver beyond the HLLE algorithm.  As an alternative, we opt for an a posteriori correction method to prevent against pathological solutions.  After solving for the dependent variables in the flavor-transformation subroutine, we check the size of $|E_{ex}|$ and decrease it if it violates the positive-definite requirement.}

To be explicit, we first define the length of the Bloch polarization vector as follows
\begin{equation}
  \ell \equiv\sqrt{|E_{ex}|^2 + \left(\frac{E_{ee} - E_{xx}}{2}\right)^2} \, . 
\end{equation}
The positive-definite requirement is equivalent to $\mathrm{Tr}(E) - 2\,\ell \geq 0$.  
If $2\,\ell>\mathrm{Tr}(E)$, then we modify $|E_{ex}|$ as follows
\begin{equation}
  |E_{ex}| \leftarrow |E_{ex}|\times\left[\sqrt{1 + \frac{[\mathrm{Tr}(E)/2]^2 - \ell^2}{|E_{ex}|^2}} - \varepsilon\right] \, , \\
\end{equation}
where $\varepsilon\sim10^{-9}$ is a small number to ensure the factor in square brackets is smaller than unity.  To preserve the flux factor, we reduce $F^{j}_{ex}$ by the same factor in square brackets.  We use an analogous procedure for $\overline{E}_{ex}$.  The phases for the energy and flux densities remain the same.

\section{Testing the Riemann solver}
\label{sec:results}

\begin{figure}[ht]
    \centering
    \includegraphics{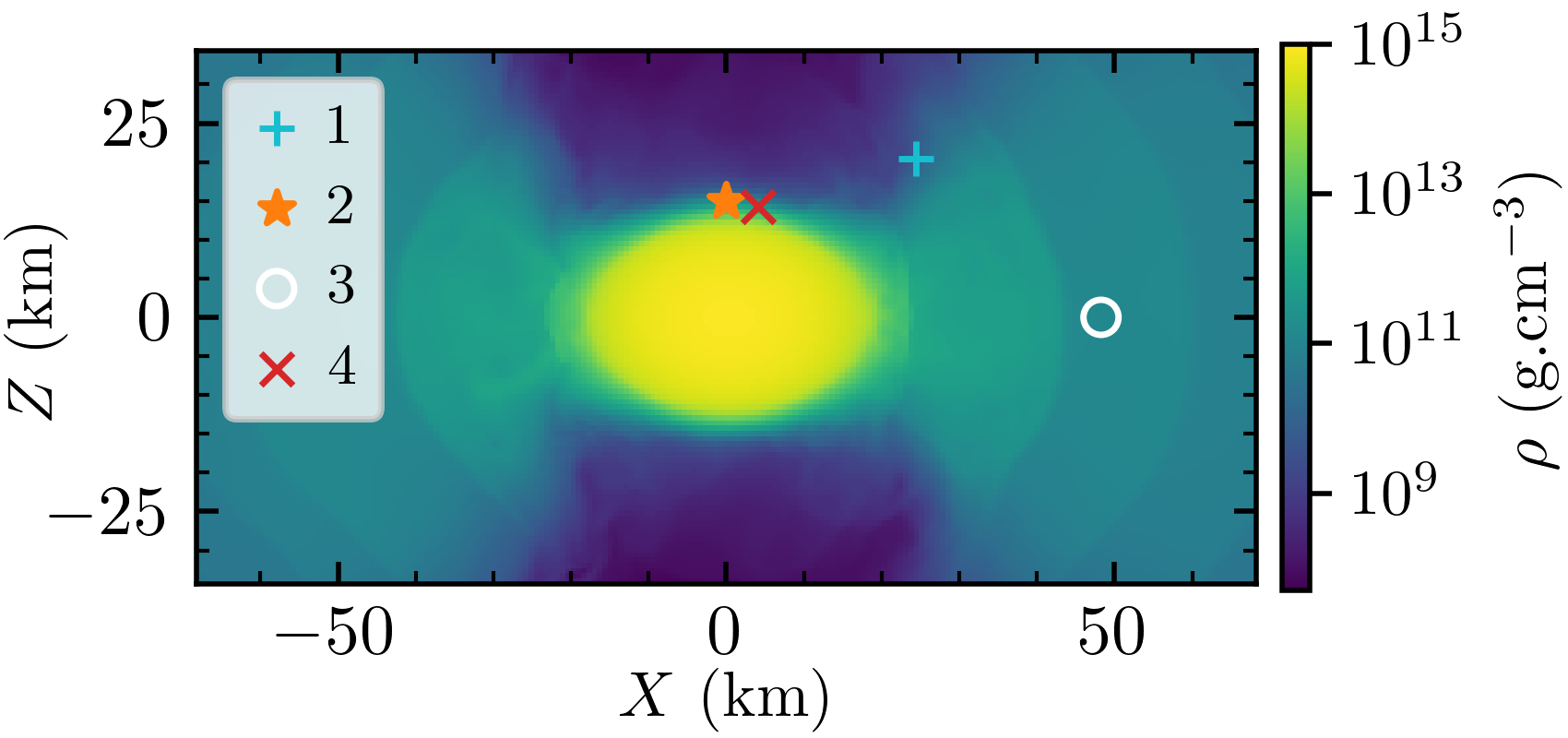}
    \caption{Poloidal slice of the 5 ms post-merger snapshot from the simulation of Ref.~\cite{Foucart:2016rxm}, showing the matter density $\rho$. We identify the four grid points studied in the text: NSM1 exhibits a deep ELN crossing, NSM2 and NSM4 exhibit shallow crossings and are located in the polar region, above the hypermassive neutron star remnant, and NSM3 lies in the equatorial plane. See Fig.~5 in~\cite{flashri} for the same plot showing the regions of fast flavor instability (without the NSM4 point).}
    \label{fig:locations_NSM}
\end{figure}

We present in Appendix~\ref{sec:beam} the results of one-dimensional calculations in “beam” configurations, for which an analytic solution exists. We show that our new algorithm leads to accurate results in these situations, but the genuine improvement compared to the \flashri algorithm can be better seen in three-dimensional calculations requiring a closure, which we present in this section. Our 3D tests are a combination of simplified models and calculations using data taken from simulation of the merger of two neutron stars (see Ref.~\cite{Foucart:2016rxm}). The three simplified test geometries are identical to those of Ref.\ \cite{flashri}: Fiducial, 90Degree, and TwoThirds.  We describe the simplified models in more detail in Appendix~\ref{app:fid_90d}, and defer the TwoThirds test to Sec.~\ref{ssec:2_3}.
We will consider 4 cases from different locations in a 5-ms post-merger snapshot of the NSM simulation, represented on Fig.~\ref{fig:locations_NSM}. Three are the same as the ones described of Ref.~\cite{flashri}, and called there NSM1, NSM2, NSM3. In addition, we include a fourth point which we label NSM4. It exhibits similar conditions to the NSM2 point and we include it here because we have found the results from the moment calculation are qualitatively different.  Appendix~\ref{app:NSM4} contains information on the ELN crossing at this particular location. Table~\ref{tab:NSM_parameters1} gives the initial number densities, flux factors, mass density, and electron fraction from the output of the simulation of Ref.\ \cite{Foucart:2016rxm}. 

We compare the numerical results of our quantum moment code, \flash, with multiangle calculations performed with the code \emu. \emu is an open-source multidimensional particle-in-cell neutrino quantum kinetics code~\cite{2021PhRvD.103h3013R}. It represents the neutrino radiation field using discrete computational particles, each of which represents a large number of physical neutrinos. Each particle stores two density matrices representing the quantum states of the neutrinos and antineutrinos contained in the particle. The background grid accumulates and averages the neutrino density and number flux in each grid cell, and each particle then interpolates from these quantities from the grid to form its own Hamiltonian. The particle positions are advected at the speed of light, and the density matrices are evolved according to the Schr\"odinger equation using a globally fourth order explicit time integrator.

We remind the reader that even with a perfect advection scheme in \flash we should still expect some differences between \flash and \emu due to a necessarily imperfect closure in moment-based methods. We hope to obtain a sufficiently accurate advection scheme so that the differences between \emu and \flash can provide an estimate of the efficacy of the closure.

\subsection{Setup}

\subsubsection{Simulation geometries and parameters}

Our 3D calculations are all done in a rectangular geometry using cartesian coordinates.  We have the flexibility to specify the number of grid points and domain length for each dimension.  Unlike Ref.\ \cite{flashri}, we do not limit our simulations to those with cubic geometries.  We will always pick the same length for each dimension in the simulation domain, but we will select a different number of grid points for each dimension.  Based on the procedure in Appendix B of Ref.\ \cite{flashri}, we orient the $\z$-axis parallel to the vector difference of the $e$-flavor and $\overline{e}$-flavor fluxes, i.e., the net ELN flux (we recall that, in all our simulations, the distributions of $\nu_x$ and $\bar{\nu}_x$ are initially identical).  We vary the number of grid points in the $\z$ direction to monitor convergence, while keeping the number of grid points in the $\x$ and $\y$ directions constant at 16.  In this manner, we are able to run simulations with finer resolution in the direction of the ELN flux, without incurring a large burden on computational cost.  We have verified that the rectangular geometries show nearly identical qualitative and quantitative behavior compared to simulations with cubic geometries.  We will use the rectangular geometry for all NSM and test simulation setups, except for NSM1.  For that particular point, we use a cubic geometry with 128 grid points on each side.

\paragraph*{Parameters for \emu simulations —} In all simulations comparing to \flash results, we assume an identical domain size and periodic boundary conditions. Our grid dimensions and number of particles per cell (ppc) vary for each simulation.
We use a grid of $128^3$ cells and ppc=1506 for an angular resolution of $5.6^\circ$ in the 3D Test Cases (Sec.~\ref{ssec:2_3} and Appendix~\ref{app:fid_90d}).
The NSM simulations in this section employ ppc=378, corresponding to an angular resolution of about 11.3$^\circ$.  Except for NSM4, the grid covers a 3D cube with a variable number of cells to ensure numerical resolution of the linear phase of the FFI.  For the NSM4 simulation, we use a 1D spatial calculation with 256 cells in the $\z$ direction.

We use periodic boundary conditions for all calculations. The domain side length $L$ and the number of grid points used for the \flash and \emu simulations are provided in Table~\ref{tab:NSM_parameters2}.

\renewcommand{\arraystretch}{1.3}

\begin{table*}
    \centering
    \begin{tabular}{c|cccccccc}
        \hspace{0.1cm} \multirow{2}{*}{Name} \hspace{0.1cm} & $\mathcal{N}_{ee}$ & $\overline{\mathcal{N}}_{ee}$ & $\sum \mathcal{N}_{(x)}$ & $\mathbf{f}_{ee}$ & $\overline{\mathbf{f}}_{ee}$ & $\mathbf{f}_{xx}=\overline{\mathbf{f}}_{xx}$ & $\rho$ & $Y_e$\\
        & $(10^{32} \, \mathrm{cm}^{-3})$ & $(10^{32} \, \mathrm{cm}^{-3})$ & $(10^{32} \, \mathrm{cm}^{-3})$ & &&& ($10^{10} \, \mathrm{g.cm^{-3}}$) & \\\hline
        NSM1 & 14.22 & 19.15 & 19.65
        & $\begin{pmatrix}\phantom{-}0.0974\\ \phantom{-}0.0421\\ -0.1343\end{pmatrix}$
        & $\begin{pmatrix}\phantom{-}0.0723\\ \phantom{-}0.0313\\ -0.3446\end{pmatrix}$
        & $\begin{pmatrix}-0.0216\\ \phantom{-}0.0743\\ -0.5354\end{pmatrix}$
        & 0.692 & 0.281\\
        NSM2 & 23.29 & 28.53 & 60.11
        & $\begin{pmatrix}\phantom{-}0.0086\\ -0.0174\\ -0.1635\end{pmatrix}$
        & $\begin{pmatrix}\phantom{-}0.0070\\ -0.0142\\ -0.2338\end{pmatrix}$
        & $\begin{pmatrix}-0.0476\\ -0.0231\\ -0.2679\end{pmatrix}$
        & 22.0 & 0.192\\
        NSM3 & 28.80 & 37.42 & 19.32
        & $\begin{pmatrix}\phantom{-}0.0004\\ -0.0033\\ \phantom{-}0.0044\end{pmatrix}$
        & $\begin{pmatrix}\phantom{-}0.0003\\ -0.0025\\ -0.1306\end{pmatrix}$
        & $\begin{pmatrix}-0.0008\\ -0.0051\\ -0.1292\end{pmatrix}$
        & 13.1 & 0.235\\
        NSM4 & 30.89 & 33.27 & 84.12
        & $\begin{pmatrix}\phantom{-}0.0213\\ -0.0142\\ -0.1301\end{pmatrix}$
        & $\begin{pmatrix}\phantom{-}0.0197\\ -0.0132\\ -0.1683\end{pmatrix}$
        & $\begin{pmatrix}\hphantom{-}0.0599\\ -0.0357\\ -0.2004\end{pmatrix}$
        & 55.3 & 0.132\\
    \end{tabular}\\
    \caption{Physical quantities associated to each NSM point, obtained from Ref.~\cite{Foucart:2016rxm}.  The first three columns show the number densities of each anti/neutrino flavor. For clarity, the third column shows the sum of all four heavy lepton anti/neutrino densities. Our two-flavor simulations assume $\mathcal{N}_{xx}=\overline{\mathcal{N}}_{xx}=\sum \mathcal{N}_{(x)}/4$, where the other half of the heavy-lepton flavor neutrinos do not participate in flavor mixing. The next three columns show the flux factor vectors, the norm of which are the flux factors. The seventh and eighth columns give the matter mass density and electron fraction, respectively.
    }
    \label{tab:NSM_parameters1}
\end{table*}

\begin{table}
    \centering
    \begin{tabular}{c|cc|cc}
        & \multicolumn{2}{c|}{\flash} & \multicolumn{2}{c}{\emu} \\ 
        $\ $ \multirow{2}{*}{Name} $\ $ & $L$ & $N_{gp}$ & $L$ & $N_{gp}$\\
        & (cm) & & (cm) & \\\hline
        NSM1 & 7.87 &  $128^3$ &7.87 & $128^3$\\
        NSM2 & 16.53 & $16\times16\times1024$ & 8.27 & $128^3$\\
        NSM3 & 5.80 & $16\times16\times512$ & 5.80 & $128^3$\\
        NSM4 & 24.52 & $16\times16\times256$ & 24.52 & $1\times1\times256$\\
    \end{tabular}\\
    \caption{List of baseline simulation parameters for the \flash and \emu NSM simulations. For each algorithm, the first column shows the length of each side of the domain and the second column the number of grid points ($N_{gp}$).
    }
    \label{tab:NSM_parameters2}
\end{table}

\subsubsection{Quantum kinetic equations}

In this section, the QKEs~\eqref{eq:qke_general} are solved at the mean-field level (no collisions, $\mathcal{C} = 0$), and the Hamiltonian-like operator includes three terms: $H = H_V + H_M + H_\nu$. The vacuum Hamiltonian $H_V$ results from the noncongruency of the flavor and mass eigenbases
\begin{subequations}
\begin{equation}\label{eq:h_v}
    H_V \equiv \frac{1}{2\,p}\, U \, \mathbb{M}^2 \, U^\dagger \, ,
\end{equation}
with
\begin{equation}
    U = \begin{pmatrix} \cos \theta & \sin \theta \\ - \sin \theta & \cos \theta \end{pmatrix} \ , \quad  \mathbb{M}^2 = \begin{pmatrix} 0 & 0 \\ 0 & \delta m^2 \end{pmatrix} \, ,
\end{equation}
\end{subequations}
where $\theta$ is the mixing angle and $\delta m^2$ the mass-squared difference between the two neutrino mass eigenstates. The matter mean-field term $H_M$, related to the flavor-dependent weak interactions with background electrons and positrons, reads
\begin{equation}
    H_M \equiv \sqrt{2}\, G_F\, n_e \begin{pmatrix} 1 & 0 \\ 0 & 0 \end{pmatrix} \, ,
\end{equation}
where $G_F \simeq 1.166 \times 10^{-11} \, \mathrm{MeV}^{-2}$ is the Fermi constant, and $n_e$ is the difference of electron and positron number densities. This expression is valid in the frame comoving with the fluid, in which we will work in this paper. Given the matter density $\rho$ and the electron fraction $Y_e$, the number density $n_e$ is given by
\begin{equation}
    n_e = Y_e \frac{\rho}{m_u} \, ,
\end{equation}
with $m_u \simeq 1.661 \times 10^{-24} \, \mathrm{g}$ the atomic mass unit. Finally, the self-interaction mean-field term is
\begin{equation}\label{eq:hnu_dens_mat}
    H_\nu \equiv \frac{\sqrt{2} G_F}{(2 \pi)^3} \int{\dd^3 \vec{q} \, (1- \cos \vartheta) \left[\varrho(t,\vec{x},\vec{q})-\bar{\varrho}^*(t,\vec{x},\vec{q})\right]} \, ,
\end{equation}
where $\vartheta$ is the angle between the free variable $\vec{p}$ and the integration variable $\vec{q}$. The source term for antineutrinos is similar, with $\overline{H} = H_V - H_M - H_\nu^\ast$.

With this Hamiltonian the right-hand sides of the moment QKEs~\eqref{eq:moment_QKEs} then read in this situation:
\begin{subequations}
\label{eq:moment_QKEs_3D}
\begin{align}
    \mathcal{S}_E &= - \imath \left[H_V + H_M + H_E, E\right] + \imath \left[H_{F,j},F^j\right] \, , \\
    \mathcal{S}_F^j &= - \imath \left[H_V + H_M + H_E, F^j\right] + \imath \left[H_{F,k},P^{jk}\right] \, , \\
    {\overline{\mathcal{S}}}_E &= - \imath \left[H_V - H_M - H_E^*, \bE\right] - \imath \left[H_{F,j}^*,\bF^j\right] \, , \\
    {\overline{\mathcal{S}}}_F^j &= - \imath \left[H_V - H_M - H_E^*, \bF^j\right] - \imath \left[H_{F,k}^*,\bP^{jk}\right] \, ,
\end{align}
\end{subequations}
where the self-interaction Hamiltonian~\eqref{eq:hnu_dens_mat} has been rewritten as $H_\nu = H_E - \hat{\vec{p}}\cdot \vec{H_F}$ with
\begin{align}
    H_E &\equiv \sqrt{2} G_F \left(\N - \bcN^*\right) \, , \\
    H_F^j &\equiv \sqrt{2} G_F \left(\F^j - \bcF^{j*}\right) \, .
\end{align}
The calligraphic quantities $\N$ and $\F$ are the energy-integrated number density and number flux moments defined to be:
\begin{equation}
\label{eq:energy_integrated_moments}
\begin{aligned}
    \N(t,\vec{x}) &\equiv  \int{\dd p \, N(t,\vec{x},p)}  = \int{\dd p \, \frac{E(t,\vec{x},p)}{p}} \, , \\
    \F^j (t,\vec{x}) &\equiv \int{\dd p \, \frac{F^j(t,\vec{x},p)}{p}} \, .
\end{aligned}
\end{equation}

\subsubsection{Closure} 
\label{subsec:closure}

To close the system of equations~\eqref{eq:moment_QKEs_3D}, we need to express the pressure moments as a function of the energy and flux moments, which corresponds to setting a \emph{closure}. 
The pressure moment in the fluid frame is expressed as an interpolation between the optically thick and thin limits (we omit the flavor indices)
\begin{equation}
\label{eq:P_class}
    P^{jk} = \frac{3(1-\chi)}{2} \frac{E}{3} \delta^{jk} + \frac{3 \chi -1}{2} \frac{F^j F^k}{|\vec{F}|^2} E \, ,
\end{equation}
where the Eddington factor $\chi$ is determined by an appropriate closure relation. Specifically, in the classical Minerbo closure~\cite{Murchikova:2017zsy,minerbo_maximum_1978,Smit_closure} one can approximate $\chi$ as
\begin{equation}
\label{eq:chi_MEC}
  \chi(\hat{f}) = \frac{1}{3} + \frac{2 \hat{f}^2}{15}\left(3-\hat{f}+3 \hat{f}^2\right)\, ,
\end{equation}
where $\hat{f} = \abs{\vec{F}}/E$ is the flux factor. We choose this closure as it is the one used in the classical NSM simulation from which we extracted the test cases NSM 1,2,3,4.

This classical Minerbo closure is directly used for the flavor-diagonal elements of the pressure tensor [i.e., compute $P_{aa}^{jk}$ via Eq.~\eqref{eq:P_class} with $E_{aa}$, $\vec{F}_{aa}$ and $\chi(\hat{f}_{aa})$]. For the off-diagonal elements we adopt the same form~\eqref{eq:P_class}, with a “flavor-traced” interpolant $\chi(\hat{f}^{(\mathrm{FT})})$ given by Eq.~\eqref{eq:chi_MEC}, with
\begin{equation}
  \hat{f}^{(\mathrm{FT})} \equiv \frac{|\mathbf{F}_{ee} + \mathbf{F}_{xx}|}{E_{ee} + E_{xx}} \, . \label{eq:f_FT}
\end{equation}
Note that this choice of off-diagonal closure is arbitrary, but it has shown good qualitative and quantitative performance in previous FFI calculations~\cite{flashri,2024PhRvD.109d3046F}. See~\cite{Froustey:2024sgz,Kneller:2024buy} for recent theoretical progress on quantum closures, which could be implemented in future work.

\subsubsection{Initial perturbations}

A common factor in all our tests is the need to `perturb' the initial flavor-space off-diagonal elements of the energy and flux density.
To devise these entries, in previous work~\cite{flashri} we seeded the perturbations using random numbers at a scale of one part in $10^6$ of the diagonal elements and did so separately for the real and imaginary components.  We give the explicit expression from Ref.~\cite{flashri}
\begin{equation}\label{eq:od_seed}
  \delta E_{ex}(\mathbf{x}) 
  = 10^{-6}\,p\,\max_{c=e,x}\{N_{cc}\}[A(\mathbf{x})+\imath B(\mathbf{x})],
\end{equation}
where $-1<A,B<1$ are uniform random numbers at each location. A similar expression exists for the antineutrinos.

In this section, we also include simulations with diagonal perturbations, i.e., instead of seeding off-diagonals we seed perturbations using the diagonal terms of the energy density matrix
\begin{equation}
\label{eq:diag_seed}
  \delta E_{aa}(\vec{x}) = \langle E_{aa}(t=0)\rangle\left\{1+2\times10^{-6}\left[R(\vec{x})-\frac{1}{2}\right]\right\} \, ,
\end{equation}
for $a=e,x$, and where $R(\vec{x})$ is a uniform random number in $[0,1]$ and depends on location.  The averaged value $\langle E_{aa}(t=0)\rangle$ is the initial unperturbed value of the energy density, either designated for the test simulations or taken from the output of the NSM simulations of Ref.~\cite{Foucart:2016rxm}.  The fluxes change in the same manner so as to preserve the flux factors at every location.  

Before proceeding, we explain the ramifications of using Eq.~\eqref{eq:diag_seed} for the perturbations as opposed to the off-diagonal perturbations of Eq.~\eqref{eq:od_seed}.
To reiterate, the FFI can only be sourced by the self-interaction term with nonzero off-diagonal elements.   If we seed only the diagonal elements, then we must include the vacuum Hamiltonian to populate the off-diagonal elements of $E$ and $F^j$.
$H_V$ in Eq.~\eqref{eq:h_v} has no angular dependence and will populate the off-diagonal elements of $E$ in the same manner for each location $\vec{x}$, modulo the one part in $10^6$ difference.  The implication is that we seed the homogeneous mode of $E_{ex}$ --- and by extension $N_{ex}$ --- with a power much larger than any other scale.  This is in stark contrast to using Eq.~\eqref{eq:od_seed} which does not select a preferred scale for the power.  Part of the research program for using moments to model FFC was to determine the fastest growing mode.  If we seed predominantly the $\vec{k}=\vec{0}$ mode and that mode is also unstable to FFC for a given simulation setup, then the interpretation of our results requires more nuance than the past approach.


Table~\ref{tab:NSM_results} gives a summary of the FFI results for the four NSM points. The first column gives the name of the NSM point and the method of calculation.  The Linear Stability Analysis (LSA) method is from Ref.~\cite{2024PhRvD.109d3046F}, using the closure outlined in Sec.~\ref{subsec:closure}, except for an additional test with the NSM2 point (see details in Sec.~\ref{ssec:NSM2}).  The second and third columns give the growth rate and fastest growing mode during the linear growth phase of FFC. Overall, the moment method is capable of replicating the qualitative and often quantitative behavior of the exact solution, and the improvements to the advection scheme in Section~\ref{sec:methods} significantly improve the performance of the moment scheme. In the following subsections, we delve into the four test cases to expose the performance of the method in different environments.

\begin{table}[b]
    \caption{Growth rates and fastest growing mode wavevector values for the four NSM points. We compare the results obtained with moment LSA (see Ref.~\cite{2024PhRvD.109d3046F}), the previous implementation \flashri (see Table 4 in \cite{flashri}), the new modulus-phase implementation of \flash, and \emu calculations.}
    \begin{ruledtabular}
    \begin{tabular}{ccc}
        \multirow{2}{*}{Name} & \imo & \kmax \\ 
         & $(10^{10}\,\mathrm{s}^{-1})$ & $(\mathrm{cm}^{-1})$ \\\hline 
        \multicolumn{3}{c}{\textit{NSM1}}           \\
        LSA             &  $7.19$ & $5.64$   \\
        \flash ($ri$)   &  $8.1$  & $6.4(4)$ \\
        \flash          &  $6.4$  & $4.8(4)$ \\
        \emu            &  $5.6$  & $4.8(4)$ \\ \vspace{4pt}\\
        \multicolumn{3}{c}{\textit{NSM2}} \\
        LSA             & (stable)  & (stable)      \\
        LSA$^\dagger$   & $1.22$  & $3.70$   \\
        \flash ($ri$)   & $5.2$   & $6.1(4)$ \\
        \texttt{FLASH}$^\dagger$  & $1.3$   & $3.8(2)$   \\
        \emu            & $1.1$   & $3.8(4)$ \\ \vspace{4pt}\\
        \multicolumn{3}{c}{\textit{NSM3}} \\
        LSA             & $2.45$  & $6.60$  \\
        \flash ($ri$)   & $10.7$ & $13.0(5)$ \\
        \flash          & $2.8$ & $6.5(5)$ \\
        \emu            & $4.2$ & $6.5(5)$ \\ \vspace{4pt}\\
        \multicolumn{3}{c}{\textit{NSM4}} \\
        LSA             & $1.39$ & $2.08$ \\
        \flash ($ri$)   & $-$ & $-$ \\
        \flash          & $1.39$ & $2.0(1)$ \\
        \emu            & $1.48$ & $1.8(1)$ \\
    \end{tabular}\\
    \footnotesize{$^\dagger$ Calculated with different closure relation [see Eq.\ \eqref{eq:NSM2_pressure}].}\\

    \label{tab:NSM_results}
\end{ruledtabular}
\end{table}

\subsection{General results in deep crossing configurations}

We begin with deep and medium crossing configurations (respectively, called “NSM1” and “NSM3” in \cite{flashri}) to illustrate a variety of FFI characteristics that we are able to capture with our moment code. In NSM1, we show the main properties of the FFI and how the new advection algorithm improves the previous results \flashri. We also discuss the limitations of the maximum entropy closure in light of the \emu results.

\subsubsection{Deep crossing configuration [NSM1]}
\label{ssec:NSM1}
 
\begin{figure}[!ht]
    \centering      
    \includegraphics[width=\columnwidth]{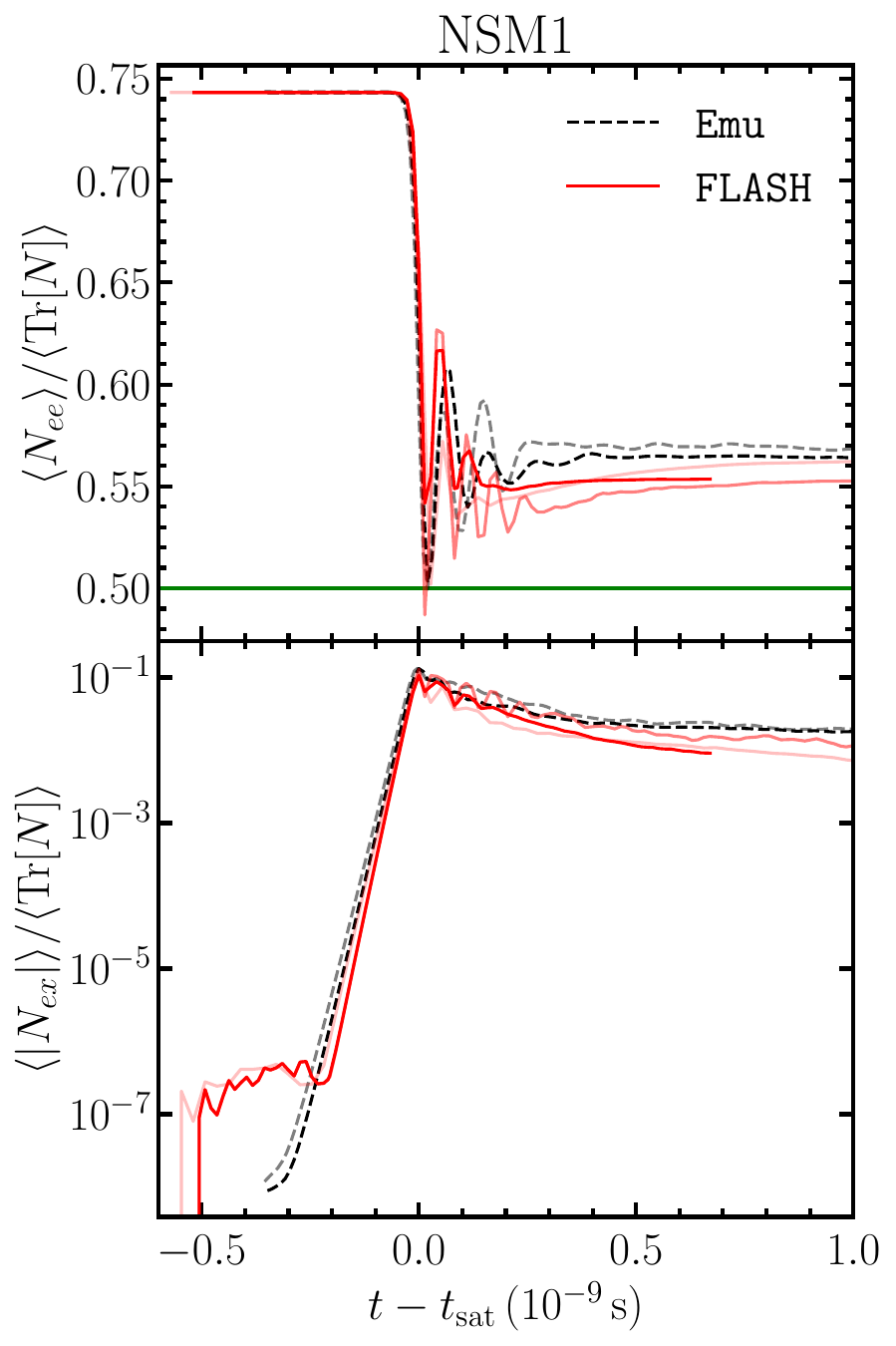}
    \caption{\textit{Top panel:} electron-flavor number density moment evolution in time. \textit{Bottom panel:} normalized modulus of the $N_{ex}$ component. The time axis is necessarily offset from the peak saturation point of $\langle N_{ex}\rangle$ to make visual comparison easier. \flash results using different grid structures are in shades of red: dark $(L=7.87\,{\rm cm},\, N_{gp}=128^3)$; medium $(L=3.93\,{\rm cm},\, N_{gp}=64^3)$; light $(L=7.87\,{\rm cm},\, N_{gp}=64^3)$.  Dashed black and gray curves show results from \emu simulations with the same resolution yet differing domain sizes.  The horizontal green line shows the value for two-flavor equilibrium.  
    \label{fig:NSM1_Mn0_N_comp}
    }
\end{figure}

Reference \cite{Grohs:2022fyq} gave a detailed account of this particular simulation point, as \flash $(ri)$ and \emu were in close quantitative agreement.  It has a relatively deep ELN crossing as compared to the other NSM points that we consider in this work. The initial conditions are specified in Table~\ref{tab:NSM_parameters1}, while the simulation grid is specified in Table~\ref{tab:NSM_parameters2}, and the results are shown in Fig. \ref{fig:NSM1_Mn0_N_comp}.  
The general features of an FFI with deep crossings are apparent. The off-diagonal components grow exponentially during the linear phase (bottom panel) until saturation, after which the diagonal components (top panel) fluctuate as they approach the final equilibrium, and the off-diagonal components decay with time. The \flash results show fluctuations very similar to those in the \emu simulations, which the \flashri results (not shown; see Ref.~\cite{flashri}) struggled to do.

\begin{figure}[!ht]
    \centering      
    \includegraphics[width=\columnwidth]{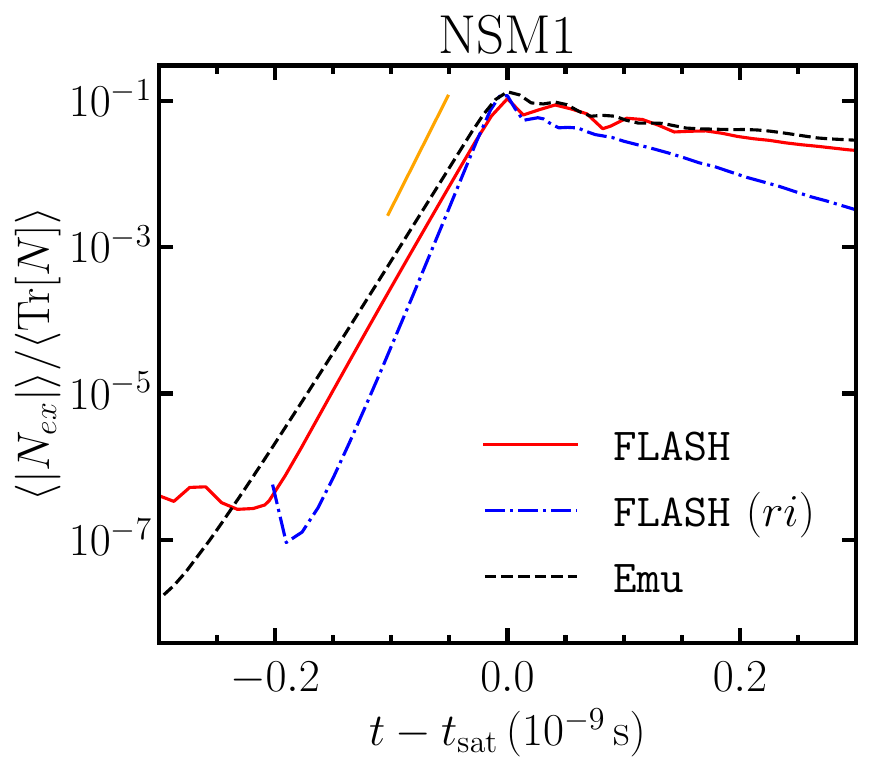}
    \caption{Domain averaged off-diagonal number density moment versus time for the NSM1 simulation. Solid red and dashed black curves correspond to the same calculations as Fig.~\ref{fig:NSM1_Mn0_N_comp} with the inclusion of the dot-dash blue curve for the \flashri simulation from Ref.~\cite{flashri}.  The orange line gives the LSA prediction for the growth rate~\cite{2024PhRvD.109d3046F}. 
    \label{fig:NSM1_Mn0_Nex_methods}
    }
\end{figure}

At early times, the \flash results in the bottom panel of Fig.\ \ref{fig:NSM1_Mn0_N_comp} show evolution different from exponential growth.  This behavior results from the vacuum and matter potentials sourcing the off-diagonal components in a homogeneous manner.  The growth rate for \flash is only slightly larger than in \emu.  This is in much better agreement with \emu results than the \flashri result at the same resolution (see Table~\ref{tab:NSM_results}).  We separately plot the off-diagonal components for the three simulation methods in Fig.\ \ref{fig:NSM1_Mn0_Nex_methods}.  Both the growth rate prior to saturation and the decoherence after saturation are slower in the \flash (solid red) results than in the  \flashri (dot-dashed blue) ones, and are more consistent with the \emu results (dashed black).  The slower decoherence rate allows more flavor oscillations prior to relaxation, and as a result, we see more peaks and troughs in the evolution of $\langle N_{ee} \rangle$ compared to the \flashri implementation (compare, for instance, the top panel of Fig.~\ref{fig:NSM1_Mn0_N_comp} with the top panel of Fig.~2 in \cite{Grohs:2022fyq}, or the top left panel of Fig.~7 in \cite{flashri}).

\begin{figure}[!ht]
    \centering      
    \includegraphics[width=\columnwidth]{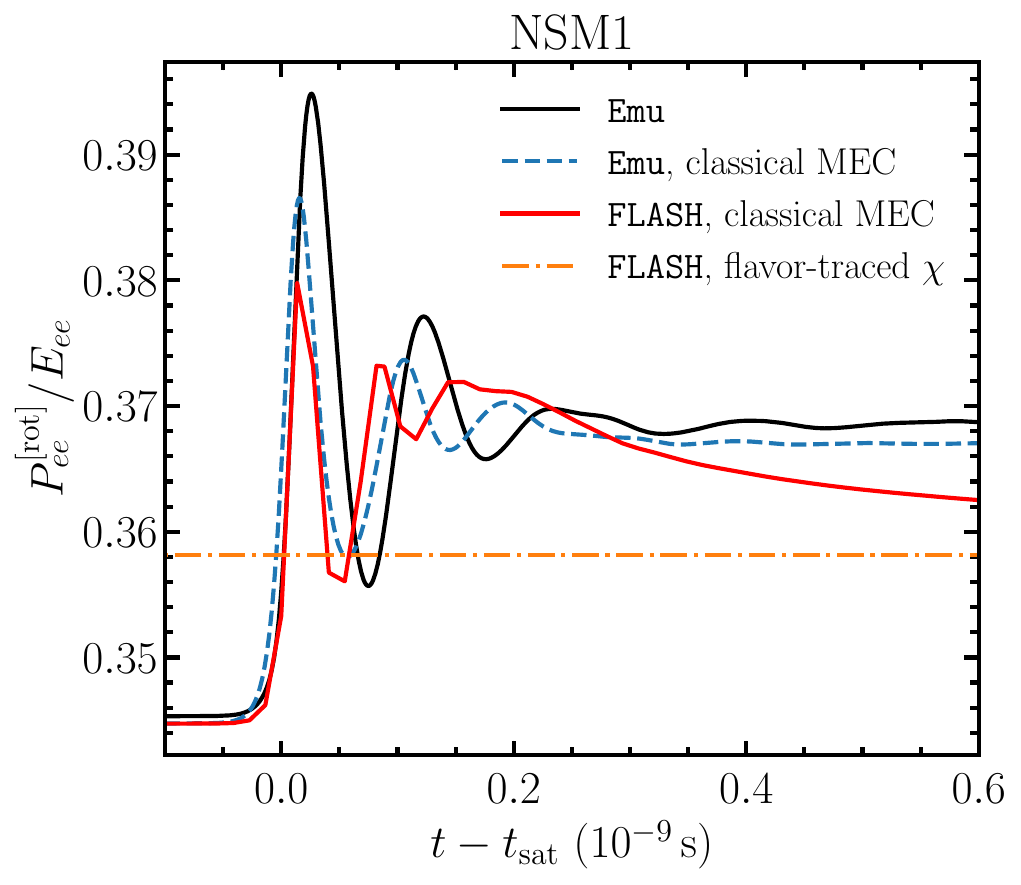}
    \caption{Rotated $e$-flavor Eddington tensor plotted as a function of time from an \emu and a \flash calculation of the NSM1 point.  All curves give the domain-averaged $ee$ component and are normalized by the energy density moment. Solid black is the Eddington tensor computed directly from \emu output.  Dashed blue gives the result of the MEC using the output number and flux moments from \emu.  Solid red is the result of the MEC using the output number and flux moments from \flash. Dash-dotted orange uses the flavor-traced number and flux components from \flash to compute the Eddington factor $\chi(\hat{f}^{(\mathrm{FT})})$.
    \label{fig:NSM1_Mn0_P_comp}
    }
\end{figure}

Our modulus-phase implementation of \flash uses classical MEC relations for the flavor-diagonal moments and an extension to \cnumbers for the off-diagonal moments.  These relations are necessary to calculate the components of the pressure tensor used in Eq.~\eqref{eq:moment_QKEs} because it is not possible to determine them with the limited information available in the moment method. Because of this, it is interesting to compare the pressure tensors assumed by various choices of moment closure with the true result determined by \emu simulations following Eq.~\eqref{eq:mom_2}. Figure~\ref{fig:NSM1_Mn0_P_comp} shows the $ee$ flavor component of the pressure tensor projected onto the direction of the flux, which we denote as $P_{ee}^\mathrm{[rot]}$. The solid red curve shows the pressure using the electron neutrino energy and flux moments from \flash data, i.e., it corresponds to the closure assumed in \flash. The pressure calculated with the flavor-traced Eddington factor (that is, $P_{ee}^\mathrm{[rot]} = \chi(\hat{f}^\mathrm{(FT)}) E_{ee}$) is shown as dash-dotted orange — note that it was the closure chosen in \flashri. It is constant since the space-averaged flavor traces of $E$ and $\vec{F}$ are conserved quantities for the system of equations~\eqref{eq:moment_QKEs_3D}. The true closure calculated from the results from \emu is shown in black, and the pressure obtained by imposing the MEC on the first two \emu moments [following Eq.~\eqref{eq:P_class}] is shown in dashed blue. The difference between the black and blue curves shows that the MEC cannot reproduce the true pressure in the NSM1 simulation, although the MEC does capture some qualitative features. Both are quite different from the \flash closure. The red curve has a similar structure to the blue curve, but the frequency appears to be larger with a smaller amplitude. The difference between these two curves is a direct measure of the difference between the first moments $(E_{ee},\vec{F}_{ee})$ calculated with \emu and \flash. Finally, the difference between the red and orange curves demonstrates that the energy and flux moments (and, therefore, the assumed pressure moment) are different between different flavors.

\begin{figure}[!ht]
    \centering      
    \includegraphics[width=\columnwidth]{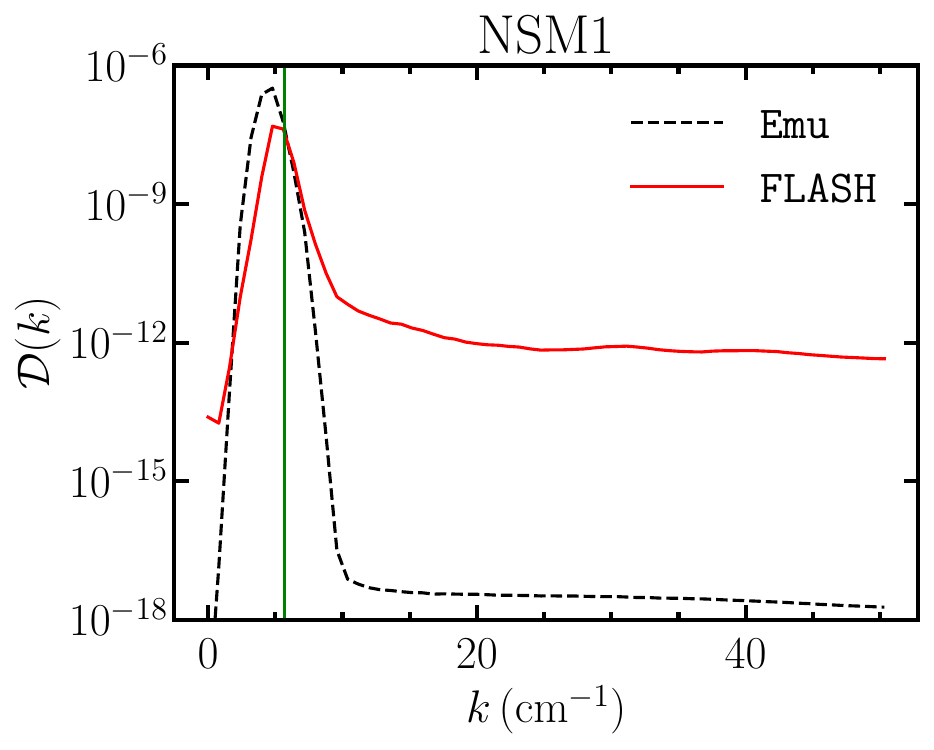}
    \caption{Power in the DFT versus wavenumber for a time snapshot of the NSM1 simulation taken during the linear growth phase ($t-\tsat\sim-0.1\times10^{-9}\,{\rm s}$). Solid red (dashed black) curve corresponds to the \flash (\emu) calculation.  The green vertical line gives the LSA prediction of \kmax.  
    The \flash fastest growing mode is very similar to that predicted by both LSA and \emu. 
    \label{fig:NSM1_Mn0_FFT_comp}
    }
\end{figure}

The spatial structure of the fastest growing mode is also improved by the \flash advection method. The DFTs [Eq.~\eqref{eq:def_DFT}] at a snapshot in time 0.1 ns before saturation for both the \emu and \flash calculations are plotted in Fig.\ \ref{fig:NSM1_Mn0_FFT_comp}, where the line colors and styles are the same as Fig.~\ref{fig:NSM1_Mn0_N_comp}, together with the wavenumber of the fastest growing mode determined by linear stability analysis (green).  The values of $k_\mathrm{max}$ (i.e., the peaks of the DFTs) in the \emu and \flash simulations are the same up to the wavenumber spacing (set by the simulation domain size), though the shape of the \flash peak may indicate a bias towards smaller physical scales. This, too, represents a significant improvement from the treatment of advection in \flashri (see Fig.~10 in \cite{flashri}).

\subsubsection{Medium crossing [NSM3]}
\label{ssec:NSM3}

\begin{figure}[!ht]
    \centering      
    \includegraphics[width=\columnwidth]{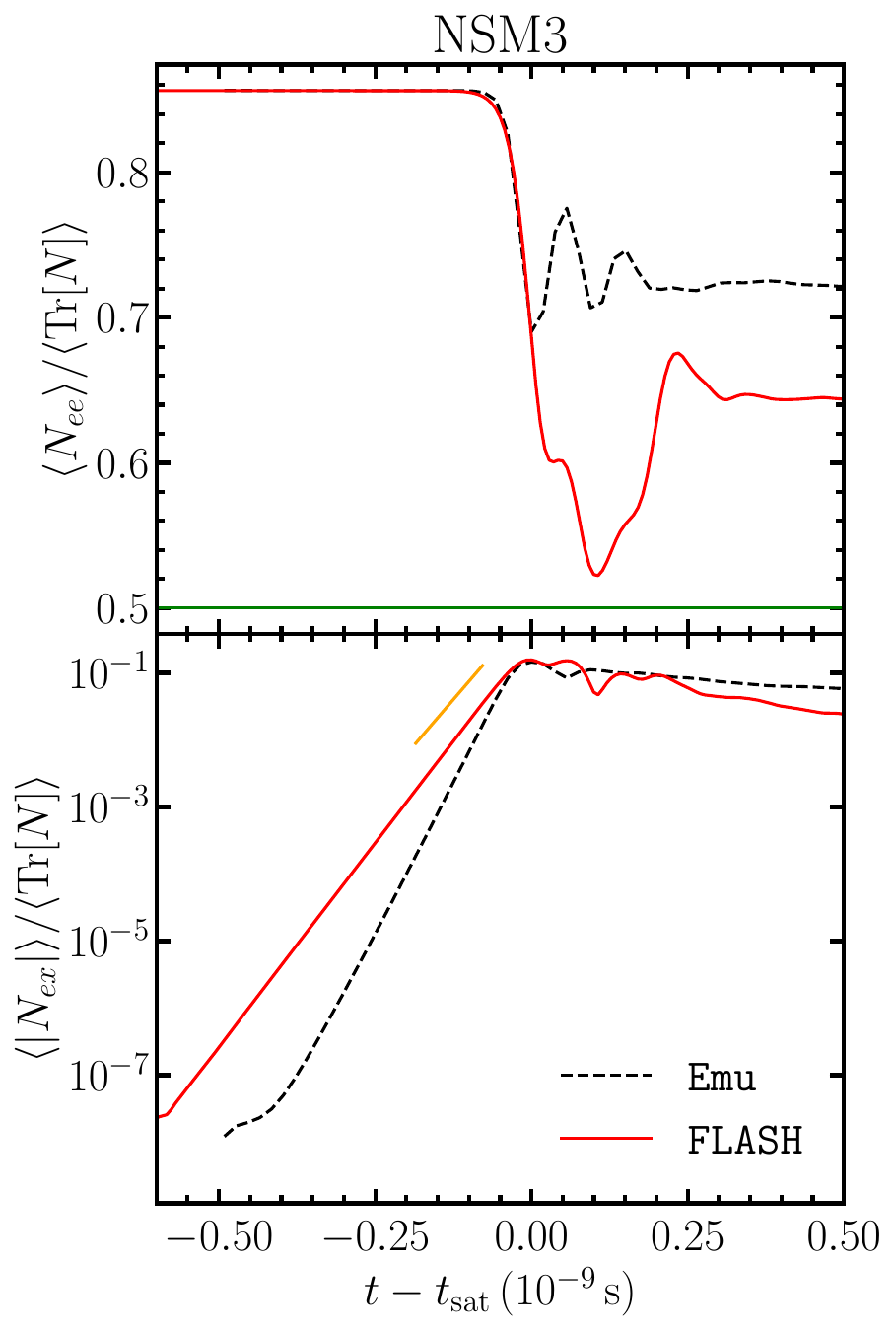}
    \caption{Domain averaged number density components versus time for the NSM3 simulation.  Line styles and colors are the same as Figs.\ \ref{fig:NSM1_Mn0_N_comp} and \ref{fig:NSM1_Mn0_Nex_methods}. 
    \label{fig:NSM3_N_comp}
    }
\end{figure}

The NSM3 point is extracted from near the equatorial plane of the merger disk, and has a neutrino distribution with an intermediate-depth ELN crossing. The initial conditions are specified in Table~\ref{tab:NSM_parameters1}, while the simulation grid is specified in Table~\ref{tab:NSM_parameters2}. Figure \ref{fig:NSM3_N_comp} shows the domain-averaged moments (normalized by the trace) with a similar labeling scheme as Fig.~\ref{fig:NSM1_Mn0_N_comp}.  
The bottom panel shows a slower growth rate for \flash than for \emu, although the agreement is significantly better than it was with \flashri (see Table~\ref{tab:NSM_results}).  Oscillations in the decoherence phase are a characteristic of fast flavor simulations.  Both the \flash and \emu simulations in Fig. \ref{fig:NSM3_N_comp} exhibit oscillations during the decoherence phase, although with a somewhat different pattern.  During the first $\sim0.2\,{\rm ns}$ of the decoherence phase, we observe a decrease in $\langle|N_{ex}|\rangle$ which is quantitatively similar for both calculations, although the variance for \flash appears to be larger in that time interval. 

Despite the differences in growth rate, the properties of the fastest growing mode are quite similar between the \flash and \emu simulations.

\begin{figure}[!ht]
    \centering      
    \includegraphics[width=\columnwidth]{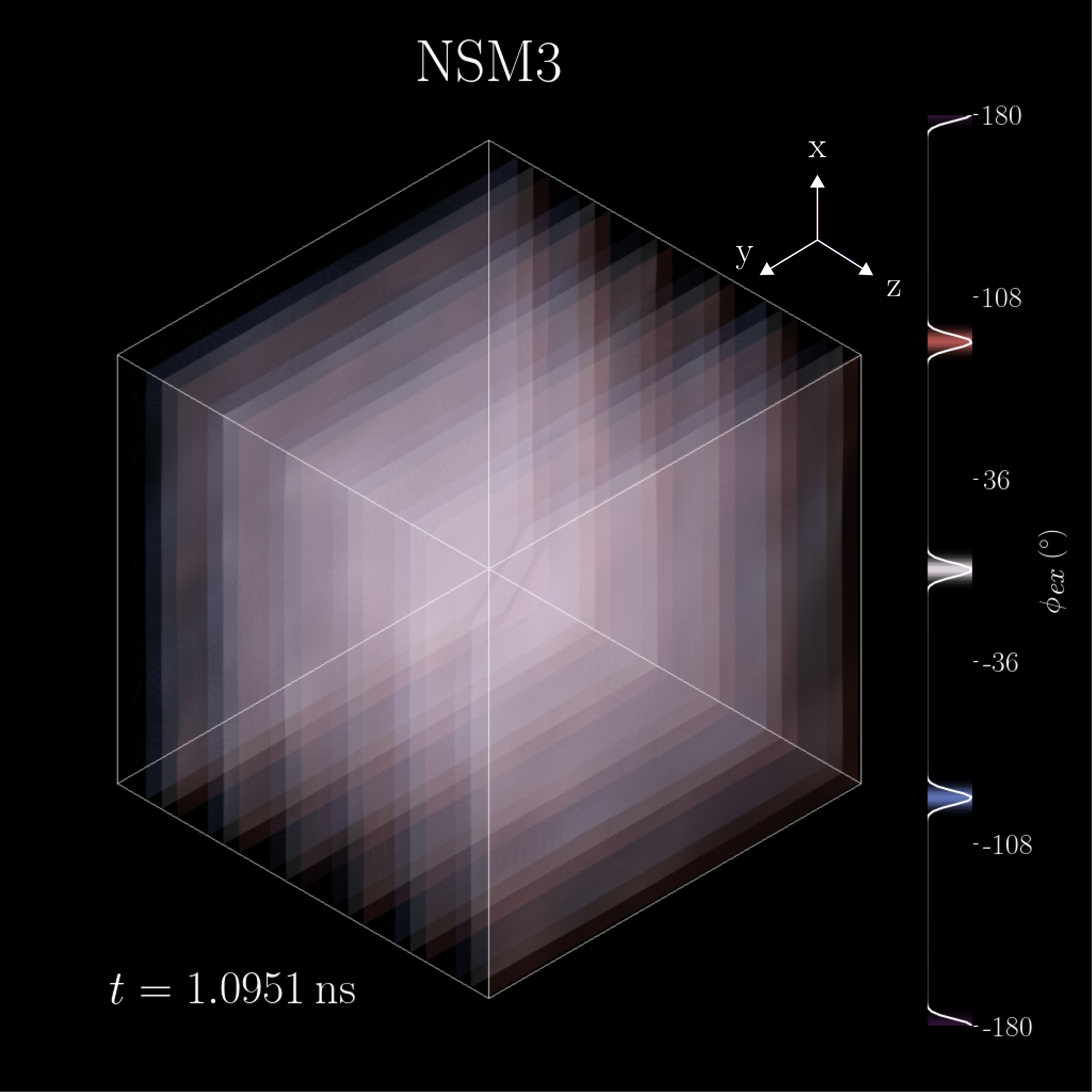}
    \caption{Time snapshot of 3D contours of constant $\phi_{ex}=\arg(N_{ex})$ in the NSM3 simulation domain.  The data is from the \flash calculation shown on Fig.~\ref{fig:NSM3_N_comp}, at a time before saturation $t-\tsat\sim-0.1\,{\rm ns}$. Red, white, and blue contours correspond to $\phi_{ex}=90^\circ,0^\circ,-90^\circ$ respectively. 
    \label{fig:NSM3_phase_3D}
    }
\end{figure}

Figure \ref{fig:NSM3_phase_3D} shows a 3D rendering of the phase of $N_{ex}$ from the \flash calculation, labeled as $\phi_{ex}$, for a time snapshot $t-\tsat\sim-0.1\,{\rm ns}$ during the linear growth phase.  The contours visible in Fig.\ \ref{fig:NSM3_phase_3D} are closely synchronized to $\x\y$-planes at sequential values of $\z$, implying a plane-wave structure for $\phi_{ex}$ across the simulation domain
that moves in the $-\z$ direction apr\`es the ELN flux.

\begin{figure}[!ht]
    \centering      
    \includegraphics[width=\columnwidth]{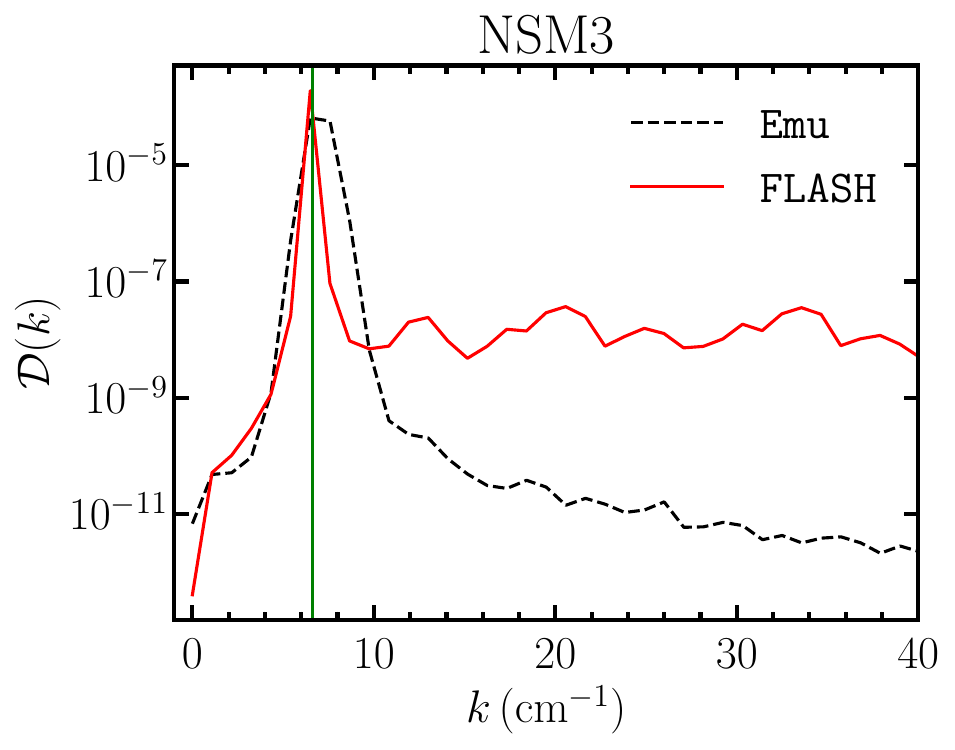}
    \caption{Power in the DFT versus wavenumber for a time $t-\tsat\sim-0.1\times10^{-9}\,{\rm s}$ in the NSM3 simulation. Line styles and colors are the same as Fig.\ \ref{fig:NSM1_Mn0_FFT_comp}. 
    \label{fig:NSM3_FFT_comp}
    }
\end{figure}

The modulus of $N_{ex}$ also varies across the simulation domain, by as large as a factor of 5.  For a sequence of cells along the $\z$-direction, however, the variation is $\sim10\%$.  Furthermore, this variation in $|N_{ex}|$ is not on the same length scale as the contours in Fig.~\ref{fig:NSM3_phase_3D}.  Therefore, the plane-wave structure in $\phi_{ex}$ leads the DFT of $N_{ex}$ to peak at the scale exhibited in Fig.\ \ref{fig:NSM3_phase_3D}.  Indeed, Fig.\ \ref{fig:NSM3_FFT_comp} gives the power in the DFT for \flash and \emu for the same time displayed in Fig.\ \ref{fig:NSM3_phase_3D}.  \kmax for \flash is only slightly smaller than the value for \emu.  The scale for \kmax is exactly the scale in the plane-wave structure of Fig.\ \ref{fig:NSM3_phase_3D}.

\subsection{TwoThirds test case and unstable homogeneous mode}
\label{ssec:2_3}

We have presented neutrino flavor transformation results for two NSM points.  For NSM1, we showed the modulus-phase implementation gave a slight improvement on an already robust $ri$ calculation.  The improvement became even more pronounced with NSM3, exemplified by the uniform $\phi_{ex}$ wavefronts shown in Fig.\ \ref{fig:NSM3_phase_3D}.
We will return to other NSM points later in Sec.\ \ref{ssec:NSM4} to discuss some new insights into moment methods.  To assist in diagnosing the causes of the new phenomenology from these other NSM simulations, we give an interregnum on a contrived test case where different $k$-modes play a role in the simulation.  Specifically, we set out to investigate the “TwoThirds” test case: isotropically distributed neutrinos versus antineutrinos with nonzero flux in the $\z$-direction (flux factor of $1/3$) and $2/3$ the number density of neutrinos.

Reference \cite{flashri} contained an $ri$ moment calculation of the TwoThirds test case along with two other test cases: Fiducial and 90Degree.  Appendix \ref{app:fid_90d} contains results from modulus phase \flash for these two other test cases,
together with the initial conditions and geometries of all three tests in Table \ref{tab:3test_parameters}.  Although the TwoThirds test case is symmetric about the $\z$-axis, we use a 3D geometry for our simulation setup.  \flash simulations use rectangular geometries for the number of grid points, whereas \emu simulations use a cubic grid.  We include the vacuum term in the \flash simulations (but not in the \emu simulations) with a solar mass splitting and mixing angle, namely, $\delta m^2=7.53\times10^{-5}\,{\rm eV}^2$, $\theta=0.587$, yielding a vacuum frequency of $\omega_{\odot}\sim10^{3}\,{\rm s}^{-1}$.  The inclusion of the vacuum term will precipitate a linear growth phase sourced by the $k=0$ mode for the \flash simulations.  We do not include a matter term in either the \flash or \emu simulations, though perturbations source other unstable modes in both cases. 

\begin{figure}[!ht]
    \centering      
    \includegraphics[width=\columnwidth]{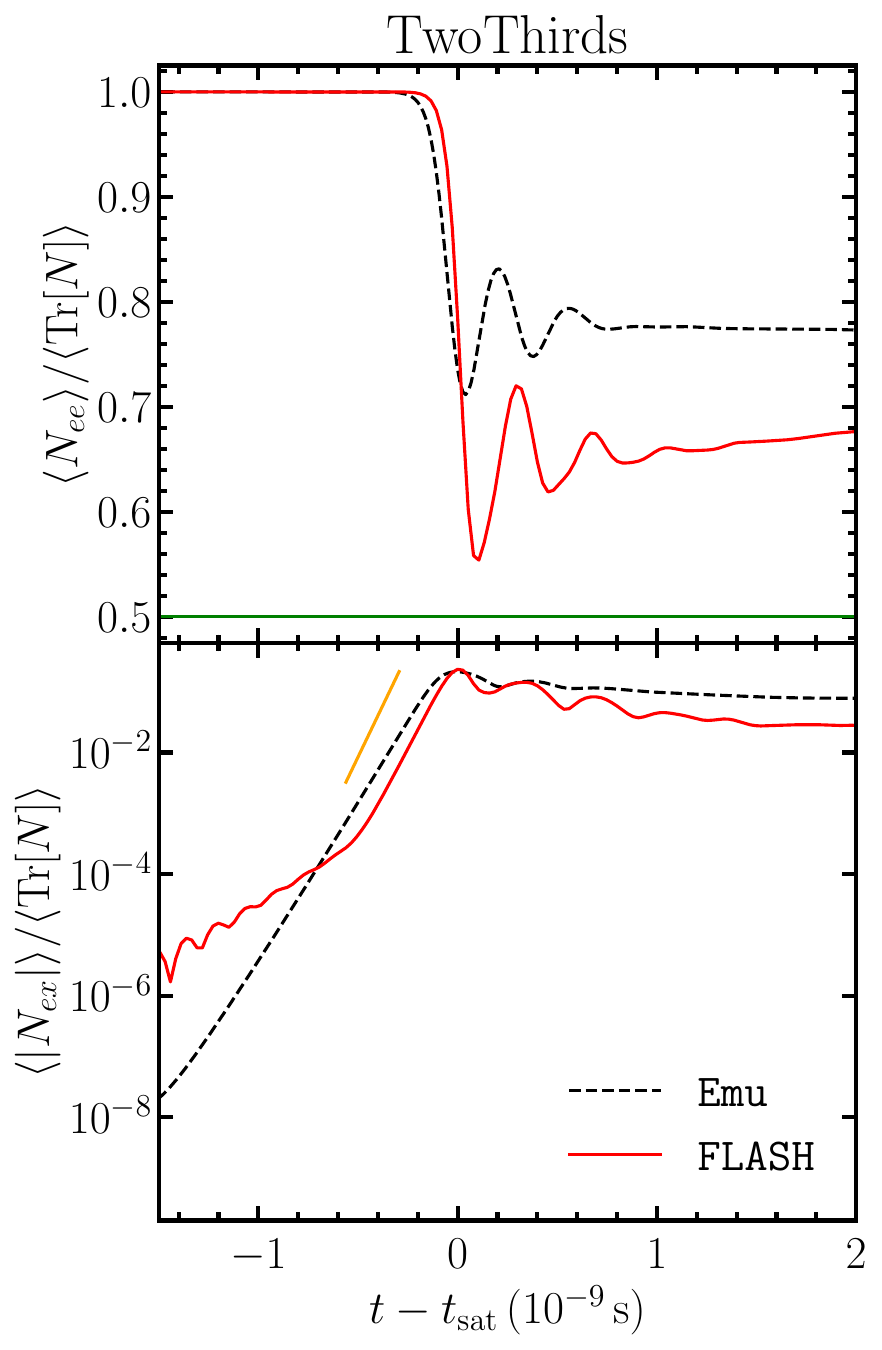}
    \caption{Domain averaged number density components versus time for the TwoThirds simulation.  Line styles and colors are the same as Figs.\ \ref{fig:NSM1_Mn0_N_comp} and \ref{fig:NSM1_Mn0_Nex_methods}. 
    \label{fig:2_3_N_comp}
    }
\end{figure}

Figure \ref{fig:2_3_N_comp} shows the results of the Two-Thirds test case for the domain-averaged number density moments.  The solid red line gives the results for the resolution of $N_{gp}=16\times16\times128$.
We observe a noticeably different evolution at early times for \flash versus \emu.  In the \flash calculation, the vacuum potential sources \avgnex starting from a value of zero.  In any individual cell, the initial evolution of $N_{ex}$ is proportional to the difference $N_{ee}-N_{xx}$.  Our initial conditions set this difference to be the same in each cell modulo the perturbations to the diagonal elements, namely one part in $10^{6}$. As a result, we expect nearly similar evolution across the whole domain, i.e., the most power in the homogeneous mode.  That $k=0$ mode was stable to the FFI for NSM1 and NSM3, but \emph{not} for TwoThirds.  A LSA calculation gives the growth rate for the $k=0$ mode as $\imo=4.6\times10^{9}\,{\rm s}^{-1}$, though $k=0$ is not the fastest growing mode.  The presence of oscillations in $\langle|N_{ex}|\rangle$ obscures the exponential growth, but nevertheless we estimate an empirical rate of $5.5\times10^{9}\,{\rm s}^{-1}$ for $-0.7\,{\rm ns}\lesssim t-\tsat\lesssim -0.6\,{\rm ns}$.  During this first period of exponential growth, the power grows in other modes.  At a time $t-\tsat\sim-0.5\,{\rm ns}$ the power in one of these modes surpasses $k=0$, and the growth rate quickens.  The orange line in the bottom panel of Fig.\ \ref{fig:2_3_N_comp} gives the prediction of \imo during this second linear phase, in close agreement with the \flash results.  In addition, the \flash and \emu growth rates are commensurate with one another at the $\sim10\%$ level, a result much more in agreement than the \flashri calculation.  After saturation, the \flash \avgnee evolution exhibits a qualitatively similar oscillation pattern as \emu, albeit with a different displacement and a phase difference.  The \flash calculation does not show a clear asymptote once the oscillations cease in Fig.\ \ref{fig:2_3_N_comp}.  However, we have verified that \avgnee does indeed asymptote to $\avgnee/\langle \Tr[N]\rangle\sim0.71$ for $t-\tsat\gtrsim7\,{\rm ns}$.

The appearance of two modes for the TwoThirds \flash calculation is a unique feature of the initial conditions, dynamical equations, and also the spatial resolution.
We note that the simulation displayed by the solid red curve in Fig.\ \ref{fig:2_3_N_comp} is indeed numerically resolved --- a result we verified by running even higher resolution simulations. 

\begin{figure}[!ht]
    \centering      
    \includegraphics[width=\columnwidth]{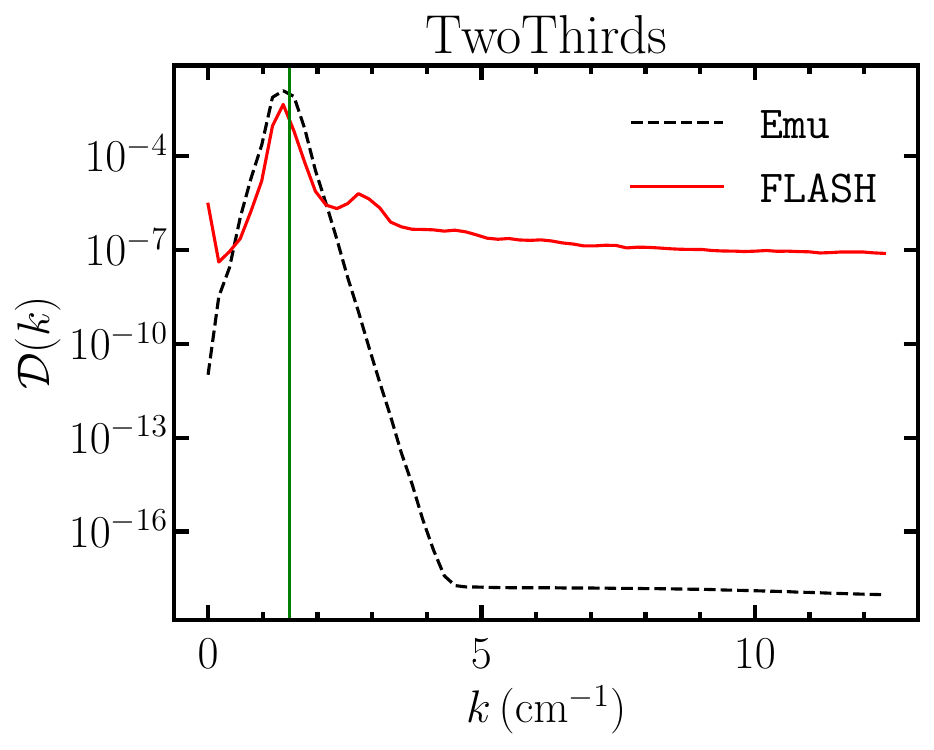}
    \caption{Power in the DFT versus wavenumber for a time $t-\tsat\sim-0.1\times10^{-9}\,{\rm s}$ in the TwoThirds simulation. Line styles and colors are the same as Fig.\ \ref{fig:NSM1_Mn0_FFT_comp}. 
    \label{fig:2_3_FFT_comp}
    }
\end{figure}

To wit,  Figure \ref{fig:2_3_FFT_comp} shows the DFTs for the TwoThirds test case, with a similar nomenclature borrowed from past figures.
The solid red curve shows peak power at a nearly identical \kmax compared to \emu, but we see features in \flash which are absent in \emu.  There is still a residual power at $k=0$ which drove the FFI in the first phase of the linear evolution in Fig.\ \ref{fig:2_3_N_comp}.
The \emu calculation uses off-diagonal perturbation seeds implying a negligible amount of power at $k=0$.  Furthermore, we observe a smaller peak at $k\sim3\,{\rm cm}^{-1}$ in \flash.  The sequence of peaks fits a harmonic structure and is absent in all of the other calculations presented so far.  We emphasize that the harmonic structure is a numerical artifact in \flash and the result of two concomitant conditions: power initially seeded for the homogeneous mode; an unstable homogeneous mode to FFI.

Table \ref{tab:3test_results} in App.\ \ref{app:fid_90d} gives a summary of the growth rates and fastest growing modes for each calculation of the three test cases.  \flash \imo values are all larger than the \emu values, but we see better agreement than we do with \flashri.
The growth rates for \flash are all smaller than \flashri with the largest shift being the TwoThirds test case.
\kmax is also uniformly lower for \flash than \flashri.
We conclude that \flash can better match \emu results than \flashri for the three test cases, and in particular the TwoThirds test presented in this section.

\subsection{Impact of the homogeneous mode [NSM4]}
\label{ssec:NSM4}

We return to Ref.\ \cite{Foucart:2016rxm} to find our next NSM simulation point.  In this section, we pick a point which was not included in Ref.\ \cite{flashri}, and term this point NSM4.  We specify the initial conditions for flavor transformation from the output of Ref.\ \cite{Foucart:2016rxm}, which are given in Table \ref{tab:NSM_parameters1}.  Appendix \ref{app:NSM4} gives more information and diagrams on the ELN crossing for this point.
For the comparison calculations with \flash, we use a 1D \emu simulation with 256 cells in the $\z$ direction.  Although the geometry is 1D, the particle directions are in 3D and weighted using the MEC with the flux factors in Table \ref{tab:NSM_parameters1}.  Therefore, there exists an ELN crossing and we expect to see FFC under these computational conditions.
For the initial perturbations, we seed the flavor diagonal elements of each particle and allow the vacuum potential to source the off-diagonal elements. We use the atmospheric mass splitting and reactor mixing angle in $H_V$, which is the same as the \flash calculation.  The matter Hamiltonian is also identical between the two methods of calculation.

\begin{figure}[!ht]
    \centering      
    \includegraphics[width=\columnwidth]{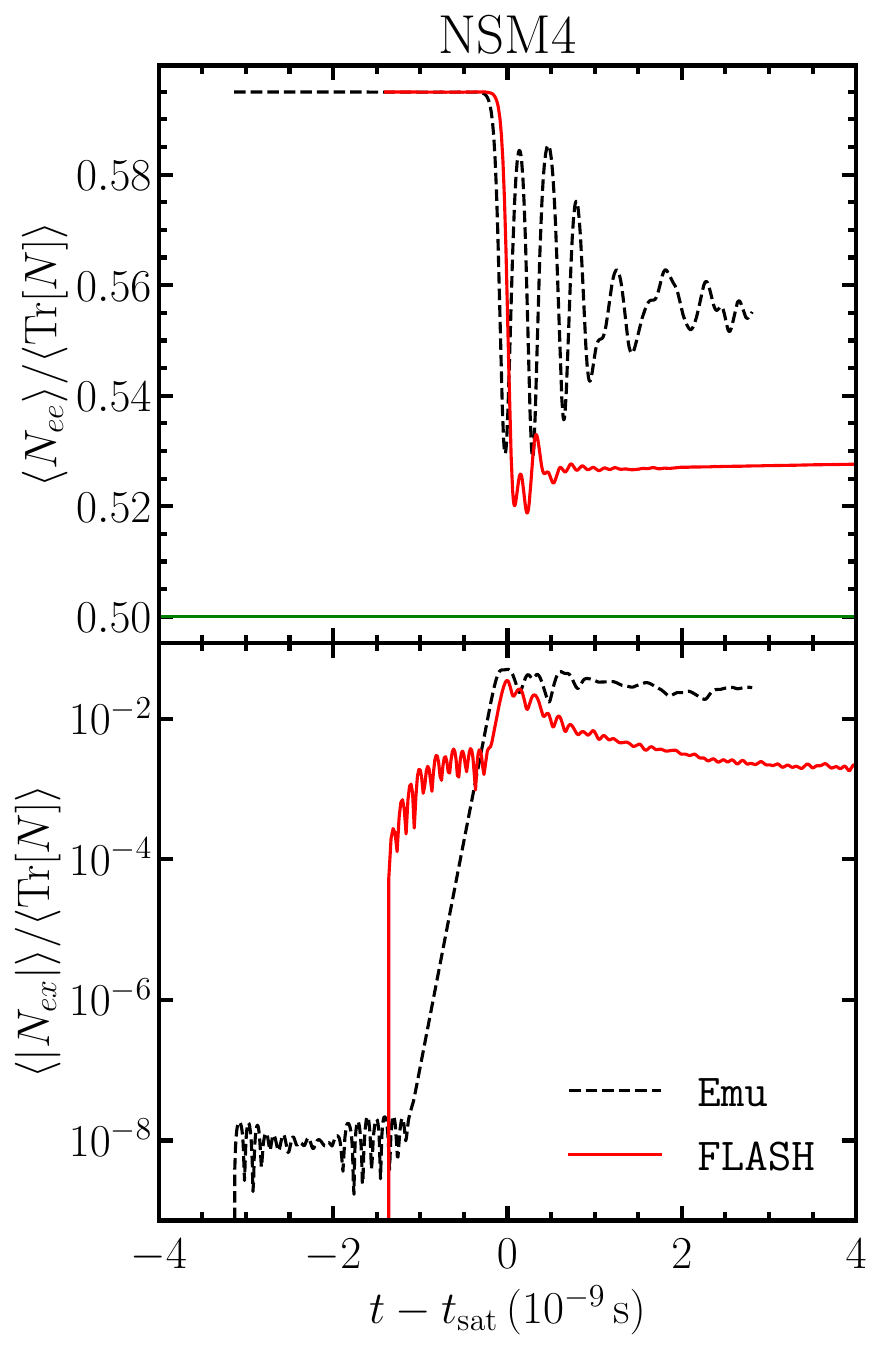}
    \caption{Domain averaged number density components versus time for the NSM4 simulation. Line styles and colors are the same as Fig.~\ref{fig:NSM1_Mn0_N_comp}.
    \label{fig:NSM4_N_comp}
    }
\end{figure}

Figure \ref{fig:NSM4_N_comp} shows the domain-averaged number density moments plotted against time.  Both \emu and \flash show rapid oscillations in \avgnex at early times, where the amplitude of the oscillations is orders of magnitude larger for \flash.  These oscillations are sourced from the homogeneous mode which is stable to FFI.  Eventually, the power grows in another $k$ mode and the linear growth phase begins.  After saturation, the \emu and \flash calculations exhibit overall electron neutrino transformation on a similar scale — note that the figure emphasizes asymptotic values of \avgnee that differ by only a few percent. However, we observe markedly different behavior when examining the results in detail.  The rapid decrease in \avgnex in \flash gives rise to small amplitude oscillations in \avgnee, whereas the persistent coherence in \emu provides ample opportunity for oscillations after saturation of the instability.  

\begin{figure}[!ht]
    \centering      
    \includegraphics[width=\columnwidth]{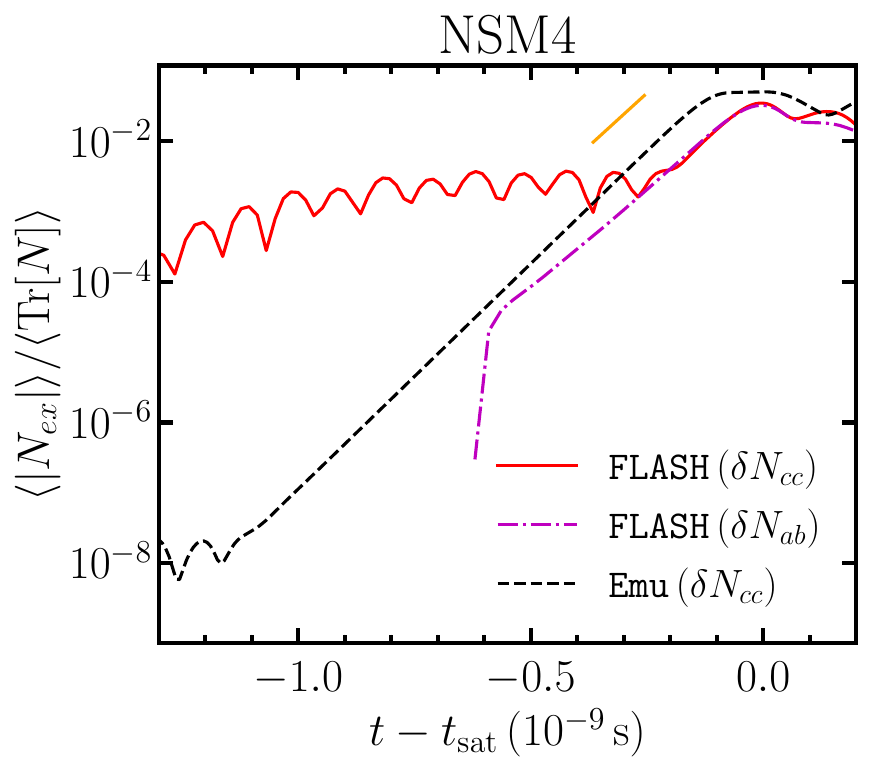}
    \caption{Domain averaged off-diagonal number density moment versus time for the NSM4 simulation.  Solid red and dashed black curves correspond to the same calculations as Fig.\ \ref{fig:NSM4_N_comp}.  The dash-dotted magenta line is a \flash calculation with computational parameters identical to the solid red curve and initial perturbations seeded on the off-diagonal terms of $N$.  The orange line gives the prediction of \imo from a moment LSA.
    \label{fig:NSM4_Nex_pcomp}
    }
\end{figure}

We show another view of \avgnex in Fig.~\ref{fig:NSM4_Nex_pcomp}, zoomed in around the saturation point.  Red and black curves are the same as Fig.\ \ref{fig:NSM4_N_comp}.  The magenta dash-dotted line is the result from another \flash calculation where we used Eq.~\eqref{eq:od_seed} to seed the off-diagonal elements of the angular moments, i.e., eliminating the bias in the homogeneous mode.  The lack of power in the $k=0$ mode yields an evolution mostly dominated by the FFI. The fastest growing mode quickly sets in, leading to a linear growth phase that lasts $\sim 0.5 \, \mathrm{ns}$.
We include an orange line with a slope equal to the LSA prediction for \imo.  Again, there is good agreement between all four sets of calculations, indicating that the red curve is following the expected FFI behavior for the period of time immediately preceding \tsat. 
The rapid oscillations in \avgnex for the red \flash curve are at a frequency $\sim6.6\times10^{10}\,{\rm s}^{-1}$.  This is similar to the oscillations in the \emu calculation seen at early times in Fig.\ \ref{fig:NSM4_N_comp}.  This frequency scale is much larger than the vacuum potential and smaller than the matter potential.  

\begin{figure}[!ht]
    \centering      
    \includegraphics[width=\columnwidth]{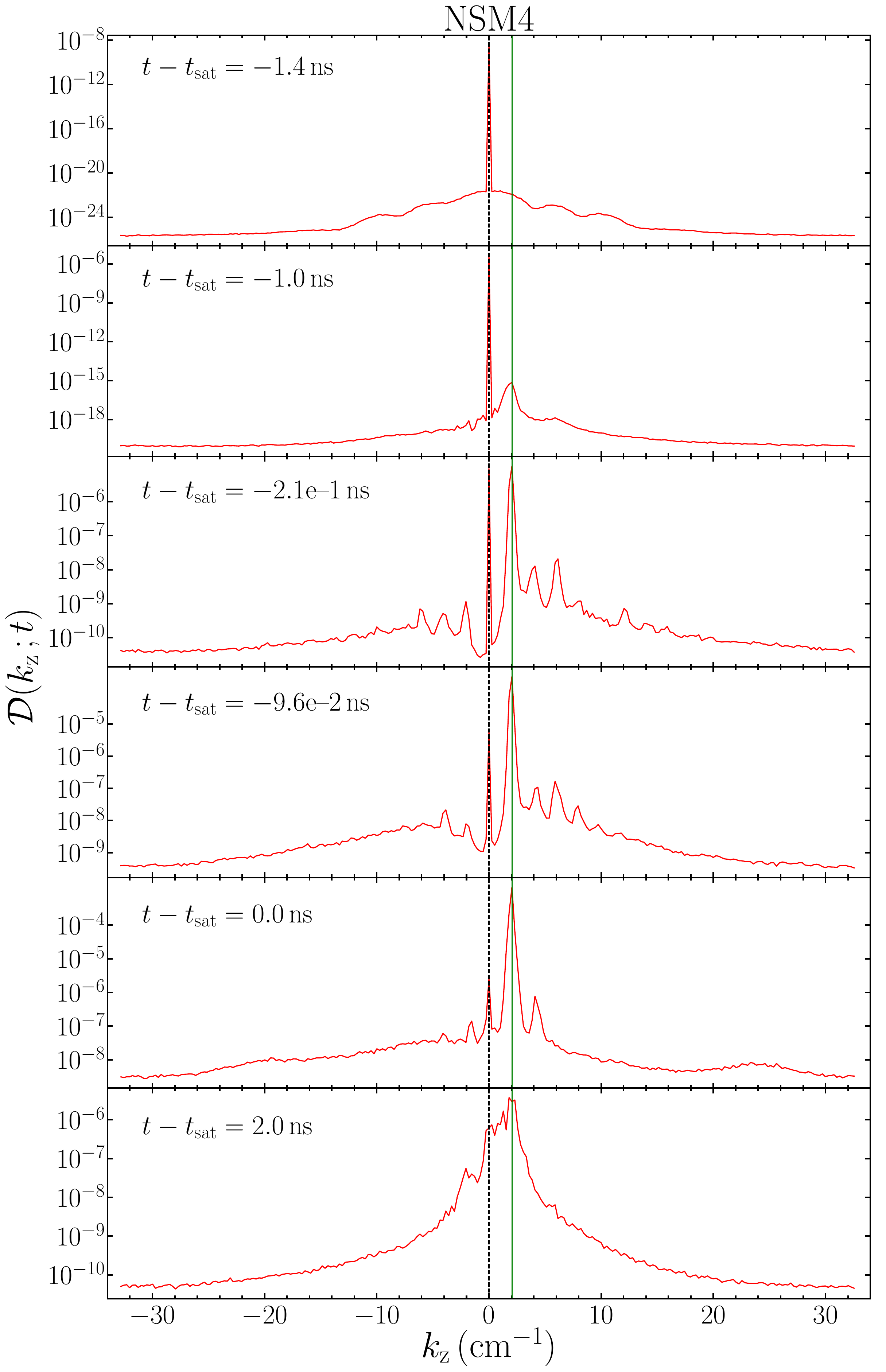}
    \caption{Sequence in time of power in the DFT versus wavenumber in the NSM4 simulation.  Only the \flash results are shown.  Times range from early to late from top to bottom.  Vertical green line gives prediction for \kmax from LSA. 
    \label{fig:NSM4_FFTz_tpanels}
    }
\end{figure}

We next show a sequence of DFTs in time for the homogeneously perturbed \flash calculation (i.e., the red curves on Figs.~\ref{fig:NSM4_N_comp} and \ref{fig:NSM4_Nex_pcomp}) in Fig.~\ref{fig:NSM4_FFTz_tpanels}.  For this figure, we plot the DFT as a function of $k_{\z}$ over positive and negative values.  Notice that the vertical axis scale changes with time as the overall power grows.  The top panel shows the power at a time immediately after $t=0$ when the vacuum potential has sourced $N_{ex}$.  The second panel shows a time snapshot when \avgnex is experiencing oscillations.  The power in $k=0$ dominates, but we observe a small peak growing at a wavevector $k_z=1.8\,{\rm cm}^{-1}$, which corresponds to the fastest growing mode.  The third panel shows the time when the power in \kmax first supersedes the power in $k=0$, corresponding to the start of the exponential growth at $t-\tsat\sim-0.2\,{\rm ns}$ in Fig.\ \ref{fig:NSM4_Nex_pcomp}.  From the second to the third panel, one can clearly see harmonics in the DFT, which we believe to be numerical artifacts.  In the fourth panel, we show a time snapshot during the linear growth phase where the power in $k_z=1.8\,{\rm cm}^{-1}$ dominates over the homogeneous mode.  The fifth panel shows the DFT at saturation, while the sixth panel shows a time snapshot well after the ELN-XLN has dissipated.  The harmonic structure disappears by this late time.

The harmonic structure visible in panels 3-5 of Fig.\ \ref{fig:NSM4_FFTz_tpanels} is not present in the \emu calculation shown in Figs.\ \ref{fig:NSM4_N_comp} and \ref{fig:NSM4_Nex_pcomp}, despite the fact that \emu sources the homogeneous mode at early times.  Although the harmonics are reminiscent of the TwoThirds test calculation in Sec.~\ref{ssec:2_3}, NSM4 differs in two respects: the $k=0$ mode is stable to FFI; and we included the matter term.  We have done a number of tests to isolate the cause of the harmonics.  First, we implemented higher resolution simulations for this particular case and still observe the harmonic structure, indicating that the harmonics are not a result of spatial resolution.  Next, we ran the same geometry but excluded the matter term and do not observe any harmonics.  Lastly, we checked the DFT for the calculation which used off-diagonal seeds (magenta dash-dotted line in Fig.\ \ref{fig:NSM4_Nex_pcomp}) and do not observe any harmonics.

For a more thorough analysis, we ran a series of tests where we used a sinusoidal perturbation to seed the off-diagonal components with a wavelength $\lambda=2\pi/\kmax$ akin to Eqs.\ \eqref{eq:sin_pert1} and \eqref{eq:sin_pert2} but using $\z$ as the coordinate to vary $\delta N_{ex}$. This initial condition for $N_{ex}$ looks much like the setup in Fig.\ \ref{fig:mm_v_MP}, with values of $N_{ex}$ uniformly distributed around a circle in the complex plane. In these tests, we observe that the primary cause for the development of harmonics is the advection term.  Harmonics develop almost immediately when we begin evolving the system in the absence of flavor-transformation terms.  In addition, harmonics are present across the entire $k_\z$ domain of Fig.\ \ref{fig:NSM4_FFTz_tpanels}.  For an ideal advection scheme, the initial sinusoidal perturbation will move along the $\z$ direction. The modulus $\lvert N_{ex} \rvert$ will be constant, whereas the phase $\arg(N_{ex})$ will exhibit a ``sawtooth'' pattern. Over the first 50 time steps in the NSM4 geometry using our advection scheme, the sawtooth pattern in $\arg(N_{ex})$ persists and moves slightly in the $-\z$ direction as expected. $|N_{ex}|$ develops errors which are periodically spaced on the sinusoid.  We notice that these errors in $|N_{ex}|$ lead to prominent harmonics in the DFT.  The power in each harmonic is small when we only set the $\z$-component of the flux to be nonzero, and become larger each time we add another nonzero flux component.  Importantly, the harmonics appear at each $\pm$ integer multiple of \kmax except for one value, namely $k=0$.  When we include the vacuum potential in the test, we see the harmonic features are still present and power now resides in the $k=0$ mode.

We conjecture that these harmonic features are a result of our first-order Riemann solver for \cnumbers.  As shown in Fig.\ \ref{fig:mm_v_MP}, the difference between $\mathcal{M}^{(L)}_{i+\half}$ and $\mathcal{M}^{(R)}_{i-\half}$ is a vector tangent to the circle at $\mathcal{M}_i$.  When adding that difference to the moment value in the particular cell, the effect would be to increase the modulus.  In addition, any numerical error present in the differencing scheme will change the moduli in different amounts, thereby destroying the circular symmetry.  These errors do lead to harmonics at first, but we have verified that the harmonics dissipate if enough time elapses and numerical diffusion distributes power to the homogeneous mode.

In the FFI cases we are interested in studying here, $|N_{ex}|$ is growing due to the flavor transformation as well as the advection.  These numerical errors may be obscured by the underlying physical mechanisms changing $N_{ex}$.
Indeed, we do not discern any harmonic features in most of our simulations.  NSM1, NSM3, Fiducial, and 90Degree do not show a clean sequence of harmonics. We suspect that this is due to the lack of a substantial
homogeneous mode.  The harmonics do not influence the primary outcomes of our simulations. However, we have not surveyed all possible conditions for the fast flavor simulations, so it is important to keep them in mind when using the magnitude phase implementation.

\subsection{Example of a dramatic closure dependence [NSM2]}
\label{ssec:NSM2}

Our final simulation in this work is the NSM2 point.  This point exhibits a shallow ELN crossing in comparison with NSM1 and NSM3.  In Ref.\ \cite{flashri}, both \flashri and \emu showed a linear growth phase compatible with an FFI for the NSM2 geometry, albeit with growth rates different by a factor $\sim5$.
However, the moment LSA~\cite{2024PhRvD.109d3046F} showed that a moment calculation should not be unstable to FFI for the closure prescription which uses Eqs.\ \eqref{eq:P_class}, \eqref{eq:chi_MEC}, and~\eqref{eq:f_FT} for $P_{ex}$. When we use the improved advection scheme developed in the modulus-phase implementation of \flash, we find consistency between the LSA and the simulation results: NSM2 is stable to FFC.

This leaves us with the question of why the modulus-phase implementation does not better match \emu, since NSM2 has all of the characteristics of an FFI and \emu calculations clearly show an instability. We will take this opportunity to discuss improvements to \flash, which can give better agreement with \emu.  Specifically, we focus on choosing a different closure relation for $P_{ex}$, one which does not rely on flavor-traced flux factors and the MEC directly.  We have no reason to believe that \flash cannot accurately describe the flavor-diagonal distributions with the MEC and corresponding flux factors of the individual species, at least in the linear phase of the instability (in the decoherence phase, the MEC for on-diagonal elements is not accurate anymore, see for instance Fig.~\ref{fig:NSM1_Mn0_P_comp}). Therefore, we preserve the flavor-diagonal closure relations which we have used for all other simulations, i.e., Eqs.~\eqref{eq:P_class} and \eqref{eq:chi_MEC}, and we write $\chi_e \equiv \chi(|\vec{F}_{ee}|/E_{ee})$ and similarly for $\chi_x$.

For the off-diagonal pressure moment however, we adopt the following prescription, which comes from the general study of quantum closures carried out in Ref.~\cite{Kneller:2024buy}.\footnote{Using the notations from~\cite{Kneller:2024buy}, this prescription corresponds, in the linear phase of the instability, to set $\theta_P/\theta_E = 1$ and $\phi_P - \phi_E = 0$.}   We first define $\kappa$ in terms of the diagonal energy density moments and Eddington factors 
\begin{equation}
  \kappa\equiv \frac{\chi_eE_{ee} - \chi_xE_{xx}}{E_{ee} - E_{xx}}.
\end{equation}
Using only $\kappa$, we calculate the diagonal Cartesian moments for $P_{ex}$ and set the off-diagonal spatial components to zero
\begin{subequations}
\label{eq:NSM2_pressure}
\begin{align}
  \frac{P^{\z\z}_{ex}}{E_{ex}} &= \kappa,\\
  \frac{P^{\x\x}_{ex}}{E_{ex}} &= \frac{1-\kappa}{2},\\
  P^{\y\y}_{ex} &= P^{\x\x}_{ex},\\
  P^{jk}_{ex} &= 0\qquad j\ne k \, .
\end{align}
\end{subequations}
An analogous relationship exists for $\overline{P}_{ex}$ using $\overline{\kappa}$.
We take $P_{ex}$ to be axially symmetry around the $\z$-axis as that is the direction of the net ELN flux.
The above expression is obtained from considerations valid for the initial conditions and the linear growth phase, such that we do not expect Eq.~\eqref{eq:NSM2_pressure} to provide an accurate description of $P_{ex}$ in the nonlinear phase.

\begin{figure}[!ht]
    \centering      
    \includegraphics[width=\columnwidth]{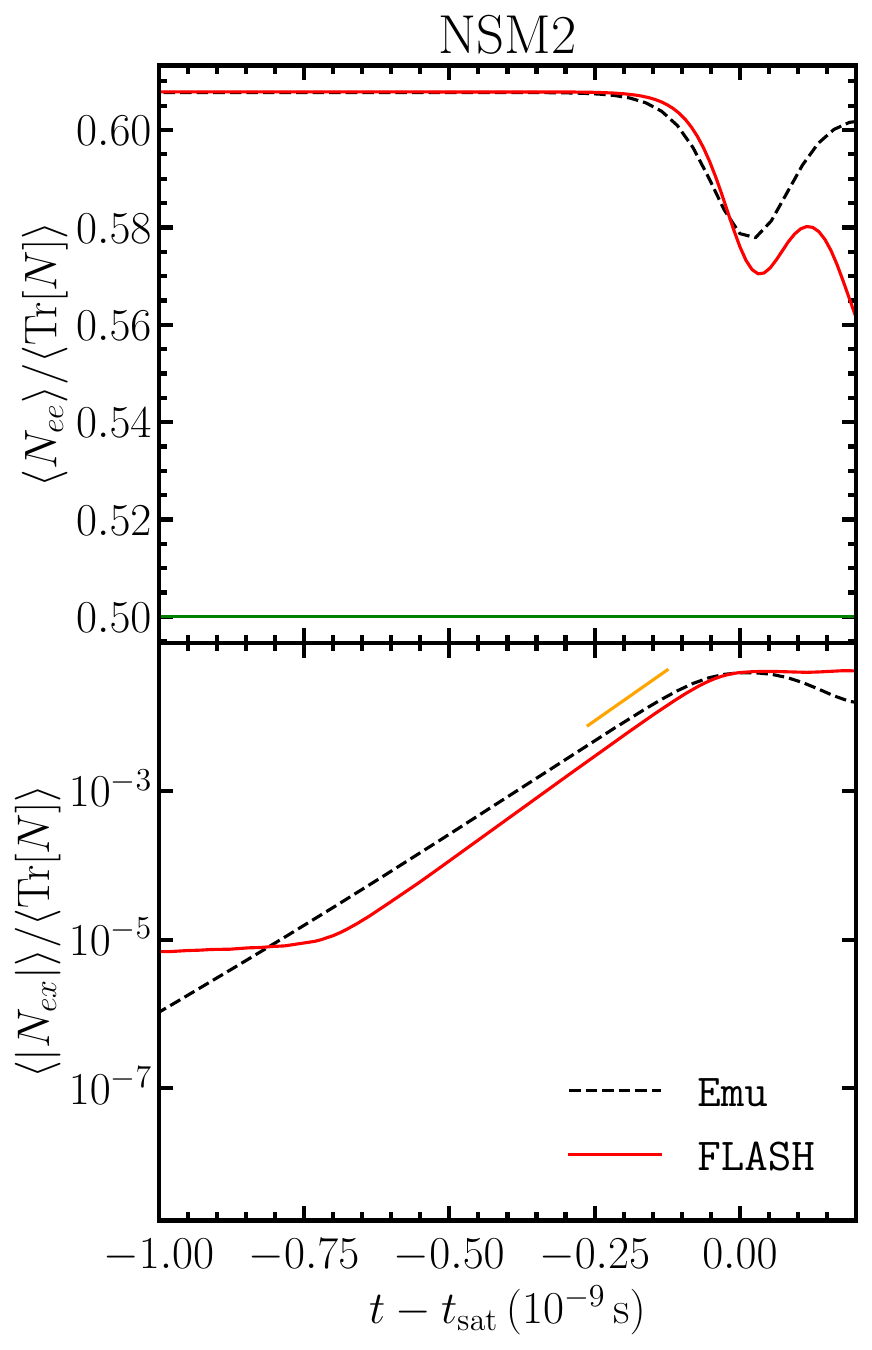}
    \caption{Domain averaged number density components versus time for the NSM2 simulation.  Line styles and colors are the same as Figs.\ \ref{fig:NSM1_Mn0_N_comp} and \ref{fig:NSM1_Mn0_Nex_methods}, but the \flash calculation uses here the closure~\eqref{eq:NSM2_pressure}. 
    \label{fig:NSM2_N_comp}
    }
\end{figure}

We perform a simulation of the NSM2 point using this prescription, and a mass density of $2.20\times10^{11}\,{\rm g/cm}^3$ and an electron fraction $0.192$.  
Figure \ref{fig:NSM2_N_comp} shows the number density moments plotted against time for \flash and the \emu simulation from Ref.~\cite{flashri}.  During the linear growth phase, we observe a faster growth rate in \flash as compared to \emu, but \imo only differs by $\sim20\%$, instead of the factor of 5 discrepancy between \emu and \flashri (and we highlight once again that the \flashri calculation did not show the behavior predicted by LSA for the same parameters).

The calculation using the new closure relation necessitated a larger spatial resolution.  The \emu calculation used a grid of $128^3$ cells with a box length $L=8.27\,{\rm cm}$.
With the pressure moment expressions in Eq.\ \eqref{eq:NSM2_pressure}, the \flash simulations were unresolved until we specified 1024 cells in the $\z$-direction with a box length $L=16.53\,{\rm cm}$, implying four times the resolution in $\z$ as \emu.

\begin{figure}[!ht]
    \centering      
    \includegraphics[width=\columnwidth]{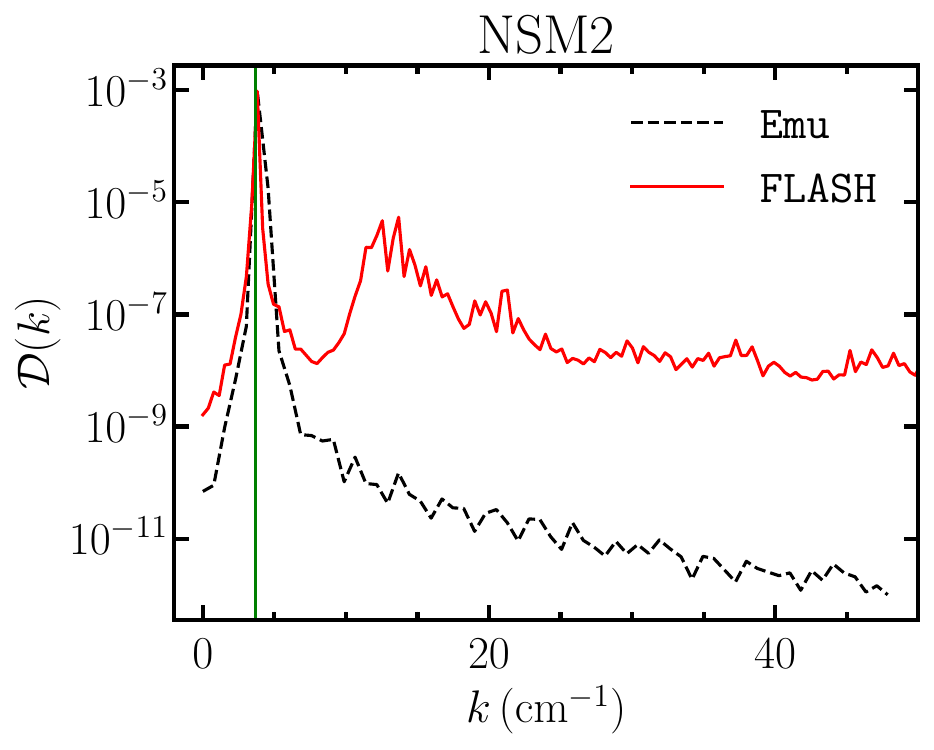}
    \caption{Power in the DFT versus wavenumber for a time $t-\tsat\sim-0.1\times10^{-9}\,{\rm s}$ in the NSM2 simulation. Line styles and colors are the same as Fig.\ \ref{fig:NSM1_Mn0_FFT_comp}. 
    \label{fig:NSM2_FFT_comp}
    }
\end{figure}

We show the DFT at a time $0.1\,{\rm ns}$ before saturation in Fig.\ \ref{fig:NSM2_FFT_comp}. The higher resolution in the \flash calculation gives a larger range of $k$ values for the DFT.  There is no additional structure for $k>50\,{\rm cm}^{-1}$ in the \flash calculation. Although there is excellent agreement between \flash and \emu for the fastest growing mode, the DFT for \flash exhibits other features.  We do not believe the second peak in the \flash curve for $k\sim15\,{\rm cm}^{-1}$ is physical as there is no analog in the \emu curve.  
This peak is not a harmonic of \kmax as the features seen in the TwoThirds and NSM4 simulations were. However, as the power in this peak is two orders of magnitude lower than the fastest growing mode it has minimal effect on the main characteristics of the neutrino evolution. We conclude that care must be taken with the closure when describing the evolution of configurations with shallow crossings.

\section{Conclusions \& Discussion}
\label{sec:conclusion}

Evolving the angular moments of particle distributions is a time-tested approach for solving the equations of radiation hydrodynamics.
A common method for solving the moment evolution equations is to discretize the moments in space/time/momentum and use a Riemann solver for the Riemann problem across the cell boundaries. Generalizing a  Riemann solver for the transport of quantum moments — a necessity to deal with flavor transformation phenomena — is not a straightforward endeavor. We have identified two crucial issues that a Riemann solver needs to address: how to indicate the direction of flow of the flavor space coherence of the flux, and how to reconstruct the off-diagonal elements of the moments at the cell interfaces. Our solution for the first issue is to use a signed modulus to describe the flux moment, and to address the second issue, we designed a new reconstruction algorithm in the spirit of the well-known “minmod” limiter generalized to the complex plane. 

In previous work \cite{flashri}, we had straightforwardly extended the standard minmod approach for classical neutrino transport separately to real and imaginary parts of the flavor off-diagonal moments. Thus, we are able to compare results from our new Riemann solver using the modulus phase implementation (\flash) in a variety of flavor transformation scenarios, not only to a particle-in-cell multi-angle method (\emu) but also to this previous approach (\flashri).  We tested our modulus phase implementation in a variety of scenarios focused on the fast flavor instability. Comparing the new algorithm with \flashri, we see systematic decreases in the errors in key diagnostic quantities, the growth rate \imo, and the wavenumber \kmax in all test cases. Tables~\ref{tab:NSM_results} and \ref{tab:3test_results} list these results.  In all cases, the modulus phase implementation produces results that are closer to those of the particle-in-cell method.  

The modulus phase implementation is more successful because it is inspired by the physical situation at hand. Eqs.~\eqref{eq:lambda_thin_min_smod} and \eqref{eq:lambda_thin_max_smod} for the characteristic speeds $\lambda^j_{\rm thin, min}$ and $\lambda^j_{\rm thin,max}$ represent a particularly important change.
These new expressions do a better job of keeping the characteristic speeds in the optically thin limit in the same range as the particle velocities.   
In the real imaginary formulation this was not the case, and the incongruous characteristic speeds can lead to incorrect spatial fluxes and pressures.
The end result was that $|N_{ex}|$ grew faster than expected, which manifested in a large \imo.  The faster growth rate correlated with a smaller length scale, so the \flashri scheme also generally led to larger values of \kmax.

A glance at Tables \ref{tab:NSM_results} and \ref{tab:3test_results} shows that while the method presented here substantially improves agreement with the particle-in-cell results, some differences remain. An important target for future investigation is the \emph{closure}.
In this paper we have mainly used the same Maximum Entropy Closure (MEC) as in Ref.~\cite{flashri}. The classical MEC is based on a principle of extremizing an entropy functional over solid angle for the distribution, and we extended this method to include the off-diagonal elements of the moments. However, as stressed previously in this work and Ref.\ \cite{flashri}, using Eq.~\eqref{eq:chi_MEC} with the “flavor-traced flux factor” \eqref{eq:f_FT}
is an ad hoc extension using an Eddington factor based on numerical experiments. 
As the results show, this closure works reasonably well, but it is not well motivated from a theoretical perspective. Reference \cite{Froustey:2024sgz} proposes a closure based on extremizing the \emph{von Neumann} entropy functional to couple the distribution for $\varrho_{ex}$ to that of $\varrho_{ee}$ and $\varrho_{xx}$ in the linear regime. The extremization of the von Neumann entropy is constrained by the number and flux moments. Using this ansatz, Ref.\ \cite{Froustey:2024sgz} obtains a closure, which they then use in a linear stability analysis and find regions of instability and associated growth rates in better agreement with a more exact linear stability analysis than the closure relations used in this work. The use of such a closure (which should be generalized to the non-linear regime) in dynamical calculations is left for future work. More generally, it is paramount we develop better closures in particular in the non-linear regime (see in particular Fig.~\ref{fig:NSM1_Mn0_P_comp}).

Finally, we point out that in a large-scale astrophysical simulation that includes flavor transformation of neutrinos, the simulation will need to run for much longer timescales than the simulations presented here.  The results of our simulations suggest that flavor equilibration is not achieved on very short timescales; thus, we have no expectation that an instantaneous flavor equilibration approximation in a classical simulation will adequately describe the physics of, say, neutron star mergers or core-collapse supernovae.   In a dynamical simulation of a compact object, the off-diagonal terms will be nonzero, and could persist as they advect to different locations within the system. We anticipate that an accounting of the flavor off-diagonal neutrino physics will need to be kept, and we hope that the signed modulus phase representation will be useful in this context as well as in the type of shorter timescale phenomena presented here.


\acknowledgments

The authors thank Baha Balantekin, George Fuller, and Erick Urquilla for useful discussions.  This work was supported at NC State by the DOE grant DE-FG02-02ER41216 and DE-SC00268442 (ENAF).  J.F. is supported by the Network for Neutrinos, Nuclear Astrophysics and Symmetries (N3AS), through the National Science Foundation Physics Frontier Center award No. PHY-2020275. F.F. gratefully acknowledges support from the Department of Energy, Office of Science, Office of Nuclear Physics, under contract number DE-AC02-05CH11231, from NASA through grant 80NSSC22K0719, and NSF through grant AST-2107932. S.R. was supported by a National Science Foundation Astronomy and Astrophysics Postdoctoral Fellowship under Award No. AST-2001760. This work was partially supported by the Office of Defense Nuclear Nonproliferation Research \& Development (DNN R\&D), National Nuclear Security Administration, U.S. Department of Energy (GCM). This work is performed in part under the auspices of the U.S. Department of Energy by Lawrence Livermore National Laboratory with support from LDRD project 24-ERD02 (GCM).
\appendix


\section{1D Beam tests}
\label{sec:beam}

In this Appendix, we consider a series of 1D test problems to show the moment-based approach to transport gives results which are a good match to those from the multi-angle \emu code and to analytic results. We shall also compare two versions of \flash: one using the signed modulus-phase representation for \cnumbers, and the other which uses real-imaginary. 

\subsection{Setup}

For these tests we only include in the QKEs the neutrino self-interaction Hamiltonian potential~\eqref{eq:hnu_dens_mat}. The right-hand sides of Eqs.~\eqref{eq:moment_QKEs} are then simplified compared to Eqs.~\eqref{eq:moment_QKEs_3D}, and read
\begin{subequations}
\label{eq:moment_QKEs_1D}
\begin{align}
    \mathcal{S}_E &= - \imath \left[H_E, E\right] + \imath \left[H_{F,\x},F^\x\right] \, , \\
    \mathcal{S}_F^\x &= - \imath \left[H_E, F^\x\right] + \imath \left[H_{F,j},P^{\x j}\right] \, , \\
    {\overline{\mathcal{S}}}_E &= - \imath \left[-H_E^*, \bE\right] - \imath \left[H_{F,\x}^*,\bF^\x\right] \, , \\
    {\overline{\mathcal{S}}}_F^\x &= - \imath \left[-H_E^*, \bF^\x\right] - \imath \left[H_{F,j}^*,\bP^{\x j}\right] \, ,
\end{align}
\end{subequations}
Note that, for these tests, we restrict the neutrino propagation along a single direction ($\hat{\x}$). All our tests in 1D are for a  domain size $L=8.0\,{\rm cm}$ and both \flash and \emu set the number of cells to $N_{gp}=1024$. We use periodic boundary conditions for all calculations. Our initial conditions are such that the neutrinos propagate in the positive $\hat{\x}$ direction in this geometry and have unit flux factors for all flavors.  Conversely, the antineutrinos propagate in the minus $\hat{\x}$ direction and also have unit flux factor.  These ``beam'' configurations imply that all moments of the distribution are either identical to plus/minus the energy density moment or zero. Namely, the first non-zero moments are:
\begin{equation}
\label{eq:moments_1D}
    \begin{aligned}
        P^{\x \x}_{ab} &= E_{ab} \ , \quad  &&F^{\x}_{ab} = E_{ab} \, , \\
        \overline{P}^{\x \x}_{ab} &= \overline{E}_{ab} \ , \quad &&\overline{F}^{\x}_{ab} = - \overline{E}_{ab} \, .
    \end{aligned}
\end{equation}
There is thus no closure needed in that—or, more precisely, the closure is uniquely determined and is given by Eq.~\eqref{eq:moments_1D}.

Initially, all neutrinos and antineutrinos are $e$-flavor.  We will set the initial neutrino number density to be $\N_{ee}(t=0)=4.89\times10^{32}\,{\rm cm}^{-3}$, following the Fiducial tests in \cite{2021PhRvD.104j3023R}.  We use a model parameter to set the electron antineutrino density
\begin{equation}
  \alpha \equiv \frac{\overline{\N}_{ee}(t=0)}{\N_{ee}(t=0)} \, ,
\end{equation}
where $\alpha$ will vary for each test. We further define the asymmetry value and the strength of the self-interaction as follows
\begin{align}
  a &\equiv \frac{1-\alpha}{1+\alpha} \, , \\
  \mu &= \frac{\sqrt{2}\,G_F \,\N_{ee}(t=0)}{2}(1+\alpha) \, .
\end{align}
Ref.~\cite{2016JCAP...03..042C} gives analytic predictions for the growth and fastest growing mode during the linear growth phase in such a configuration:
\begin{align}
  {\rm Im}(\Omega)_{\rm max} &= 2\,\mu\,\sqrt{1-a^2} \, , \label{eq:Beam_ImOmax_pred} \\
  k_{\rm max} &= 2\,\mu\label{eq:Beam_kmax_pred} \, .
\end{align}

\subsection{Results}

\subsubsection{Tests using random initial perturbations}

In order to start the instability, we must seed perturbations for $N_{ex}(t=0)$.\footnote{Note that in the NSM simulations of Sec.~\ref{sec:results}, we test the inclusion of a vacuum term in the Hamiltonian, which has off-diagonal elements that are able to induce the initial flavor transformation.} For our first set of \flash simulations, we randomly seed the modulus of the off-diagonal neutrino component at a scale of one part in $10^6$ compared to $N_{ee}$.  In addition, we randomly seed the phases. The result is as follows
\begin{multline}
\label{eq:mod_phi_seeds}
  N_{ex}(\x_i;t=0) = U_{1,i}\times 10^{-6}\,N_{ee}(\x_i;t=0) \\ \times \exp\{\imath\pi[2\,U_{2,i} - 1]\},
\end{multline}
where $U_{1,i}$ and $U_{2,i}$ are uniformly sampled random numbers in the range $[0,1]$ for cell $i$. 
We seed perturbations in antineutrino moments in the same manner as Eq.~\eqref{eq:mod_phi_seeds}.
The \emu simulations separately perturb the real and imaginary parts of the density matrices of each particle at the level of one part in $10^6$ for each particle.

\begin{figure}[!ht]
    \centering      
    \includegraphics[width=\columnwidth]{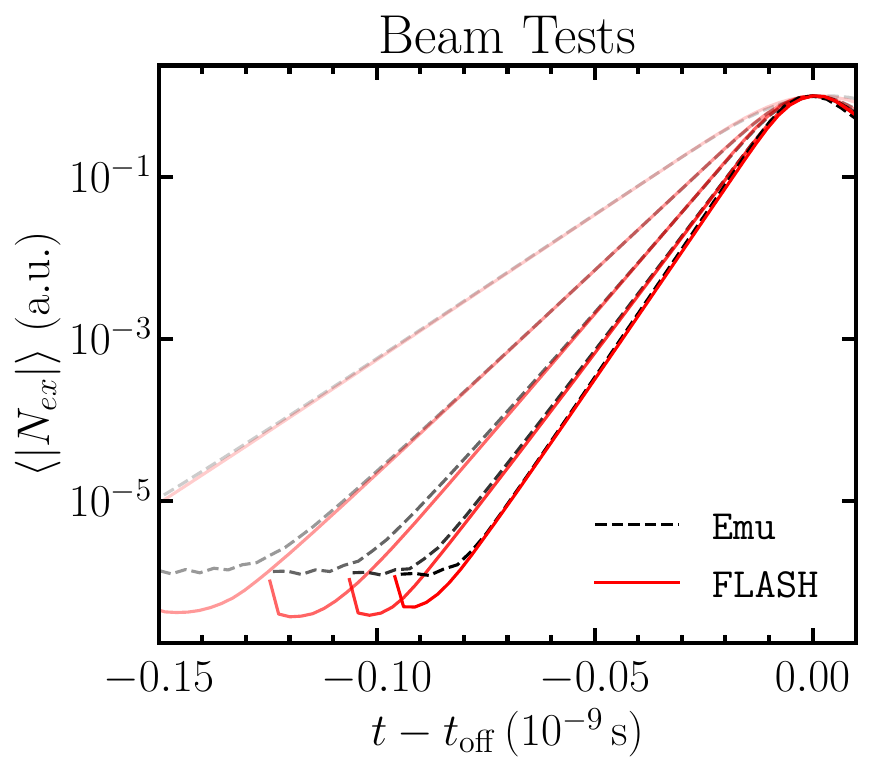}
    \caption{Domain averaged off-diagonal number density moment versus time for the 1D beam tests with random seeding.  Solid red (dashed black) curves give the results from \flash (\emu) calculations. Darker shading indicates increasing values of $\alpha \in \{0.2,0.4,0.6,0.8,1.0\}$.  $t_{\rm off}$ is an offset time applied to the results of each simulation to give alignment in the linear growth phase.
    \label{fig:Beam_N_comp}
    }
\end{figure}

\begin{figure}[!ht]
    \centering      
    \includegraphics[width=\linewidth]{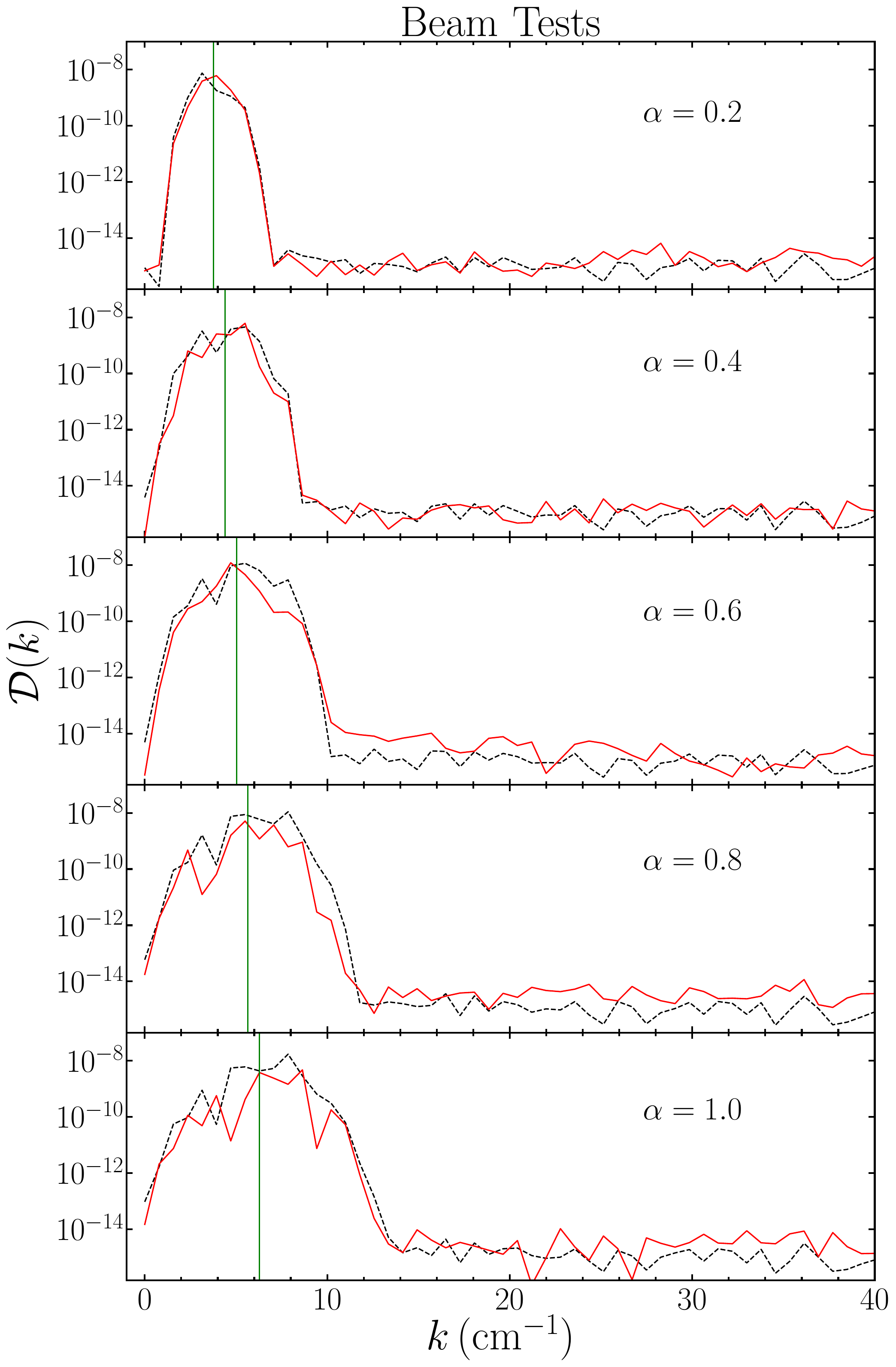}
    \caption{Discrete Fourier Transforms [see Eq.~\eqref{eq:def_DFT}] of off-diagonal number density moment versus wavevector magnitude for the 1D beam tests with random seeding.  Solid red (dashed black) curves correspond to \flash (\emu) calculations. The antineutrino number density increases from top panel ($\alpha=0.2$) to bottom ($\alpha=1.0$). The DFT is calculated at $t\sim t_{\rm off}/2$ during the linear growth phase. The green vertical lines show the fastest growing mode predicted by Eq.~\eqref{eq:Beam_kmax_pred}.  
    \label{fig:Beam_FFT_comp}
    }
\end{figure}

Figure \ref{fig:Beam_N_comp} plots the domain-averaged value of the off-diagonal number density moment as a function of time for the \flash modulus-phase implementation (solid red) and \emu (dashed black) simulations.  Plotted are the results using 5 different values of $\alpha$, namely $\{0.2,0.4,0.6,0.8,1.0\}$ which yield increasing growth rates.  The horizontal axis gives the simulation time offset from a time near the first peak. This arbitrary offset $t_\mathrm{off}$ is chosen to make the visual comparison of the linear phases for \emu and \flash straightforward. The vertical axis gives the domain-averaged magnitude of the off-diagonal number density moment $\langle \lvert N_{ex} \rvert\rangle$, normalized such that $\langle \lvert N_{ex}(t=t_{\rm off}) \rvert \rangle=1$.  We see an excellent agreement between the two sets of simulations for the growth rate once the instability sets in.  \flash maintains a faster growth rate for all values of $\alpha$ except for the symmetric case.  To determine \kmax, we show the DFTs for the simulations in Fig.\ \ref{fig:Beam_FFT_comp} as a function of $k=|\vec{k}|$.  The DFTs are taken at a timestep halfway between $t=0$ and $t=t_{\rm off}$ for each simulation.  We use the same definitions as in~\cite{2021PhRvD.104j3023R}, but normalizing the power in each $k$-mode by the sum of the squares of the initial diagonal number density components, that is:
\begin{equation}
\label{eq:def_DFT}
    \mathcal{D}(k;t) \equiv \frac{\lvert\widetilde{N}_{ex}(k;t)\rvert^2}{\displaystyle\sum_{\vec{x}}\left[N^2_{ee}(\vec{x};t=0) + N^2_{xx}(\vec{x};t=0)\right]} \, ,
\end{equation}
where the Fourier transforms are defined as~\cite{2021PhRvD.104j3023R}:
\begin{align}
\label{eq:def_DFT1}
    \widetilde{N}_{ex}(\vec{k};t) &\equiv \sum_{\vec{x}} e^{-\imath \vec{k}\cdot\vec{x}}N_{ex}(\vec{x};t) \, , \\
\label{eq:def_DFT2}
    \lvert \widetilde{N}_{ex}(k;t) \rvert^2 &\equiv \int{k^2 \dd{\vec{n}} \, \lvert\widetilde{N}_{ex}(k \vec{n};t)\rvert^2} \, .
\end{align}
In~\eqref{eq:def_DFT1}, we sum over all discrete spatial grid positions, and we actually compute a discrete version of~\eqref{eq:def_DFT2} by attributing the contributions of each discrete $\vec{k}$-mode to bins in magnitude $k$.

Each panel shows \flash and \emu results from a different value of $\alpha$.  The vertical green line is the prediction of \kmax from Eq.~\eqref{eq:Beam_kmax_pred}.  Although the peak in the spectrum is broad, we see that the \flash results are consistent with \emu and the prediction from Ref.~\cite{2016JCAP...03..042C}.

\begin{figure}[!ht]
    \centering      
    \includegraphics[width=\columnwidth]{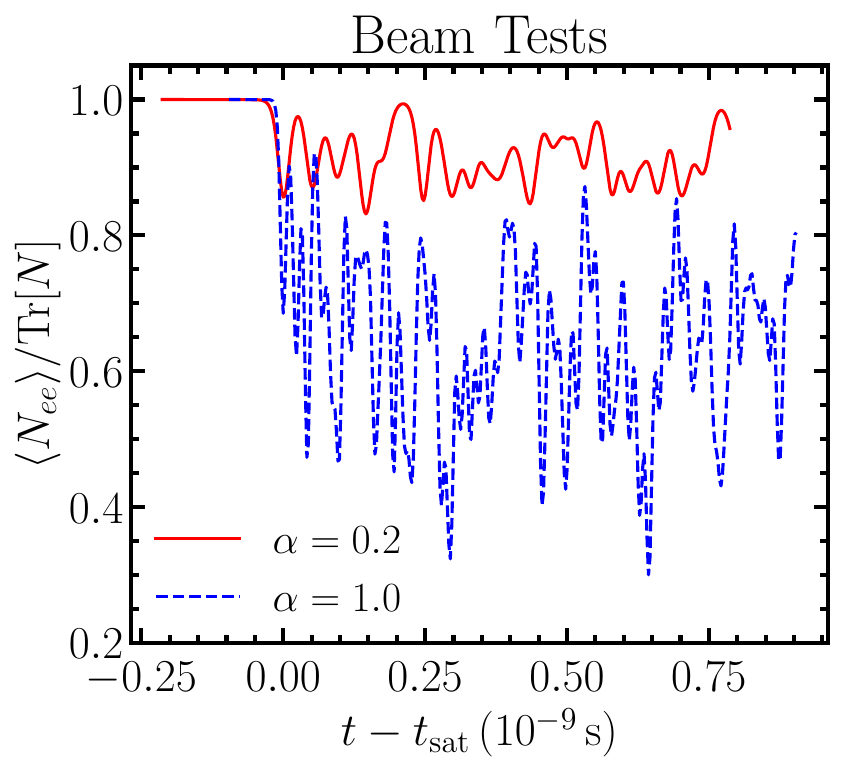}
    \caption{Domain-averaged value of the electron neutrino number density moment normalized by the trace of $N$ versus time for two \flash 1D beam tests with random seeding. $t_{\rm sat}$ is the saturation time.  
    \label{fig:2Beam_N}
    }
\end{figure}

Table \ref{tab:beam_random} summarizes the results for the beam tests with random-perturbation seeding.  We show the analytic, \flash, and \emu results, along with simulations from the real-imaginary implementation of \flash from Ref.\ \cite{flashri}.  Columns 2 and 3 of Table \ref{tab:beam_random} give \imo and \kmax, respectively.  \imo is calculated from the slope of $\langle \lvert N_{ex} \rvert\rangle$ versus $t$ in Fig.\ \ref{fig:Beam_N_comp}. 
For all values of $\alpha$, we see smaller growth rates in the simulations than the analytic prediction -- regardless of the method of calculation.  Although this is a highly idealized system designed to produce an analytic solution, our various methods of calculation rely on numerical routines to integrate the system of equations and as a result we do not expect perfect agreement between calculation and prediction.  We have verified for the modulus-phase implementation of \flash that we see worse agreement for \imo with coarser resolution at all values of $\alpha$ shown in Tab.\ \ref{tab:beam_random}. This indicates \imo does have a dependence on the numerical settings of the simulation.
In the third column of Tab.\ \ref{tab:beam_random}, we determine \kmax by halving the width of the broad peak in Fig.\ \ref{fig:Beam_FFT_comp} for \flash, \emu, and \flash ($ri$).  We estimate the peak width based on the noise floor in Fig.\ \ref{fig:Beam_FFT_comp}. Column 3 is an estimate but shows that the center of the peak roughly coincides with the prediction in Eq.\ \eqref{eq:Beam_kmax_pred}. The fourth column of Table~\ref{tab:beam_random} shows a time and domain averaged value of the $e$-flavor number density moment post-FFI, normalized by the trace. Specifically, the quantity $\langle N_{ee}\rangle_t$ is averaged from $2\tsat$ until the end of the simulation $t=1\,{\rm ns}$. Note that $\langle N_{ee}\rangle$ does not asymptotically converge in the beam geometry as the FFI is perpetual and $N_{ee}$ undergoes chaotic motion --- see in particular Refs.~\cite{2014PhRvD..90b5009H,2024PhRvD.109j3040U} and Fig.~(4d) of Ref.~\cite{2023PhRvD.108d3007X}.  As an aside, we show $\langle N_{ee}\rangle$ versus time for $\alpha=0.2,1.0$ in Fig.\ \ref{fig:2Beam_N} which clearly shows a lack of asymptotic behavior.  We have verified that perturbing the system slightly, i.e., picking new random numbers in Eq.\ \eqref{eq:mod_phi_seeds}, gives a set of results diverging from the curves in Fig.\ \ref{fig:2Beam_N} and indicating chaos. However, the time-averaged moments post-instability have well-defined values, independently of the initial conditions, allowing us to quote these values in Table~\ref{tab:beam_random}. In addition, we observe that the amplitude of oscillations and average about those oscillations in Fig.\ \ref{fig:2Beam_N} does scale with $\alpha$. Results for \flash and \flash ($ri$) are nearly identical, but results for \emu show slightly higher values, indicating that our moment methods slightly overestimate the amount of flavor conversion in this test.

\begin{table}
    \begin{tabular}{c|ccc}
        \multirow{2}{*}{Name} & \imo & \kmax & \multirow{2}{*}{$\langle N_{ee}\rangle_t/{\rm Tr}[N]$} \\
         & $(10^{10}\,\mathrm{s}^{-1})$ & $(\mathrm{cm}^{-1})$ &  \\\hline
        $\alpha=0.2$    & & & \\
        Analytic        &  $8.42$ & $3.8$ &         \\
        \flash ($ri$)   &  $8.22$ & $3.9(8)$ & $0.91$ \\
        \flash          &  $8.23$ & $3.5(8)$ & $0.91$ \\
        \emu            &  $8.08$ & $3.5(8)$ & $0.92$ \\ \hline
        $\alpha=0.4$    &  &  &  \\
        Analytic        & $11.91$ & $4.4$ &         \\
        \flash ($ri$)   & $11.69$ & $4.3(8)$ & $0.81$ \\
        \flash          & $11.60$ & $4.7(8)$ & $0.82$ \\
        \emu            & $11.50$ & $4.3(8)$ & $0.84$ \\ \hline
        $\alpha=0.6$    &  &  &  \\
        Analytic        & $14.59$ & $5.0$ &         \\
        \flash ($ri$)   & $14.06$ & $5.1(8)$ & $0.72$ \\
        \flash          & $14.48$ & $5.5(8)$ & $0.74$ \\
        \emu            & $14.06$ & $5.1(8)$ & $0.75$ \\ \hline
        $\alpha=0.8$    &  &  &  \\
        Analytic        & $16.85$ & $5.7$ &         \\
        \flash ($ri$)   & $16.12$ & $6.3(8)$ & $0.68$ \\
        \flash          & $16.43$ & $6.3(8)$ & $0.68$ \\
        \emu            & $16.17$ & $5.9(8)$ & $0.69$ \\ \hline
        $\alpha=1.0$    &  &  &  \\
        Analytic        & $18.84$ & $6.3$ &         \\
        \flash ($ri$)   & $18.41$ & $7.0(8)$ & $0.63$ \\
        \flash          & $18.10$ & $7.0(8)$ & $0.63$ \\
        \emu            & $18.23$ & $7.0(8)$ & $0.65$ \\ \hline
    \end{tabular}
    \caption{Results for the analytic prediction of Ref.\ \cite{2016JCAP...03..042C}, the real-imaginary implementation of \flash ($ri$), the modulus-phase implementation of \flash, and \emu calculations for the beam tests with random seeding.  The domain size is $L=8.0$ cm and there are 1024 cells in the simulations.  The second column gives the growth rate during the linear phase and the third column the fastest growing mode wavevector.  For both sets of \flash and the \emu simulations, \kmax is determined by taking half of the width of the peak in the DFT.  The uncertainty in the last digit of \kmax is shown in parentheses.  The last column is the domain-and-time-averaged value of $N_{ee}$ after saturation of the instability, normalized by the trace.
    }
    \label{tab:beam_random}
\end{table}

\subsubsection{Tests with sinusoidal initial perturbations}

We consider another series of beam tests with periodic perturbations.  Using the analytic expression in Eq.\ \eqref{eq:Beam_kmax_pred}, we seed a sinusoidal perturbation for the off-diagonal number density moments
\begin{align}
  N_{ex}(t=0) &= 10^{-6}N_{ee}(t=0)\exp[+\imath\kmax \x],\label{eq:sin_pert1}\\
  \overline{N}_{ex}(t=0) &= 10^{-6}\overline{N}_{ee}(t=0)\exp[-\imath\kmax \x],\label{eq:sin_pert2}
\end{align}
assuming the beams propagate in the $\pm \hat{\x}$ direction.  We modify the size of the domain so that we can fit 8 wavelengths on the domain while keeping the number of cells at 1024.  The plots for number density versus time are nearly identical to Fig.~\ref{fig:Beam_N_comp} in the linear growth phase.  The DFTs do exhibit well-defined peaks at the expected \kmax.
Table~\ref{tab:beam_periodic} gives the growth rates for the \flash simulations with the periodic perturbations (\emu results not included).  We observe excellent agreement between the analytic prediction and both implementations of \flash.

\begin{table}[!ht]
    \begin{tabular}{c|ccc}
         $\alpha$ & Analytic & \flash & \flash ($ri$) \\ \hline
        $0.2$ & $8.42$  & $8.41$  &  $8.41$ \\
        $0.4$ & $11.91$ & $11.91$ & $11.88$\\
        $0.6$ & $14.59$ & $14.59$ & $14.55$\\
        $0.8$ & $16.85$ & $16.85$ & $16.79$\\
        $1.0$ & $18.84$ & $18.83$ & $18.78$\\
    \end{tabular}
    \\
    \caption{Results for the growth rates for \flashri and \flash beam tests with periodic perturbations, compared with the analytic prediction~\eqref{eq:Beam_ImOmax_pred}. All rates are in units of $10^{10}\,{\rm s}^{-1}$.
    \label{tab:beam_periodic}
    }
\end{table}

In conclusion, we assert that the modulus-phase implementation does no worse than the real-imaginary implementation of \cnumbers in \flash for the beam tests.  Tables \ref{tab:beam_random} and \ref{tab:beam_periodic} show no systematic preference for one method of \cnumber implementation over the other.  In addition, comparing either implementation to \emu shows close quantitative agreement.  Our motivation for developing the modulus-phase implementation was based on the heuristic vision that the distributions for the off-diagonal neutrino density matrix have non-zero flux moments, and that moment has a specific direction in coordinate space.  To realize that vision, we implemented Eqs.\ \eqref{eq:lambda_thin_min_smod} and \eqref{eq:lambda_thin_max_smod} along with the associated infrastructure and algorithms. To show that the modulus-phase implementation is indeed a worthwhile pursuit, we need a 3D geometry, non-unity flux factors, and a closure relationship. In this more general setup, presented in Sec.~\ref{sec:results}, Eqs.\ \eqref{eq:lambda_thin_min_smod} and \eqref{eq:lambda_thin_max_smod} indeed systematically change our neutrino flavor transformation results.

\section{Fiducial and 90Degree test cases}
\label{app:fid_90d}

Reference \cite{flashri} contained three test cases with simple geometries: Fiducial, 90Degree, and TwoThirds. In Sec.~\ref{ssec:2_3}, we studied the TwoThirds test case and revealed a system with two linear growth phases.  We revisit the other two test cases in this appendix.

Table \ref{tab:3test_parameters} gives the initial conditions and geometries for all three test cases. We use the rectangular cell geometries with $16\times16\times128$ number of grid points for \flash simulations, and the cubic geometry with $128^3$ cells for \emu.  Briefly summarized: the Fiducial test case is axially symmetric about $\z$ with opposing fluxes of neutrinos versus anti-neutrinos; the 90Degree test case is identical to the Fiducial, with the antineutrino flux rotated $90^{\circ}$ into the $\y$-direction.  Although all of these tests are symmetric in at least one dimension, we use a 3D geometry for our simulation setup.  For the \flash simulations, we use solar mass splitting and mixing angles, namely, $\delta m^2=7.53\times10^{-5}\,{\rm eV}^2$, $\theta=0.587$, yielding a vacuum frequency of $\omega_{\odot}\sim10^{3}\,{\rm s}^{-1}$.  We set the matter term to zero.  The \emu simulations only utilize the self-interaction potential with initial perturbations seeded in the off-diagonal density matrix elements.

\begin{table*}[ht]
    \centering
    \begin{tabular}{c|ccccccc}
        \multirow{2}{*}{Name} & $\mathcal{N}_{ee}$ & $\overline{\mathcal{N}}_{ee}$ & $\sum \mathcal{N}_{(x)}$ & $\mathbf{f}_{ee}$ & $\overline{\mathbf{f}}_{ee}$ & $\quad \mathbf{f}_{xx}=\overline{\mathbf{f}}_{xx} \quad $ & $L$ \tabularnewline
         & $(10^{32}\,\mathrm{cm}^{-3})$ & $(10^{32}\,\mathrm{cm}^{-3})$ & $(10^{32}\,\mathrm{cm}^{-3})$ & &&& (cm)\tabularnewline \hline
        Fiducial & $4.89$ & $4.89$ & 0 & $(0, \quad   0 \quad \ ,\ 1/3 \quad )$ & $(0, \quad   0 \quad \  , \ \, -1/3 \ \ )$ & $(0,0,0)$ & 8 \tabularnewline
        90Degree & $4.89$ & $4.89$ & 0 & $(0,1/\sqrt{18},1/\sqrt{18})$ & $(0,1/\sqrt{18},-1/\sqrt{18})$ & $(0,0,0)$ & 8 \tabularnewline
        TwoThirds & $4.89$ & $3.26$ & 0 & $(0,\quad   0 \quad \  , \quad   0 \quad \ )$ & $(0, \quad   0 \quad \ , \ \, -1/3 \ \ )$ & $(0,0,0)$ & 32
    \end{tabular}
    \caption{List of simulation parameters and initial conditions for the three 3D test cases. The first three columns show the number densities of each anti/neutrino flavor. For clarity, the third column shows the sum of all four heavy lepton anti/neutrino densities. Our two-flavor simulations assume $\mathcal{N}_{xx}=\overline{\mathcal{N}}_{xx}=\sum \mathcal{N}_{(x)}/4$, where we assume that the other half of the heavy-lepton neutrinos do not participate in flavor mixing. The next three columns show the flux factor vectors, the norm of which are the flux factors. The last column shows the length of each side of the domain.
    }
    \label{tab:3test_parameters}
\end{table*}

\begin{figure*}[!ht]
    \centering      
    \includegraphics[width=0.65\linewidth]{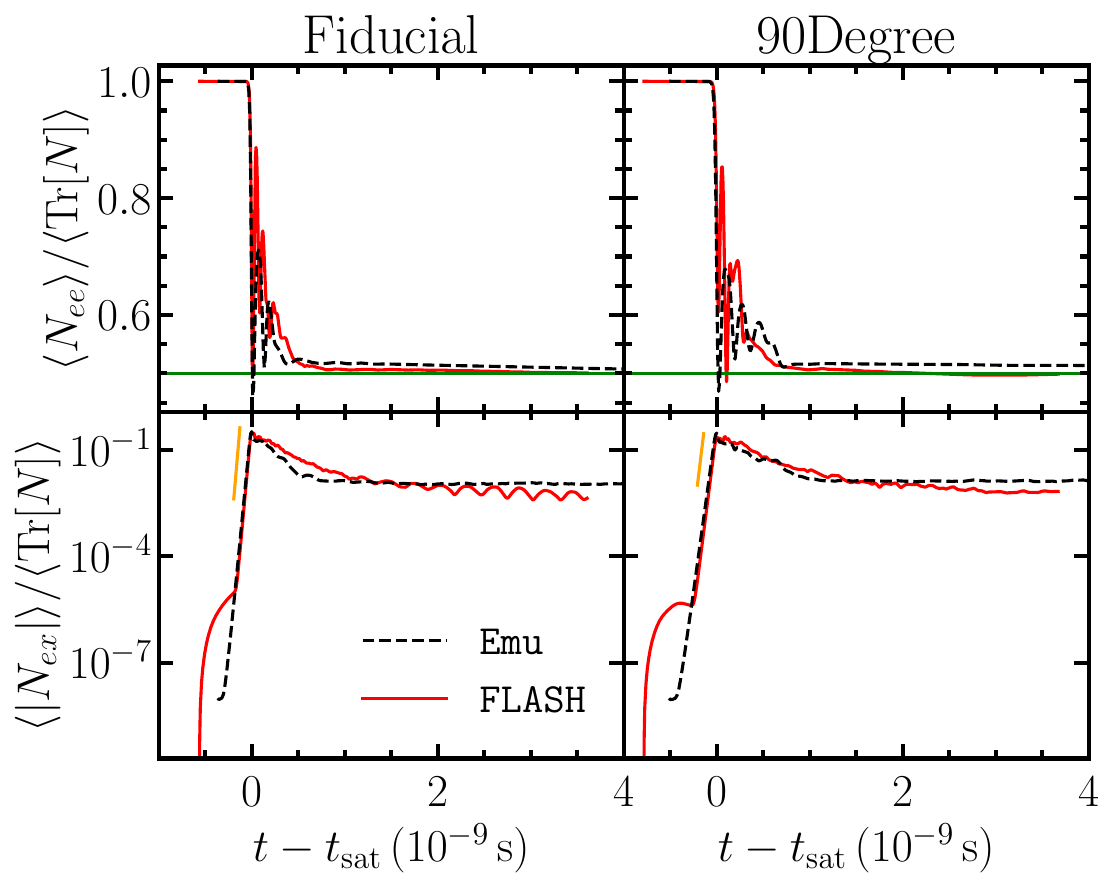}
    \caption{Domain averaged number density moments versus time for the Fiducial and 90Degree test cases.  Line styles and colors are the same as Fig.~\ref{fig:NSM1_Mn0_N_comp} and Fig.~\ref{fig:NSM1_Mn0_Nex_methods}.
    \label{fig:2tests_N_comp}
    }
\end{figure*}

Figure \ref{fig:2tests_N_comp} shows the results from the Fiducial and 90Degree simulations for the domain-averaged number density moments.  For both test cases, there is a rapid increase from zero in $|N_{ex}|$ as the vacuum potential sources the off-diagonal flavor components.  Once $\langle |N_{ex}|\rangle\sim10^{-5}\langle\Tr[N]\rangle$, the linear growth phase of FFC commences and a one-to-one comparison between \flash and \emu is appropriate.
We observe excellent agreement between \flash and \emu during this phase.  We do see a difference in the decoherence phase with the flavor oscillations for the two sets of calculations being out of alignment.  In fact, the damped oscillations appear much more distinct in the Fiducial \emu calculation despite the larger $\langle|N_{ex}|\rangle$ values in \flash.  We suspect the cause of the incongruity is due to the MEC relation we use for all of the components, which is manifestly different from the distribution found in \emu (see Fig.~3 of Ref.\ \cite{flashri}).  Even with an incorrect closure relation, we find that the asymptotic $\langle N_{ee}\rangle$ abundance is $0.5$, a necessary result given the $CP$ symmetry of the initial setup \cite{2022PhRvL.129z1101N,2023PhRvD.107j3022Z,2023PhRvD.108f3003X}.

\begin{figure}[!ht]
    \centering      
    \includegraphics[width=0.9\linewidth]{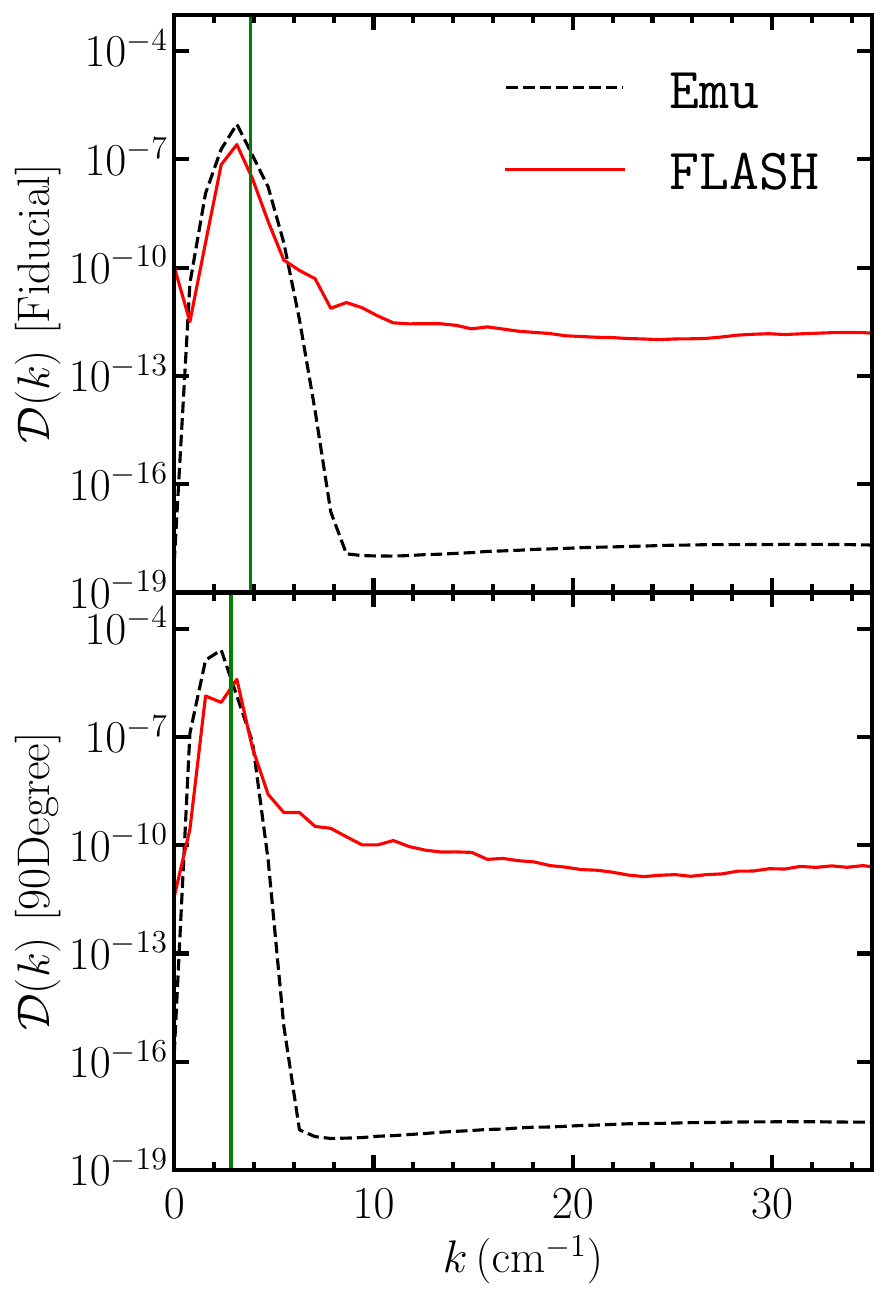}
    \caption{Power in the DFTs versus wavenumber for the Fiducial (top) and 90Degree (bottom) test cases. Line styles and colors are the same as Fig.\ \ref{fig:NSM1_Mn0_FFT_comp}.
    \label{fig:2tests_FFT_comp}
    }
\end{figure}

Figure \ref{fig:2tests_FFT_comp} shows the DFTs for the two test cases, with a similar nomenclature borrowed from past figures.  The DFTs are taken at a snapshot $t-\tsat\sim-0.1\,{\rm ns}$. The Fiducial and 90Degree simulations show power on a scale \kmax in agreement with \emu and the LSA predictions.
No clear peak in the DFT exists for the 90Degree \flash simulation, so we estimate the fastest growing mode at $\kmax\sim2.4\,{\rm cm}^{-1}$.
The Fiducial \flash simulation also shows some residual power at $k=0$ from the initial seeding of the perturbations.  No such power is visible in the 90Degree case due to a larger amount of time having elapsed for this snapshot.
We do not observe any obvious harmonics in either test case at the current time snapshot, or any other times.

Table \ref{tab:3test_results} gives a summary of the growth rates and fastest growing modes for each of the three test cases.  \flash \imo values are all larger than the \emu values, but we see better agreement than we do with \flashri.  \kmax is identical between \flash and \emu within the uncertainty.

\begin{table}
    \begin{tabular}{c|cc}
        \multirow{2}{*}{Name} & \imo & \kmax \\ 
         & $(10^{10}\,\mathrm{s}^{-1})$ & $(\mathrm{cm}^{-1})$ \\\hline 
        Fiducial        & & \\
        LSA             & $7.04$ & $3.8$ \\
        \flash ($ri$)   & $7.1$ & $3.9(4)$ \\
        \flash          & $6.6$ & $3.1(4)$ \\
        \emu            & $6.3$ & $3.1(4)$ \\ \hline
        90Degree        &  & \\
        LSA             & $5.01$ & $2.9$ \\
        \flash ($ri$)   & $5.4$ & $3.1(4)$ \\
        \flash          & $4.8$ & $2.4(4)$ \\
        \emu            & $4.4$ & $2.4(4)$ \\ \hline
        TwoThirds       &  & \\
        LSA             & $1.57$ & $1.5$ \\
        \flash ($ri$)   & $2.0$ & $1.8(1)$ \\
        \flash          & $1.4$ & $1.4(1)$ \\
        \emu            & $1.2$ & $1.4(1)$ \\ \hline
    \end{tabular}\\
    \caption{Growth rates and fastest growing mode values for the 3 test simulations. Nomenclature is the same as Table \ref{tab:NSM_results}.
    }
    \label{tab:3test_results}
\end{table}

\section{NSM4 crossing}
\label{app:NSM4}

\begin{figure}[!ht]
    \centering      
    \includegraphics[width=0.94\columnwidth]{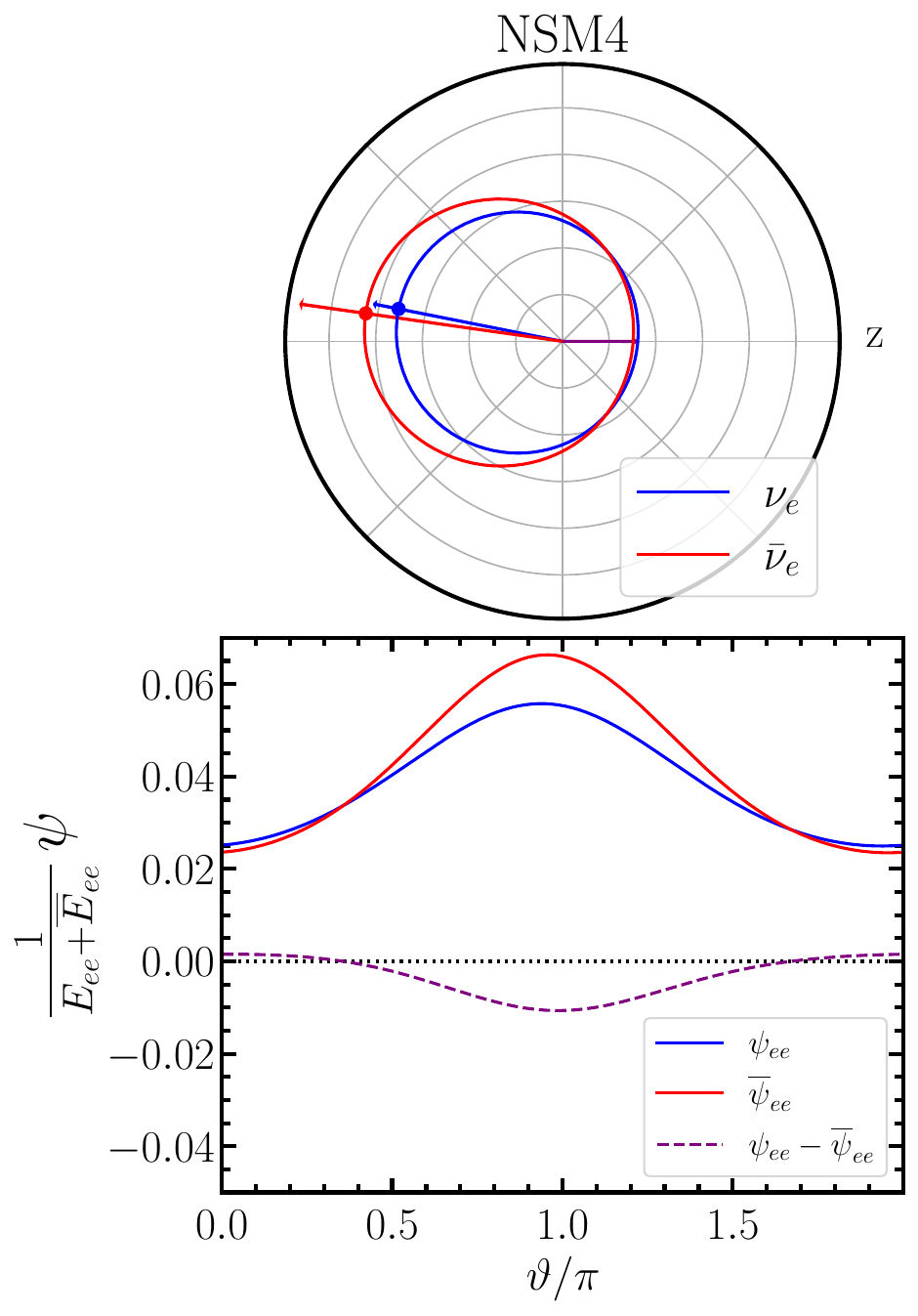}
    \caption{[Top] Polar representation of angular distributions for electron neutrinos (blue) and electron anti-neutrinos (red) for the NSM4 point at the beginning of the simulation.  Blue (red) vectors indicate the net flux direction.  Purple vector is the difference of blue and red vectors and points along the $\z$ axis by construction.  [Bottom] Angular distributions for $\nu_e$, $\overline{\nu}_e$ and the difference (purple) as a function of polar angle $\vartheta$. Angular distributions for $\nu_x$ and $\overline{\nu}_x$ are identical to one another at the beginning of the simulation.
    \label{fig:NSM4_eln}
    }
\end{figure}

In order to visualize the ELN crossing in the initial flavor configuration of the NSM4 point introduced in Sec.~\ref{ssec:NSM4}, we show a polar (top) and standard (bottom) representation for the initial electron flavor neutrino (blue) and antineutrino (red) angular distributions in Fig.~\ref{fig:NSM4_eln} (see Fig.~6 in \cite{flashri} for the same plots with NSM1,2,3). The data for the energy and flux density moments is taken from an M1 simulation in Ref.~\cite{Foucart:2016rxm}. Consistently with this simulation, the angular distributions are given by classical maximum entropy (“Minerbo”) functions—see Eq.~(34) in \cite{flashri}.  

The polar representations show 2D cross sections of the distributions such that the net flux for each distribution is in the figure plane.  That net flux direction is given by the blue and red dots on the respective polar distributions, while the flux vector is shown by the arrows. We take the difference in the flux vectors to find the net ELN flux in dashed purple and define the $\z$ axis to be coincident with this direction.

The bottom panel of Fig.\ \ref{fig:NSM4_eln} shows the same red and blue curves as the top panel plotted versus polar angle $\vartheta$.  We normalize the distributions $\psi$ by the sum $E_{ee}+\overline{E}_{ee}$ for convenience.  In addition, we plot the difference in the distributions as the dashed purple curve.  We observe a shallow ELN crossing for this particular point — the configuration is somehow similar to the NSM2 point (see Fig.~6 in \cite{flashri}, center panels), these two points being spatially close, located above the hypermassive neutron star (see Fig.~\ref{fig:locations_NSM}).

\clearpage

\bibliography{references}

\begin{thebibliography}{91}%
\makeatletter
\providecommand \@ifxundefined [1]{%
 \@ifx{#1\undefined}
}%
\providecommand \@ifnum [1]{%
 \ifnum #1\expandafter \@firstoftwo
 \else \expandafter \@secondoftwo
 \fi
}%
\providecommand \@ifx [1]{%
 \ifx #1\expandafter \@firstoftwo
 \else \expandafter \@secondoftwo
 \fi
}%
\providecommand \natexlab [1]{#1}%
\providecommand \enquote  [1]{``#1''}%
\providecommand \bibnamefont  [1]{#1}%
\providecommand \bibfnamefont [1]{#1}%
\providecommand \citenamefont [1]{#1}%
\providecommand \href@noop [0]{\@secondoftwo}%
\providecommand \href [0]{\begingroup \@sanitize@url \@href}%
\providecommand \@href[1]{\@@startlink{#1}\@@href}%
\providecommand \@@href[1]{\endgroup#1\@@endlink}%
\providecommand \@sanitize@url [0]{\catcode `\\12\catcode `\$12\catcode
  `\&12\catcode `\#12\catcode `\^12\catcode `\_12\catcode `\%12\relax}%
\providecommand \@@startlink[1]{}%
\providecommand \@@endlink[0]{}%
\providecommand \url  [0]{\begingroup\@sanitize@url \@url }%
\providecommand \@url [1]{\endgroup\@href {#1}{\urlprefix }}%
\providecommand \urlprefix  [0]{URL }%
\providecommand \Eprint [0]{\href }%
\providecommand \doibase [0]{http://dx.doi.org/}%
\providecommand \selectlanguage [0]{\@gobble}%
\providecommand \bibinfo  [0]{\@secondoftwo}%
\providecommand \bibfield  [0]{\@secondoftwo}%
\providecommand \translation [1]{[#1]}%
\providecommand \BibitemOpen [0]{}%
\providecommand \bibitemStop [0]{}%
\providecommand \bibitemNoStop [0]{.\EOS\space}%
\providecommand \EOS [0]{\spacefactor3000\relax}%
\providecommand \BibitemShut  [1]{\csname bibitem#1\endcsname}%
\let\auto@bib@innerbib\@empty
\bibitem [{\citenamefont {{Radice}}\ \emph {et~al.}(2020)\citenamefont
  {{Radice}}, \citenamefont {{Bernuzzi}},\ and\ \citenamefont
  {{Perego}}}]{2020ARNPS..70...95R}%
  \BibitemOpen
  \bibfield  {author} {\bibinfo {author} {\bibfnamefont {D.}~\bibnamefont
  {{Radice}}}, \bibinfo {author} {\bibfnamefont {S.}~\bibnamefont
  {{Bernuzzi}}}, \ and\ \bibinfo {author} {\bibfnamefont {A.}~\bibnamefont
  {{Perego}}},\ }\href {\doibase 10.1146/annurev-nucl-013120-114541} {\bibfield
   {journal} {\bibinfo  {journal} {Annual Review of Nuclear and Particle
  Science}\ }\textbf {\bibinfo {volume} {70}},\ \bibinfo {pages} {95} (\bibinfo
  {year} {2020})},\ \Eprint {http://arxiv.org/abs/2002.03863} {arXiv:2002.03863
  [astro-ph.HE]} \BibitemShut {NoStop}%
\bibitem [{\citenamefont {{Arnett}}\ \emph {et~al.}(1989)\citenamefont
  {{Arnett}}, \citenamefont {{Bahcall}}, \citenamefont {{Kirshner}},\ and\
  \citenamefont {{Woosley}}}]{1989ARA&A..27..629A}%
  \BibitemOpen
  \bibfield  {author} {\bibinfo {author} {\bibfnamefont {W.~D.}\ \bibnamefont
  {{Arnett}}}, \bibinfo {author} {\bibfnamefont {J.~N.}\ \bibnamefont
  {{Bahcall}}}, \bibinfo {author} {\bibfnamefont {R.~P.}\ \bibnamefont
  {{Kirshner}}}, \ and\ \bibinfo {author} {\bibfnamefont {S.~E.}\ \bibnamefont
  {{Woosley}}},\ }\href {\doibase 10.1146/annurev.aa.27.090189.003213}
  {\bibfield  {journal} {\bibinfo  {journal} {\araa}\ }\textbf {\bibinfo
  {volume} {27}},\ \bibinfo {pages} {629} (\bibinfo {year} {1989})}\BibitemShut
  {NoStop}%
\bibitem [{\citenamefont {M{\"u}ller}(2020)}]{muller2020hydrodynamics}%
  \BibitemOpen
  \bibfield  {author} {\bibinfo {author} {\bibfnamefont {B.}~\bibnamefont
  {M{\"u}ller}},\ }\href@noop {} {\bibfield  {journal} {\bibinfo  {journal}
  {Living Reviews in Computational Astrophysics}\ }\textbf {\bibinfo {volume}
  {6}},\ \bibinfo {pages} {3} (\bibinfo {year} {2020})}\BibitemShut {NoStop}%
\bibitem [{\citenamefont {Burrows}\ and\ \citenamefont
  {Vartanyan}(2021{\natexlab{a}})}]{burrows2021core}%
  \BibitemOpen
  \bibfield  {author} {\bibinfo {author} {\bibfnamefont {A.}~\bibnamefont
  {Burrows}}\ and\ \bibinfo {author} {\bibfnamefont {D.}~\bibnamefont
  {Vartanyan}},\ }\href@noop {} {\bibfield  {journal} {\bibinfo  {journal}
  {Nature}\ }\textbf {\bibinfo {volume} {589}},\ \bibinfo {pages} {29}
  (\bibinfo {year} {2021}{\natexlab{a}})}\BibitemShut {NoStop}%
\bibitem [{\citenamefont {Burrows}\ and\ \citenamefont
  {Vartanyan}(2021{\natexlab{b}})}]{Burrows:2020qrp}%
  \BibitemOpen
  \bibfield  {author} {\bibinfo {author} {\bibfnamefont {A.}~\bibnamefont
  {Burrows}}\ and\ \bibinfo {author} {\bibfnamefont {D.}~\bibnamefont
  {Vartanyan}},\ }\href {\doibase 10.1038/s41586-020-03059-w} {\bibfield
  {journal} {\bibinfo  {journal} {Nature}\ }\textbf {\bibinfo {volume} {589}},\
  \bibinfo {pages} {29} (\bibinfo {year} {2021}{\natexlab{b}})},\ \Eprint
  {http://arxiv.org/abs/2009.14157} {arXiv:2009.14157 [astro-ph.SR]}
  \BibitemShut {NoStop}%
\bibitem [{\citenamefont {Kyutoku}\ \emph {et~al.}(2021)\citenamefont
  {Kyutoku}, \citenamefont {Shibata},\ and\ \citenamefont
  {Taniguchi}}]{Kyutoku:2021icp}%
  \BibitemOpen
  \bibfield  {author} {\bibinfo {author} {\bibfnamefont {K.}~\bibnamefont
  {Kyutoku}}, \bibinfo {author} {\bibfnamefont {M.}~\bibnamefont {Shibata}}, \
  and\ \bibinfo {author} {\bibfnamefont {K.}~\bibnamefont {Taniguchi}},\ }\href
  {\doibase 10.1007/s41114-021-00033-4} {\bibfield  {journal} {\bibinfo
  {journal} {Living Rev. Rel.}\ }\textbf {\bibinfo {volume} {24}},\ \bibinfo
  {pages} {5} (\bibinfo {year} {2021})},\ \Eprint
  {http://arxiv.org/abs/2110.06218} {arXiv:2110.06218 [astro-ph.HE]}
  \BibitemShut {NoStop}%
\bibitem [{\citenamefont {Foucart}(2023)}]{Foucart:2022bth}%
  \BibitemOpen
  \bibfield  {author} {\bibinfo {author} {\bibfnamefont {F.}~\bibnamefont
  {Foucart}},\ }\href {\doibase 10.1007/s41115-023-00016-y} {\bibfield
  {journal} {\bibinfo  {journal} {Liv. Rev. Comput. Astrophys.}\ }\textbf
  {\bibinfo {volume} {9}},\ \bibinfo {pages} {1} (\bibinfo {year} {2023})},\
  \Eprint {http://arxiv.org/abs/2209.02538} {arXiv:2209.02538 [astro-ph.HE]}
  \BibitemShut {NoStop}%
\bibitem [{\citenamefont {Mezzacappa}(2020)}]{Mezzacappa:2020pkk}%
  \BibitemOpen
  \bibfield  {author} {\bibinfo {author} {\bibfnamefont {A.}~\bibnamefont
  {Mezzacappa}},\ }\href {\doibase 10.1017/S1743921322001831} {\bibfield
  {journal} {\bibinfo  {journal} {IAU Symp.}\ }\textbf {\bibinfo {volume}
  {362}},\ \bibinfo {pages} {215} (\bibinfo {year} {2020})},\ \Eprint
  {http://arxiv.org/abs/2205.13438} {arXiv:2205.13438 [astro-ph.SR]}
  \BibitemShut {NoStop}%
\bibitem [{\citenamefont {Kiuchi}(2024)}]{Kiuchi:2024lpx}%
  \BibitemOpen
  \bibfield  {author} {\bibinfo {author} {\bibfnamefont {K.}~\bibnamefont
  {Kiuchi}},\ }\href@noop {} {\enquote {\bibinfo {title} {{General relativistic
  magnetohydrodynamics simulations for binary neutron star mergers}},}\ }
  (\bibinfo {year} {2024}),\ \Eprint {http://arxiv.org/abs/2405.10081}
  {arXiv:2405.10081 [astro-ph.HE]} \BibitemShut {NoStop}%
\bibitem [{\citenamefont {Abdikamalov}\ \emph {et~al.}(2012)\citenamefont
  {Abdikamalov}, \citenamefont {Burrows}, \citenamefont {Ott}, \citenamefont
  {L\"offler}, \citenamefont {O'Connor}, \citenamefont {Dolence},\ and\
  \citenamefont {Schnetter}}]{Abdikamalov:2012zi}%
  \BibitemOpen
  \bibfield  {author} {\bibinfo {author} {\bibfnamefont {E.}~\bibnamefont
  {Abdikamalov}}, \bibinfo {author} {\bibfnamefont {A.}~\bibnamefont
  {Burrows}}, \bibinfo {author} {\bibfnamefont {C.~D.}\ \bibnamefont {Ott}},
  \bibinfo {author} {\bibfnamefont {F.}~\bibnamefont {L\"offler}}, \bibinfo
  {author} {\bibfnamefont {E.}~\bibnamefont {O'Connor}}, \bibinfo {author}
  {\bibfnamefont {J.~C.}\ \bibnamefont {Dolence}}, \ and\ \bibinfo {author}
  {\bibfnamefont {E.}~\bibnamefont {Schnetter}},\ }\href {\doibase
  10.1088/0004-637X/755/2/111} {\bibfield  {journal} {\bibinfo  {journal}
  {Astrophys. J.}\ }\textbf {\bibinfo {volume} {755}},\ \bibinfo {pages} {111}
  (\bibinfo {year} {2012})},\ \Eprint {http://arxiv.org/abs/1203.2915}
  {arXiv:1203.2915 [astro-ph.SR]} \BibitemShut {NoStop}%
\bibitem [{\citenamefont {Richers}\ \emph {et~al.}(2015)\citenamefont
  {Richers}, \citenamefont {Kasen}, \citenamefont {O’Connor}, \citenamefont
  {Fern{\'a}ndez},\ and\ \citenamefont {Ott}}]{richers2015monte}%
  \BibitemOpen
  \bibfield  {author} {\bibinfo {author} {\bibfnamefont {S.}~\bibnamefont
  {Richers}}, \bibinfo {author} {\bibfnamefont {D.}~\bibnamefont {Kasen}},
  \bibinfo {author} {\bibfnamefont {E.}~\bibnamefont {O’Connor}}, \bibinfo
  {author} {\bibfnamefont {R.}~\bibnamefont {Fern{\'a}ndez}}, \ and\ \bibinfo
  {author} {\bibfnamefont {C.~D.}\ \bibnamefont {Ott}},\ }\href@noop {}
  {\bibfield  {journal} {\bibinfo  {journal} {The Astrophysical Journal}\
  }\textbf {\bibinfo {volume} {813}},\ \bibinfo {pages} {38} (\bibinfo {year}
  {2015})}\BibitemShut {NoStop}%
\bibitem [{\citenamefont {Miller}\ \emph {et~al.}(2019)\citenamefont {Miller},
  \citenamefont {Ryan},\ and\ \citenamefont {Dolence}}]{Miller:2019gig}%
  \BibitemOpen
  \bibfield  {author} {\bibinfo {author} {\bibfnamefont {J.~M.}\ \bibnamefont
  {Miller}}, \bibinfo {author} {\bibfnamefont {B.~R.}\ \bibnamefont {Ryan}}, \
  and\ \bibinfo {author} {\bibfnamefont {J.~C.}\ \bibnamefont {Dolence}},\
  }\href {\doibase 10.3847/1538-4365/ab09fc} {\bibfield  {journal} {\bibinfo
  {journal} {Astrophys. J. Suppl.}\ }\textbf {\bibinfo {volume} {241}},\
  \bibinfo {pages} {30} (\bibinfo {year} {2019})},\ \Eprint
  {http://arxiv.org/abs/1903.09273} {arXiv:1903.09273 [astro-ph.IM]}
  \BibitemShut {NoStop}%
\bibitem [{\citenamefont {Foucart}\ \emph {et~al.}(2021)\citenamefont
  {Foucart}, \citenamefont {Duez}, \citenamefont {Hebert}, \citenamefont
  {Kidder}, \citenamefont {Kovarik}, \citenamefont {Pfeiffer},\ and\
  \citenamefont {Scheel}}]{Foucart:2021mcb}%
  \BibitemOpen
  \bibfield  {author} {\bibinfo {author} {\bibfnamefont {F.}~\bibnamefont
  {Foucart}}, \bibinfo {author} {\bibfnamefont {M.~D.}\ \bibnamefont {Duez}},
  \bibinfo {author} {\bibfnamefont {F.}~\bibnamefont {Hebert}}, \bibinfo
  {author} {\bibfnamefont {L.~E.}\ \bibnamefont {Kidder}}, \bibinfo {author}
  {\bibfnamefont {P.}~\bibnamefont {Kovarik}}, \bibinfo {author} {\bibfnamefont
  {H.~P.}\ \bibnamefont {Pfeiffer}}, \ and\ \bibinfo {author} {\bibfnamefont
  {M.~A.}\ \bibnamefont {Scheel}},\ }\href {\doibase 10.3847/1538-4357/ac1737}
  {\bibfield  {journal} {\bibinfo  {journal} {Astrophys. J.}\ }\textbf
  {\bibinfo {volume} {920}},\ \bibinfo {pages} {82} (\bibinfo {year} {2021})},\
  \Eprint {http://arxiv.org/abs/2103.16588} {arXiv:2103.16588 [astro-ph.HE]}
  \BibitemShut {NoStop}%
\bibitem [{\citenamefont {Sumiyoshi}\ and\ \citenamefont
  {Yamada}(2012)}]{Sumiyoshi:2012za}%
  \BibitemOpen
  \bibfield  {author} {\bibinfo {author} {\bibfnamefont {K.}~\bibnamefont
  {Sumiyoshi}}\ and\ \bibinfo {author} {\bibfnamefont {S.}~\bibnamefont
  {Yamada}},\ }\href {\doibase 10.1088/0067-0049/199/1/17} {\bibfield
  {journal} {\bibinfo  {journal} {Astrophys. J. Suppl.}\ }\textbf {\bibinfo
  {volume} {199}},\ \bibinfo {pages} {17} (\bibinfo {year} {2012})},\ \Eprint
  {http://arxiv.org/abs/1201.2244} {arXiv:1201.2244 [astro-ph.HE]} \BibitemShut
  {NoStop}%
\bibitem [{\citenamefont {Tamborra}\ and\ \citenamefont
  {Shalgar}(2021)}]{Tamborra:2020cul}%
  \BibitemOpen
  \bibfield  {author} {\bibinfo {author} {\bibfnamefont {I.}~\bibnamefont
  {Tamborra}}\ and\ \bibinfo {author} {\bibfnamefont {S.}~\bibnamefont
  {Shalgar}},\ }\href {\doibase 10.1146/annurev-nucl-102920-050505} {\bibfield
  {journal} {\bibinfo  {journal} {Ann. Rev. Nucl. Part. Sci.}\ }\textbf
  {\bibinfo {volume} {71}},\ \bibinfo {pages} {165} (\bibinfo {year} {2021})},\
  \Eprint {http://arxiv.org/abs/2011.01948} {arXiv:2011.01948 [astro-ph.HE]}
  \BibitemShut {NoStop}%
\bibitem [{\citenamefont {Volpe}(2024)}]{Volpe:2023met}%
  \BibitemOpen
  \bibfield  {author} {\bibinfo {author} {\bibfnamefont {M.~C.}\ \bibnamefont
  {Volpe}},\ }\href {\doibase 10.1103/RevModPhys.96.025004} {\bibfield
  {journal} {\bibinfo  {journal} {Rev. Mod. Phys.}\ }\textbf {\bibinfo {volume}
  {96}},\ \bibinfo {pages} {025004} (\bibinfo {year} {2024})},\ \Eprint
  {http://arxiv.org/abs/2301.11814} {arXiv:2301.11814 [hep-ph]} \BibitemShut
  {NoStop}%
\bibitem [{\citenamefont {{Haxton}}(1995)}]{1995ARA&A..33..459H}%
  \BibitemOpen
  \bibfield  {author} {\bibinfo {author} {\bibfnamefont {W.~C.}\ \bibnamefont
  {{Haxton}}},\ }\href {\doibase 10.1146/annurev.aa.33.090195.002331}
  {\bibfield  {journal} {\bibinfo  {journal} {\araa}\ }\textbf {\bibinfo
  {volume} {33}},\ \bibinfo {pages} {459} (\bibinfo {year} {1995})},\ \Eprint
  {http://arxiv.org/abs/hep-ph/9503430} {arXiv:hep-ph/9503430 [hep-ph]}
  \BibitemShut {NoStop}%
\bibitem [{\citenamefont {{Fuller}}\ \emph {et~al.}(2023)\citenamefont
  {{Fuller}}, \citenamefont {{Haxton}},\ and\ \citenamefont
  {{Grohs}}}]{2023ecnp.book..367F}%
  \BibitemOpen
  \bibfield  {author} {\bibinfo {author} {\bibfnamefont {G.~M.}\ \bibnamefont
  {{Fuller}}}, \bibinfo {author} {\bibfnamefont {W.~C.}\ \bibnamefont
  {{Haxton}}}, \ and\ \bibinfo {author} {\bibfnamefont {E.~B.}\ \bibnamefont
  {{Grohs}}},\ }in\ \href {\doibase 10.1142/9789811282645_0008} {\emph
  {\bibinfo {booktitle} {The Encyclopedia of Cosmology. Set 2: Frontiers in
  Cosmology. Volume 2: Neutrino Physics and Astrophysics}}},\ \bibinfo {editor}
  {edited by\ \bibinfo {editor} {\bibfnamefont {F.~W.}\ \bibnamefont
  {{Stecker}}}}\ (\bibinfo  {publisher} {WORLD SCIENTIFIC},\ \bibinfo {year}
  {2023})\ pp.\ \bibinfo {pages} {367--431}\BibitemShut {NoStop}%
\bibitem [{\citenamefont {{Pantaleone}}(1992)}]{1992PhLB..287..128P}%
  \BibitemOpen
  \bibfield  {author} {\bibinfo {author} {\bibfnamefont {J.}~\bibnamefont
  {{Pantaleone}}},\ }\href {\doibase 10.1016/0370-2693(92)91887-F} {\bibfield
  {journal} {\bibinfo  {journal} {Physics Letters B}\ }\textbf {\bibinfo
  {volume} {287}},\ \bibinfo {pages} {128} (\bibinfo {year}
  {1992})}\BibitemShut {NoStop}%
\bibitem [{\citenamefont {{Samuel}}(1993)}]{1993PhRvD..48.1462S}%
  \BibitemOpen
  \bibfield  {author} {\bibinfo {author} {\bibfnamefont {S.}~\bibnamefont
  {{Samuel}}},\ }\href {\doibase 10.1103/PhysRevD.48.1462} {\bibfield
  {journal} {\bibinfo  {journal} {\prd}\ }\textbf {\bibinfo {volume} {48}},\
  \bibinfo {pages} {1462} (\bibinfo {year} {1993})}\BibitemShut {NoStop}%
\bibitem [{\citenamefont {{Duan}}\ \emph
  {et~al.}(2006{\natexlab{a}})\citenamefont {{Duan}}, \citenamefont
  {{Fuller}},\ and\ \citenamefont {{Qian}}}]{2006PhRvD..74l3004D}%
  \BibitemOpen
  \bibfield  {author} {\bibinfo {author} {\bibfnamefont {H.}~\bibnamefont
  {{Duan}}}, \bibinfo {author} {\bibfnamefont {G.~M.}\ \bibnamefont
  {{Fuller}}}, \ and\ \bibinfo {author} {\bibfnamefont {Y.-Z.}\ \bibnamefont
  {{Qian}}},\ }\href {\doibase 10.1103/PhysRevD.74.123004} {\bibfield
  {journal} {\bibinfo  {journal} {\prd}\ }\textbf {\bibinfo {volume} {74}},\
  \bibinfo {eid} {123004} (\bibinfo {year} {2006}{\natexlab{a}})},\ \Eprint
  {http://arxiv.org/abs/astro-ph/0511275} {arXiv:astro-ph/0511275 [astro-ph]}
  \BibitemShut {NoStop}%
\bibitem [{\citenamefont {{Hannestad}}\ \emph {et~al.}(2006)\citenamefont
  {{Hannestad}}, \citenamefont {{Raffelt}}, \citenamefont {{Sigl}},\ and\
  \citenamefont {{Wong}}}]{2006PhRvD..74j5010H}%
  \BibitemOpen
  \bibfield  {author} {\bibinfo {author} {\bibfnamefont {S.}~\bibnamefont
  {{Hannestad}}}, \bibinfo {author} {\bibfnamefont {G.~G.}\ \bibnamefont
  {{Raffelt}}}, \bibinfo {author} {\bibfnamefont {G.}~\bibnamefont {{Sigl}}}, \
  and\ \bibinfo {author} {\bibfnamefont {Y.~Y.~Y.}\ \bibnamefont {{Wong}}},\
  }\href {\doibase 10.1103/PhysRevD.74.105010} {\bibfield  {journal} {\bibinfo
  {journal} {\prd}\ }\textbf {\bibinfo {volume} {74}},\ \bibinfo {eid} {105010}
  (\bibinfo {year} {2006})},\ \Eprint {http://arxiv.org/abs/astro-ph/0608695}
  {arXiv:astro-ph/0608695 [astro-ph]} \BibitemShut {NoStop}%
\bibitem [{\citenamefont {{Raffelt}}\ and\ \citenamefont
  {{Smirnov}}(2007{\natexlab{a}})}]{2007PhRvD..76h1301R}%
  \BibitemOpen
  \bibfield  {author} {\bibinfo {author} {\bibfnamefont {G.~G.}\ \bibnamefont
  {{Raffelt}}}\ and\ \bibinfo {author} {\bibfnamefont {A.~Y.}\ \bibnamefont
  {{Smirnov}}},\ }\href {\doibase 10.1103/PhysRevD.76.081301} {\bibfield
  {journal} {\bibinfo  {journal} {\prd}\ }\textbf {\bibinfo {volume} {76}},\
  \bibinfo {eid} {081301} (\bibinfo {year} {2007}{\natexlab{a}})},\ \Eprint
  {http://arxiv.org/abs/0705.1830} {arXiv:0705.1830 [hep-ph]} \BibitemShut
  {NoStop}%
\bibitem [{\citenamefont {{Duan}}\ \emph
  {et~al.}(2007{\natexlab{a}})\citenamefont {{Duan}}, \citenamefont {{Fuller}},
  \citenamefont {{Carlson}},\ and\ \citenamefont
  {{Qian}}}]{2007PhRvL..99x1802D}%
  \BibitemOpen
  \bibfield  {author} {\bibinfo {author} {\bibfnamefont {H.}~\bibnamefont
  {{Duan}}}, \bibinfo {author} {\bibfnamefont {G.~M.}\ \bibnamefont
  {{Fuller}}}, \bibinfo {author} {\bibfnamefont {J.}~\bibnamefont {{Carlson}}},
  \ and\ \bibinfo {author} {\bibfnamefont {Y.-Z.}\ \bibnamefont {{Qian}}},\
  }\href {\doibase 10.1103/PhysRevLett.99.241802} {\bibfield  {journal}
  {\bibinfo  {journal} {\prl}\ }\textbf {\bibinfo {volume} {99}},\ \bibinfo
  {eid} {241802} (\bibinfo {year} {2007}{\natexlab{a}})},\ \Eprint
  {http://arxiv.org/abs/0707.0290} {arXiv:0707.0290 [astro-ph]} \BibitemShut
  {NoStop}%
\bibitem [{\citenamefont {{Raffelt}}\ and\ \citenamefont
  {{Smirnov}}(2007{\natexlab{b}})}]{2007PhRvD..76l5008R}%
  \BibitemOpen
  \bibfield  {author} {\bibinfo {author} {\bibfnamefont {G.~G.}\ \bibnamefont
  {{Raffelt}}}\ and\ \bibinfo {author} {\bibfnamefont {A.~Y.}\ \bibnamefont
  {{Smirnov}}},\ }\href {\doibase 10.1103/PhysRevD.76.125008} {\bibfield
  {journal} {\bibinfo  {journal} {\prd}\ }\textbf {\bibinfo {volume} {76}},\
  \bibinfo {eid} {125008} (\bibinfo {year} {2007}{\natexlab{b}})},\ \Eprint
  {http://arxiv.org/abs/0709.4641} {arXiv:0709.4641 [hep-ph]} \BibitemShut
  {NoStop}%
\bibitem [{\citenamefont {Malkus}\ \emph {et~al.}(2012)\citenamefont {Malkus},
  \citenamefont {Kneller}, \citenamefont {McLaughlin},\ and\ \citenamefont
  {Surman}}]{Malkus:2012ts}%
  \BibitemOpen
  \bibfield  {author} {\bibinfo {author} {\bibfnamefont {A.}~\bibnamefont
  {Malkus}}, \bibinfo {author} {\bibfnamefont {J.~P.}\ \bibnamefont {Kneller}},
  \bibinfo {author} {\bibfnamefont {G.~C.}\ \bibnamefont {McLaughlin}}, \ and\
  \bibinfo {author} {\bibfnamefont {R.}~\bibnamefont {Surman}},\ }\href
  {\doibase 10.1103/PhysRevD.86.085015} {\bibfield  {journal} {\bibinfo
  {journal} {Phys. Rev. D}\ }\textbf {\bibinfo {volume} {86}},\ \bibinfo
  {pages} {085015} (\bibinfo {year} {2012})},\ \Eprint
  {http://arxiv.org/abs/1207.6648} {arXiv:1207.6648 [hep-ph]} \BibitemShut
  {NoStop}%
\bibitem [{\citenamefont {Malkus}\ \emph {et~al.}(2014)\citenamefont {Malkus},
  \citenamefont {Friedland},\ and\ \citenamefont
  {McLaughlin}}]{Malkus:2014iqa}%
  \BibitemOpen
  \bibfield  {author} {\bibinfo {author} {\bibfnamefont {A.}~\bibnamefont
  {Malkus}}, \bibinfo {author} {\bibfnamefont {A.}~\bibnamefont {Friedland}}, \
  and\ \bibinfo {author} {\bibfnamefont {G.~C.}\ \bibnamefont {McLaughlin}},\
  }\href@noop {} {\enquote {\bibinfo {title} {{Matter-Neutrino Resonance Above
  Merging Compact Objects}},}\ } (\bibinfo {year} {2014}),\ \Eprint
  {http://arxiv.org/abs/1403.5797} {arXiv:1403.5797 [hep-ph]} \BibitemShut
  {NoStop}%
\bibitem [{\citenamefont {Wu}\ \emph {et~al.}(2016)\citenamefont {Wu},
  \citenamefont {Duan},\ and\ \citenamefont {Qian}}]{Wu:2015fga}%
  \BibitemOpen
  \bibfield  {author} {\bibinfo {author} {\bibfnamefont {M.-R.}\ \bibnamefont
  {Wu}}, \bibinfo {author} {\bibfnamefont {H.}~\bibnamefont {Duan}}, \ and\
  \bibinfo {author} {\bibfnamefont {Y.-Z.}\ \bibnamefont {Qian}},\ }\href
  {\doibase 10.1016/j.physletb.2015.11.027} {\bibfield  {journal} {\bibinfo
  {journal} {Phys. Lett. B}\ }\textbf {\bibinfo {volume} {752}},\ \bibinfo
  {pages} {89} (\bibinfo {year} {2016})},\ \Eprint
  {http://arxiv.org/abs/1509.08975} {arXiv:1509.08975 [hep-ph]} \BibitemShut
  {NoStop}%
\bibitem [{\citenamefont {Frensel}\ \emph {et~al.}(2017)\citenamefont
  {Frensel}, \citenamefont {Wu}, \citenamefont {Volpe},\ and\ \citenamefont
  {Perego}}]{Frensel:2016fge}%
  \BibitemOpen
  \bibfield  {author} {\bibinfo {author} {\bibfnamefont {M.}~\bibnamefont
  {Frensel}}, \bibinfo {author} {\bibfnamefont {M.-R.}\ \bibnamefont {Wu}},
  \bibinfo {author} {\bibfnamefont {C.}~\bibnamefont {Volpe}}, \ and\ \bibinfo
  {author} {\bibfnamefont {A.}~\bibnamefont {Perego}},\ }\href {\doibase
  10.1103/PhysRevD.95.023011} {\bibfield  {journal} {\bibinfo  {journal} {Phys.
  Rev. D}\ }\textbf {\bibinfo {volume} {95}},\ \bibinfo {pages} {023011}
  (\bibinfo {year} {2017})},\ \Eprint {http://arxiv.org/abs/1607.05938}
  {arXiv:1607.05938 [astro-ph.HE]} \BibitemShut {NoStop}%
\bibitem [{\citenamefont {Tian}\ \emph {et~al.}(2017)\citenamefont {Tian},
  \citenamefont {Patwardhan},\ and\ \citenamefont {Fuller}}]{Tian:2017xbr}%
  \BibitemOpen
  \bibfield  {author} {\bibinfo {author} {\bibfnamefont {J.~Y.}\ \bibnamefont
  {Tian}}, \bibinfo {author} {\bibfnamefont {A.~V.}\ \bibnamefont
  {Patwardhan}}, \ and\ \bibinfo {author} {\bibfnamefont {G.~M.}\ \bibnamefont
  {Fuller}},\ }\href {\doibase 10.1103/PhysRevD.96.043001} {\bibfield
  {journal} {\bibinfo  {journal} {Phys. Rev. D}\ }\textbf {\bibinfo {volume}
  {96}},\ \bibinfo {pages} {043001} (\bibinfo {year} {2017})},\ \Eprint
  {http://arxiv.org/abs/1703.03039} {arXiv:1703.03039 [astro-ph.HE]}
  \BibitemShut {NoStop}%
\bibitem [{\citenamefont {{Vlasenko}}\ and\ \citenamefont
  {{McLaughlin}}(2018)}]{2018PhRvD..97h3011V}%
  \BibitemOpen
  \bibfield  {author} {\bibinfo {author} {\bibfnamefont {A.}~\bibnamefont
  {{Vlasenko}}}\ and\ \bibinfo {author} {\bibfnamefont {G.~C.}\ \bibnamefont
  {{McLaughlin}}},\ }\href {\doibase 10.1103/PhysRevD.97.083011} {\bibfield
  {journal} {\bibinfo  {journal} {\prd}\ }\textbf {\bibinfo {volume} {97}},\
  \bibinfo {eid} {083011} (\bibinfo {year} {2018})},\ \Eprint
  {http://arxiv.org/abs/1801.07813} {arXiv:1801.07813 [astro-ph.HE]}
  \BibitemShut {NoStop}%
\bibitem [{\citenamefont {{Duan}}\ \emph
  {et~al.}(2006{\natexlab{b}})\citenamefont {{Duan}}, \citenamefont {{Fuller}},
  \citenamefont {{Carlson}},\ and\ \citenamefont
  {{Qian}}}]{2006PhRvL..97x1101D}%
  \BibitemOpen
  \bibfield  {author} {\bibinfo {author} {\bibfnamefont {H.}~\bibnamefont
  {{Duan}}}, \bibinfo {author} {\bibfnamefont {G.~M.}\ \bibnamefont
  {{Fuller}}}, \bibinfo {author} {\bibfnamefont {J.}~\bibnamefont {{Carlson}}},
  \ and\ \bibinfo {author} {\bibfnamefont {Y.-Z.}\ \bibnamefont {{Qian}}},\
  }\href {\doibase 10.1103/PhysRevLett.97.241101} {\bibfield  {journal}
  {\bibinfo  {journal} {\prl}\ }\textbf {\bibinfo {volume} {97}},\ \bibinfo
  {eid} {241101} (\bibinfo {year} {2006}{\natexlab{b}})},\ \Eprint
  {http://arxiv.org/abs/astro-ph/0608050} {arXiv:astro-ph/0608050 [astro-ph]}
  \BibitemShut {NoStop}%
\bibitem [{\citenamefont {{Duan}}\ \emph
  {et~al.}(2007{\natexlab{b}})\citenamefont {{Duan}}, \citenamefont {{Fuller}},
  \citenamefont {{Carlson}},\ and\ \citenamefont
  {{Qian}}}]{2007PhRvD..75l5005D}%
  \BibitemOpen
  \bibfield  {author} {\bibinfo {author} {\bibfnamefont {H.}~\bibnamefont
  {{Duan}}}, \bibinfo {author} {\bibfnamefont {G.~M.}\ \bibnamefont
  {{Fuller}}}, \bibinfo {author} {\bibfnamefont {J.}~\bibnamefont {{Carlson}}},
  \ and\ \bibinfo {author} {\bibfnamefont {Y.-Z.}\ \bibnamefont {{Qian}}},\
  }\href {\doibase 10.1103/PhysRevD.75.125005} {\bibfield  {journal} {\bibinfo
  {journal} {\prd}\ }\textbf {\bibinfo {volume} {75}},\ \bibinfo {eid} {125005}
  (\bibinfo {year} {2007}{\natexlab{b}})},\ \Eprint
  {http://arxiv.org/abs/astro-ph/0703776} {arXiv:astro-ph/0703776 [astro-ph]}
  \BibitemShut {NoStop}%
\bibitem [{\citenamefont {{Duan}}\ \emph {et~al.}(2010)\citenamefont {{Duan}},
  \citenamefont {{Fuller}},\ and\ \citenamefont
  {{Qian}}}]{2010ARNPS..60..569D}%
  \BibitemOpen
  \bibfield  {author} {\bibinfo {author} {\bibfnamefont {H.}~\bibnamefont
  {{Duan}}}, \bibinfo {author} {\bibfnamefont {G.~M.}\ \bibnamefont
  {{Fuller}}}, \ and\ \bibinfo {author} {\bibfnamefont {Y.-Z.}\ \bibnamefont
  {{Qian}}},\ }\href {\doibase 10.1146/annurev.nucl.012809.104524} {\bibfield
  {journal} {\bibinfo  {journal} {Annual Review of Nuclear and Particle
  Science}\ }\textbf {\bibinfo {volume} {60}},\ \bibinfo {pages} {569}
  (\bibinfo {year} {2010})},\ \Eprint {http://arxiv.org/abs/1001.2799}
  {arXiv:1001.2799 [hep-ph]} \BibitemShut {NoStop}%
\bibitem [{\citenamefont {Johns}(2023)}]{Johns:2021qby}%
  \BibitemOpen
  \bibfield  {author} {\bibinfo {author} {\bibfnamefont {L.}~\bibnamefont
  {Johns}},\ }\href {\doibase 10.1103/PhysRevLett.130.191001} {\bibfield
  {journal} {\bibinfo  {journal} {Phys. Rev. Lett.}\ }\textbf {\bibinfo
  {volume} {130}},\ \bibinfo {pages} {191001} (\bibinfo {year} {2023})},\
  \Eprint {http://arxiv.org/abs/2104.11369} {arXiv:2104.11369 [hep-ph]}
  \BibitemShut {NoStop}%
\bibitem [{\citenamefont {Johns}\ and\ \citenamefont
  {Xiong}(2022)}]{Johns:2022yqy}%
  \BibitemOpen
  \bibfield  {author} {\bibinfo {author} {\bibfnamefont {L.}~\bibnamefont
  {Johns}}\ and\ \bibinfo {author} {\bibfnamefont {Z.}~\bibnamefont {Xiong}},\
  }\href {\doibase 10.1103/PhysRevD.106.103029} {\bibfield  {journal} {\bibinfo
   {journal} {Phys. Rev. D}\ }\textbf {\bibinfo {volume} {106}},\ \bibinfo
  {pages} {103029} (\bibinfo {year} {2022})},\ \Eprint
  {http://arxiv.org/abs/2208.11059} {arXiv:2208.11059 [hep-ph]} \BibitemShut
  {NoStop}%
\bibitem [{\citenamefont {Xiong}\ \emph {et~al.}(2023)\citenamefont {Xiong},
  \citenamefont {Wu}, \citenamefont {Martinez-Pinedo}, \citenamefont {Fischer},
  \citenamefont {George}, \citenamefont {Lin},\ and\ \citenamefont
  {Johns}}]{Xiong:2022vsy}%
  \BibitemOpen
  \bibfield  {author} {\bibinfo {author} {\bibfnamefont {Z.}~\bibnamefont
  {Xiong}}, \bibinfo {author} {\bibfnamefont {M.-R.}\ \bibnamefont {Wu}},
  \bibinfo {author} {\bibfnamefont {G.}~\bibnamefont {Martinez-Pinedo}},
  \bibinfo {author} {\bibfnamefont {T.}~\bibnamefont {Fischer}}, \bibinfo
  {author} {\bibfnamefont {M.}~\bibnamefont {George}}, \bibinfo {author}
  {\bibfnamefont {C.-Y.}\ \bibnamefont {Lin}}, \ and\ \bibinfo {author}
  {\bibfnamefont {L.}~\bibnamefont {Johns}},\ }\href {\doibase
  10.1103/PhysRevD.107.083016} {\bibfield  {journal} {\bibinfo  {journal}
  {Phys. Rev. D}\ }\textbf {\bibinfo {volume} {107}},\ \bibinfo {pages}
  {083016} (\bibinfo {year} {2023})},\ \Eprint
  {http://arxiv.org/abs/2210.08254} {arXiv:2210.08254 [astro-ph.HE]}
  \BibitemShut {NoStop}%
\bibitem [{\citenamefont {{Xiong}}\ \emph
  {et~al.}(2023{\natexlab{a}})\citenamefont {{Xiong}}, \citenamefont {{Johns}},
  \citenamefont {{Wu}},\ and\ \citenamefont {{Duan}}}]{Xiong:2022zqz}%
  \BibitemOpen
  \bibfield  {author} {\bibinfo {author} {\bibfnamefont {Z.}~\bibnamefont
  {{Xiong}}}, \bibinfo {author} {\bibfnamefont {L.}~\bibnamefont {{Johns}}},
  \bibinfo {author} {\bibfnamefont {M.-R.}\ \bibnamefont {{Wu}}}, \ and\
  \bibinfo {author} {\bibfnamefont {H.}~\bibnamefont {{Duan}}},\ }\href
  {\doibase 10.1103/PhysRevD.108.083002} {\bibfield  {journal} {\bibinfo
  {journal} {Phys. Rev. D}\ }\textbf {\bibinfo {volume} {108}},\ \bibinfo {eid}
  {083002} (\bibinfo {year} {2023}{\natexlab{a}})},\ \Eprint
  {http://arxiv.org/abs/2212.03750} {arXiv:2212.03750 [hep-ph]} \BibitemShut
  {NoStop}%
\bibitem [{\citenamefont {Liu}\ \emph {et~al.}(2023)\citenamefont {Liu},
  \citenamefont {Nagakura}, \citenamefont {Akaho}, \citenamefont {Ito},
  \citenamefont {Zaizen},\ and\ \citenamefont {Yamada}}]{Liu:2023vtz}%
  \BibitemOpen
  \bibfield  {author} {\bibinfo {author} {\bibfnamefont {J.}~\bibnamefont
  {Liu}}, \bibinfo {author} {\bibfnamefont {H.}~\bibnamefont {Nagakura}},
  \bibinfo {author} {\bibfnamefont {R.}~\bibnamefont {Akaho}}, \bibinfo
  {author} {\bibfnamefont {A.}~\bibnamefont {Ito}}, \bibinfo {author}
  {\bibfnamefont {M.}~\bibnamefont {Zaizen}}, \ and\ \bibinfo {author}
  {\bibfnamefont {S.}~\bibnamefont {Yamada}},\ }\href {\doibase
  10.1103/PhysRevD.108.123024} {\bibfield  {journal} {\bibinfo  {journal}
  {Phys. Rev. D}\ }\textbf {\bibinfo {volume} {108}},\ \bibinfo {pages}
  {123024} (\bibinfo {year} {2023})},\ \Eprint
  {http://arxiv.org/abs/2310.05050} {arXiv:2310.05050 [astro-ph.HE]}
  \BibitemShut {NoStop}%
\bibitem [{\citenamefont {Akaho}\ \emph {et~al.}(2024)\citenamefont {Akaho},
  \citenamefont {Liu}, \citenamefont {Nagakura}, \citenamefont {Zaizen},\ and\
  \citenamefont {Yamada}}]{Akaho:2023brj}%
  \BibitemOpen
  \bibfield  {author} {\bibinfo {author} {\bibfnamefont {R.}~\bibnamefont
  {Akaho}}, \bibinfo {author} {\bibfnamefont {J.}~\bibnamefont {Liu}}, \bibinfo
  {author} {\bibfnamefont {H.}~\bibnamefont {Nagakura}}, \bibinfo {author}
  {\bibfnamefont {M.}~\bibnamefont {Zaizen}}, \ and\ \bibinfo {author}
  {\bibfnamefont {S.}~\bibnamefont {Yamada}},\ }\href {\doibase
  10.1103/PhysRevD.109.023012} {\bibfield  {journal} {\bibinfo  {journal}
  {Phys. Rev. D}\ }\textbf {\bibinfo {volume} {109}},\ \bibinfo {pages}
  {023012} (\bibinfo {year} {2024})},\ \Eprint
  {http://arxiv.org/abs/2311.11272} {arXiv:2311.11272 [astro-ph.HE]}
  \BibitemShut {NoStop}%
\bibitem [{\citenamefont {Sawyer}(2005)}]{Sawyer:2005jk}%
  \BibitemOpen
  \bibfield  {author} {\bibinfo {author} {\bibfnamefont {R.~F.}\ \bibnamefont
  {Sawyer}},\ }\href {\doibase 10.1103/PhysRevD.72.045003} {\bibfield
  {journal} {\bibinfo  {journal} {Phys. Rev. D}\ }\textbf {\bibinfo {volume}
  {72}},\ \bibinfo {pages} {045003} (\bibinfo {year} {2005})},\ \Eprint
  {http://arxiv.org/abs/hep-ph/0503013} {arXiv:hep-ph/0503013} \BibitemShut
  {NoStop}%
\bibitem [{\citenamefont {Dasgupta}\ \emph {et~al.}(2017)\citenamefont
  {Dasgupta}, \citenamefont {Mirizzi},\ and\ \citenamefont
  {Sen}}]{Dasgupta:2016dbv}%
  \BibitemOpen
  \bibfield  {author} {\bibinfo {author} {\bibfnamefont {B.}~\bibnamefont
  {Dasgupta}}, \bibinfo {author} {\bibfnamefont {A.}~\bibnamefont {Mirizzi}}, \
  and\ \bibinfo {author} {\bibfnamefont {M.}~\bibnamefont {Sen}},\ }\href
  {\doibase 10.1088/1475-7516/2017/02/019} {\bibfield  {journal} {\bibinfo
  {journal} {JCAP}\ }\textbf {\bibinfo {volume} {02}},\ \bibinfo {pages} {019}
  (\bibinfo {year} {2017})},\ \Eprint {http://arxiv.org/abs/1609.00528}
  {arXiv:1609.00528 [hep-ph]} \BibitemShut {NoStop}%
\bibitem [{\citenamefont {Izaguirre}\ \emph {et~al.}(2017)\citenamefont
  {Izaguirre}, \citenamefont {Raffelt},\ and\ \citenamefont
  {Tamborra}}]{Izaguirre:2016gsx}%
  \BibitemOpen
  \bibfield  {author} {\bibinfo {author} {\bibfnamefont {I.}~\bibnamefont
  {Izaguirre}}, \bibinfo {author} {\bibfnamefont {G.}~\bibnamefont {Raffelt}},
  \ and\ \bibinfo {author} {\bibfnamefont {I.}~\bibnamefont {Tamborra}},\
  }\href {\doibase 10.1103/PhysRevLett.118.021101} {\bibfield  {journal}
  {\bibinfo  {journal} {Phys. Rev. Lett.}\ }\textbf {\bibinfo {volume} {118}},\
  \bibinfo {pages} {021101} (\bibinfo {year} {2017})},\ \Eprint
  {http://arxiv.org/abs/1610.01612} {arXiv:1610.01612 [hep-ph]} \BibitemShut
  {NoStop}%
\bibitem [{\citenamefont {Wu}\ and\ \citenamefont
  {Tamborra}(2017)}]{Wu:2017qpc}%
  \BibitemOpen
  \bibfield  {author} {\bibinfo {author} {\bibfnamefont {M.-R.}\ \bibnamefont
  {Wu}}\ and\ \bibinfo {author} {\bibfnamefont {I.}~\bibnamefont {Tamborra}},\
  }\href {\doibase 10.1103/PhysRevD.95.103007} {\bibfield  {journal} {\bibinfo
  {journal} {Phys. Rev. D}\ }\textbf {\bibinfo {volume} {95}},\ \bibinfo
  {pages} {103007} (\bibinfo {year} {2017})},\ \Eprint
  {http://arxiv.org/abs/1701.06580} {arXiv:1701.06580 [astro-ph.HE]}
  \BibitemShut {NoStop}%
\bibitem [{\citenamefont {Abbar}\ \emph {et~al.}(2019)\citenamefont {Abbar},
  \citenamefont {Duan}, \citenamefont {Sumiyoshi}, \citenamefont {Takiwaki},\
  and\ \citenamefont {Volpe}}]{Abbar:2018shq}%
  \BibitemOpen
  \bibfield  {author} {\bibinfo {author} {\bibfnamefont {S.}~\bibnamefont
  {Abbar}}, \bibinfo {author} {\bibfnamefont {H.}~\bibnamefont {Duan}},
  \bibinfo {author} {\bibfnamefont {K.}~\bibnamefont {Sumiyoshi}}, \bibinfo
  {author} {\bibfnamefont {T.}~\bibnamefont {Takiwaki}}, \ and\ \bibinfo
  {author} {\bibfnamefont {M.~C.}\ \bibnamefont {Volpe}},\ }\href {\doibase
  10.1103/PhysRevD.100.043004} {\bibfield  {journal} {\bibinfo  {journal}
  {Phys. Rev. D}\ }\textbf {\bibinfo {volume} {100}},\ \bibinfo {pages}
  {043004} (\bibinfo {year} {2019})},\ \Eprint
  {http://arxiv.org/abs/1812.06883} {arXiv:1812.06883 [astro-ph.HE]}
  \BibitemShut {NoStop}%
\bibitem [{\citenamefont {Nagakura}\ \emph {et~al.}(2021)\citenamefont
  {Nagakura}, \citenamefont {Johns}, \citenamefont {Burrows},\ and\
  \citenamefont {Fuller}}]{Nagakura:2021hyb}%
  \BibitemOpen
  \bibfield  {author} {\bibinfo {author} {\bibfnamefont {H.}~\bibnamefont
  {Nagakura}}, \bibinfo {author} {\bibfnamefont {L.}~\bibnamefont {Johns}},
  \bibinfo {author} {\bibfnamefont {A.}~\bibnamefont {Burrows}}, \ and\
  \bibinfo {author} {\bibfnamefont {G.~M.}\ \bibnamefont {Fuller}},\ }\href
  {\doibase 10.1103/PhysRevD.104.083025} {\bibfield  {journal} {\bibinfo
  {journal} {Phys. Rev. D}\ }\textbf {\bibinfo {volume} {104}},\ \bibinfo
  {pages} {083025} (\bibinfo {year} {2021})},\ \Eprint
  {http://arxiv.org/abs/2108.07281} {arXiv:2108.07281 [astro-ph.HE]}
  \BibitemShut {NoStop}%
\bibitem [{\citenamefont {Richers}\ and\ \citenamefont
  {Sen}(2022)}]{Richers:2022zug}%
  \BibitemOpen
  \bibfield  {author} {\bibinfo {author} {\bibfnamefont {S.}~\bibnamefont
  {Richers}}\ and\ \bibinfo {author} {\bibfnamefont {M.}~\bibnamefont {Sen}},\
  }\enquote {\bibinfo {title} {{Fast Flavor Transformations}},}\ in\ \href
  {\doibase 10.1007/978-981-15-8818-1_125-1} {\emph {\bibinfo {booktitle}
  {{Handbook of Nuclear Physics}}}},\ \bibinfo {editor} {edited by\ \bibinfo
  {editor} {\bibfnamefont {I.}~\bibnamefont {Tanihata}}, \bibinfo {editor}
  {\bibfnamefont {H.}~\bibnamefont {Toki}}, \ and\ \bibinfo {editor}
  {\bibfnamefont {T.}~\bibnamefont {Kajino}}}\ (\bibinfo {year} {2022})\ pp.\
  \bibinfo {pages} {1--17},\ \Eprint {http://arxiv.org/abs/2207.03561}
  {arXiv:2207.03561 [astro-ph.HE]} \BibitemShut {NoStop}%
\bibitem [{\citenamefont {Nagakura}(2023)}]{Nagakura:2023mhr}%
  \BibitemOpen
  \bibfield  {author} {\bibinfo {author} {\bibfnamefont {H.}~\bibnamefont
  {Nagakura}},\ }\href {\doibase 10.1103/PhysRevLett.130.211401} {\bibfield
  {journal} {\bibinfo  {journal} {Phys. Rev. Lett.}\ }\textbf {\bibinfo
  {volume} {130}},\ \bibinfo {pages} {211401} (\bibinfo {year} {2023})},\
  \Eprint {http://arxiv.org/abs/2301.10785} {arXiv:2301.10785 [astro-ph.HE]}
  \BibitemShut {NoStop}%
\bibitem [{\citenamefont {Israel}\ and\ \citenamefont
  {Stewart}(1979)}]{ISRAEL1979341}%
  \BibitemOpen
  \bibfield  {author} {\bibinfo {author} {\bibfnamefont {W.}~\bibnamefont
  {Israel}}\ and\ \bibinfo {author} {\bibfnamefont {J.}~\bibnamefont
  {Stewart}},\ }\href {\doibase https://doi.org/10.1016/0003-4916(79)90130-1}
  {\bibfield  {journal} {\bibinfo  {journal} {Annals of Physics}\ }\textbf
  {\bibinfo {volume} {118}},\ \bibinfo {pages} {341} (\bibinfo {year}
  {1979})}\BibitemShut {NoStop}%
\bibitem [{\citenamefont {{Thorne}}(1981)}]{1981MNRAS.194..439T}%
  \BibitemOpen
  \bibfield  {author} {\bibinfo {author} {\bibfnamefont {K.~S.}\ \bibnamefont
  {{Thorne}}},\ }\href {\doibase 10.1093/mnras/194.2.439} {\bibfield  {journal}
  {\bibinfo  {journal} {\mnras}\ }\textbf {\bibinfo {volume} {194}},\ \bibinfo
  {pages} {439} (\bibinfo {year} {1981})}\BibitemShut {NoStop}%
\bibitem [{\citenamefont {{O'Connor}}\ and\ \citenamefont
  {{Couch}}(2018)}]{2018ApJ...854...63O}%
  \BibitemOpen
  \bibfield  {author} {\bibinfo {author} {\bibfnamefont {E.~P.}\ \bibnamefont
  {{O'Connor}}}\ and\ \bibinfo {author} {\bibfnamefont {S.~M.}\ \bibnamefont
  {{Couch}}},\ }\href {\doibase 10.3847/1538-4357/aaa893} {\bibfield  {journal}
  {\bibinfo  {journal} {\apj}\ }\textbf {\bibinfo {volume} {854}},\ \bibinfo
  {eid} {63} (\bibinfo {year} {2018})},\ \Eprint
  {http://arxiv.org/abs/1511.07443} {arXiv:1511.07443 [astro-ph.HE]}
  \BibitemShut {NoStop}%
\bibitem [{\citenamefont {Minerbo}(1978)}]{minerbo_maximum_1978}%
  \BibitemOpen
  \bibfield  {author} {\bibinfo {author} {\bibfnamefont {G.~N.}\ \bibnamefont
  {Minerbo}},\ }\href {\doibase 10.1016/0022-4073(78)90024-9} {\bibfield
  {journal} {\bibinfo  {journal} {J. Quant. Spectrosc. Radiat. Transfer}\
  }\textbf {\bibinfo {volume} {20}},\ \bibinfo {pages} {541} (\bibinfo {year}
  {1978})}\BibitemShut {NoStop}%
\bibitem [{\citenamefont {{Smit}}\ \emph {et~al.}(2000)\citenamefont {{Smit}},
  \citenamefont {{van den Horn}},\ and\ \citenamefont
  {{Bludman}}}]{Smit_closure}%
  \BibitemOpen
  \bibfield  {author} {\bibinfo {author} {\bibfnamefont {J.~M.}\ \bibnamefont
  {{Smit}}}, \bibinfo {author} {\bibfnamefont {L.~J.}\ \bibnamefont {{van den
  Horn}}}, \ and\ \bibinfo {author} {\bibfnamefont {S.~A.}\ \bibnamefont
  {{Bludman}}},\ }\href@noop {} {\bibfield  {journal} {\bibinfo  {journal}
  {Astron. Astrophys.}\ }\textbf {\bibinfo {volume} {356}},\ \bibinfo {pages}
  {559} (\bibinfo {year} {2000})}\BibitemShut {NoStop}%
\bibitem [{\citenamefont {Shibata}\ \emph {et~al.}(2011)\citenamefont
  {Shibata}, \citenamefont {Kiuchi}, \citenamefont {Sekiguchi},\ and\
  \citenamefont {Suwa}}]{Shibata:2011kx}%
  \BibitemOpen
  \bibfield  {author} {\bibinfo {author} {\bibfnamefont {M.}~\bibnamefont
  {Shibata}}, \bibinfo {author} {\bibfnamefont {K.}~\bibnamefont {Kiuchi}},
  \bibinfo {author} {\bibfnamefont {Y.-i.}\ \bibnamefont {Sekiguchi}}, \ and\
  \bibinfo {author} {\bibfnamefont {Y.}~\bibnamefont {Suwa}},\ }\href {\doibase
  10.1143/PTP.125.1255} {\bibfield  {journal} {\bibinfo  {journal} {Prog.
  Theor. Phys.}\ }\textbf {\bibinfo {volume} {125}},\ \bibinfo {pages} {1255}
  (\bibinfo {year} {2011})},\ \Eprint {http://arxiv.org/abs/1104.3937}
  {arXiv:1104.3937 [astro-ph.HE]} \BibitemShut {NoStop}%
\bibitem [{\citenamefont {Murchikova}\ \emph {et~al.}(2017)\citenamefont
  {Murchikova}, \citenamefont {Abdikamalov},\ and\ \citenamefont
  {Urbatsch}}]{Murchikova:2017zsy}%
  \BibitemOpen
  \bibfield  {author} {\bibinfo {author} {\bibfnamefont {L.~M.}\ \bibnamefont
  {Murchikova}}, \bibinfo {author} {\bibfnamefont {E.}~\bibnamefont
  {Abdikamalov}}, \ and\ \bibinfo {author} {\bibfnamefont {T.}~\bibnamefont
  {Urbatsch}},\ }\href {\doibase 10.1093/mnras/stx986} {\bibfield  {journal}
  {\bibinfo  {journal} {Mon. Not. Roy. Astron. Soc.}\ }\textbf {\bibinfo
  {volume} {469}},\ \bibinfo {pages} {1725} (\bibinfo {year} {2017})},\ \Eprint
  {http://arxiv.org/abs/1701.07027} {arXiv:1701.07027 [astro-ph.HE]}
  \BibitemShut {NoStop}%
\bibitem [{\citenamefont {{Richers}}(2020)}]{2020PhRvD.102h3017R}%
  \BibitemOpen
  \bibfield  {author} {\bibinfo {author} {\bibfnamefont {S.}~\bibnamefont
  {{Richers}}},\ }\href {\doibase 10.1103/PhysRevD.102.083017} {\bibfield
  {journal} {\bibinfo  {journal} {\prd}\ }\textbf {\bibinfo {volume} {102}},\
  \bibinfo {eid} {083017} (\bibinfo {year} {2020})},\ \Eprint
  {http://arxiv.org/abs/2009.09046} {arXiv:2009.09046 [astro-ph.HE]}
  \BibitemShut {NoStop}%
\bibitem [{\citenamefont {{Wang}}\ and\ \citenamefont
  {{Burrows}}(2023)}]{2023ApJ...943...78W}%
  \BibitemOpen
  \bibfield  {author} {\bibinfo {author} {\bibfnamefont {T.}~\bibnamefont
  {{Wang}}}\ and\ \bibinfo {author} {\bibfnamefont {A.}~\bibnamefont
  {{Burrows}}},\ }\href {\doibase 10.3847/1538-4357/aca75c} {\bibfield
  {journal} {\bibinfo  {journal} {\apj}\ }\textbf {\bibinfo {volume} {943}},\
  \bibinfo {eid} {78} (\bibinfo {year} {2023})},\ \Eprint
  {http://arxiv.org/abs/2210.01824} {arXiv:2210.01824 [astro-ph.HE]}
  \BibitemShut {NoStop}%
\bibitem [{\citenamefont {Strack}\ and\ \citenamefont
  {Burrows}(2005)}]{strack:2005}%
  \BibitemOpen
  \bibfield  {author} {\bibinfo {author} {\bibfnamefont {P.}~\bibnamefont
  {Strack}}\ and\ \bibinfo {author} {\bibfnamefont {A.}~\bibnamefont
  {Burrows}},\ }\href {\doibase 10.1103/PhysRevD.71.093004} {\bibfield
  {journal} {\bibinfo  {journal} {Phys. Rev. D}\ }\textbf {\bibinfo {volume}
  {71}},\ \bibinfo {pages} {093004} (\bibinfo {year} {2005})},\ \Eprint
  {http://arxiv.org/abs/hep-ph/0504035} {arXiv:hep-ph/0504035} \BibitemShut
  {NoStop}%
\bibitem [{\citenamefont {Zhang}\ and\ \citenamefont
  {Burrows}(2013)}]{Zhang:2013lka}%
  \BibitemOpen
  \bibfield  {author} {\bibinfo {author} {\bibfnamefont {Y.}~\bibnamefont
  {Zhang}}\ and\ \bibinfo {author} {\bibfnamefont {A.}~\bibnamefont
  {Burrows}},\ }\href {\doibase 10.1103/PhysRevD.88.105009} {\bibfield
  {journal} {\bibinfo  {journal} {Phys. Rev. D}\ }\textbf {\bibinfo {volume}
  {88}},\ \bibinfo {pages} {105009} (\bibinfo {year} {2013})},\ \Eprint
  {http://arxiv.org/abs/1310.2164} {arXiv:1310.2164 [hep-ph]} \BibitemShut
  {NoStop}%
\bibitem [{\citenamefont {Johns}\ \emph {et~al.}(2020)\citenamefont {Johns},
  \citenamefont {Nagakura}, \citenamefont {Fuller},\ and\ \citenamefont
  {Burrows}}]{johns2020neutrino}%
  \BibitemOpen
  \bibfield  {author} {\bibinfo {author} {\bibfnamefont {L.}~\bibnamefont
  {Johns}}, \bibinfo {author} {\bibfnamefont {H.}~\bibnamefont {Nagakura}},
  \bibinfo {author} {\bibfnamefont {G.~M.}\ \bibnamefont {Fuller}}, \ and\
  \bibinfo {author} {\bibfnamefont {A.}~\bibnamefont {Burrows}},\ }\href@noop
  {} {\bibfield  {journal} {\bibinfo  {journal} {Physical Review D}\ }\textbf
  {\bibinfo {volume} {101}},\ \bibinfo {pages} {043009} (\bibinfo {year}
  {2020})}\BibitemShut {NoStop}%
\bibitem [{\citenamefont {{Myers}}\ \emph {et~al.}(2022)\citenamefont
  {{Myers}}, \citenamefont {{Cooper}}, \citenamefont {{Warren}}, \citenamefont
  {{Kneller}}, \citenamefont {{McLaughlin}}, \citenamefont {{Richers}},
  \citenamefont {{Grohs}},\ and\ \citenamefont
  {{Fr{\"o}hlich}}}]{2022PhRvD.105l3036M}%
  \BibitemOpen
  \bibfield  {author} {\bibinfo {author} {\bibfnamefont {M.}~\bibnamefont
  {{Myers}}}, \bibinfo {author} {\bibfnamefont {T.}~\bibnamefont {{Cooper}}},
  \bibinfo {author} {\bibfnamefont {M.}~\bibnamefont {{Warren}}}, \bibinfo
  {author} {\bibfnamefont {J.}~\bibnamefont {{Kneller}}}, \bibinfo {author}
  {\bibfnamefont {G.}~\bibnamefont {{McLaughlin}}}, \bibinfo {author}
  {\bibfnamefont {S.}~\bibnamefont {{Richers}}}, \bibinfo {author}
  {\bibfnamefont {E.}~\bibnamefont {{Grohs}}}, \ and\ \bibinfo {author}
  {\bibfnamefont {C.}~\bibnamefont {{Fr{\"o}hlich}}},\ }\href {\doibase
  10.1103/PhysRevD.105.123036} {\bibfield  {journal} {\bibinfo  {journal}
  {\prd}\ }\textbf {\bibinfo {volume} {105}},\ \bibinfo {eid} {123036}
  (\bibinfo {year} {2022})},\ \Eprint {http://arxiv.org/abs/2111.13722}
  {arXiv:2111.13722 [hep-ph]} \BibitemShut {NoStop}%
\bibitem [{\citenamefont {Harten}\ \emph {et~al.}(1983)\citenamefont {Harten},
  \citenamefont {Lax},\ and\ \citenamefont {Leer}}]{HLL_1983}%
  \BibitemOpen
  \bibfield  {author} {\bibinfo {author} {\bibfnamefont {A.}~\bibnamefont
  {Harten}}, \bibinfo {author} {\bibfnamefont {P.~D.}\ \bibnamefont {Lax}}, \
  and\ \bibinfo {author} {\bibfnamefont {B.~V.}\ \bibnamefont {Leer}},\ }\href
  {http://www.jstor.org/stable/2030019} {\bibfield  {journal} {\bibinfo
  {journal} {SIAM Review}\ }\textbf {\bibinfo {volume} {25}},\ \bibinfo {pages}
  {35} (\bibinfo {year} {1983})}\BibitemShut {NoStop}%
\bibitem [{\citenamefont {{Einfeldt}}(1988)}]{1988SJNA...25..294E}%
  \BibitemOpen
  \bibfield  {author} {\bibinfo {author} {\bibfnamefont {B.}~\bibnamefont
  {{Einfeldt}}},\ }\href {\doibase 10.1137/0725021} {\bibfield  {journal}
  {\bibinfo  {journal} {SIAM Journal on Numerical Analysis}\ }\textbf {\bibinfo
  {volume} {25}},\ \bibinfo {pages} {294} (\bibinfo {year} {1988})}\BibitemShut
  {NoStop}%
\bibitem [{\citenamefont {{Grohs}}\ \emph {et~al.}(2024)\citenamefont
  {{Grohs}}, \citenamefont {{Richers}}, \citenamefont {{Couch}}, \citenamefont
  {{Foucart}}, \citenamefont {{Froustey}}, \citenamefont {{Kneller}},\ and\
  \citenamefont {{McLaughlin}}}]{flashri}%
  \BibitemOpen
  \bibfield  {author} {\bibinfo {author} {\bibfnamefont {E.}~\bibnamefont
  {{Grohs}}}, \bibinfo {author} {\bibfnamefont {S.}~\bibnamefont {{Richers}}},
  \bibinfo {author} {\bibfnamefont {S.~M.}\ \bibnamefont {{Couch}}}, \bibinfo
  {author} {\bibfnamefont {F.}~\bibnamefont {{Foucart}}}, \bibinfo {author}
  {\bibfnamefont {J.}~\bibnamefont {{Froustey}}}, \bibinfo {author}
  {\bibfnamefont {J.~P.}\ \bibnamefont {{Kneller}}}, \ and\ \bibinfo {author}
  {\bibfnamefont {G.~C.}\ \bibnamefont {{McLaughlin}}},\ }\href {\doibase
  10.3847/1538-4357/ad13f2} {\bibfield  {journal} {\bibinfo  {journal} {\apj}\
  }\textbf {\bibinfo {volume} {963}},\ \bibinfo {eid} {11} (\bibinfo {year}
  {2024})},\ \Eprint {http://arxiv.org/abs/2309.00972} {arXiv:2309.00972
  [astro-ph.HE]} \BibitemShut {NoStop}%
\bibitem [{\citenamefont {{Barbieri}}\ and\ \citenamefont
  {{Dolgov}}(1991)}]{1991NuPhB.349..743B}%
  \BibitemOpen
  \bibfield  {author} {\bibinfo {author} {\bibfnamefont {R.}~\bibnamefont
  {{Barbieri}}}\ and\ \bibinfo {author} {\bibfnamefont {A.}~\bibnamefont
  {{Dolgov}}},\ }\href {\doibase 10.1016/0550-3213(91)90396-F} {\bibfield
  {journal} {\bibinfo  {journal} {Nuclear Physics B}\ }\textbf {\bibinfo
  {volume} {349}},\ \bibinfo {pages} {743} (\bibinfo {year}
  {1991})}\BibitemShut {NoStop}%
\bibitem [{\citenamefont {Sigl}\ and\ \citenamefont
  {Raffelt}(1993)}]{Sigl:1993ctk}%
  \BibitemOpen
  \bibfield  {author} {\bibinfo {author} {\bibfnamefont {G.}~\bibnamefont
  {Sigl}}\ and\ \bibinfo {author} {\bibfnamefont {G.}~\bibnamefont {Raffelt}},\
  }\href {\doibase 10.1016/0550-3213(93)90175-O} {\bibfield  {journal}
  {\bibinfo  {journal} {Nucl. Phys. B}\ }\textbf {\bibinfo {volume} {406}},\
  \bibinfo {pages} {423} (\bibinfo {year} {1993})}\BibitemShut {NoStop}%
\bibitem [{\citenamefont {{Balantekin}}\ and\ \citenamefont
  {{Pehlivan}}(2007)}]{2007JPhG...34...47B}%
  \BibitemOpen
  \bibfield  {author} {\bibinfo {author} {\bibfnamefont {A.~B.}\ \bibnamefont
  {{Balantekin}}}\ and\ \bibinfo {author} {\bibfnamefont {Y.}~\bibnamefont
  {{Pehlivan}}},\ }\href {\doibase 10.1088/0954-3899/34/1/004} {\bibfield
  {journal} {\bibinfo  {journal} {Journal of Physics G Nuclear Physics}\
  }\textbf {\bibinfo {volume} {34}},\ \bibinfo {pages} {47} (\bibinfo {year}
  {2007})},\ \Eprint {http://arxiv.org/abs/astro-ph/0607527}
  {arXiv:astro-ph/0607527 [astro-ph]} \BibitemShut {NoStop}%
\bibitem [{\citenamefont {Volpe}\ \emph {et~al.}(2013)\citenamefont {Volpe},
  \citenamefont {V\"a\"an\"anen},\ and\ \citenamefont
  {Espinoza}}]{Volpe:2013uxl}%
  \BibitemOpen
  \bibfield  {author} {\bibinfo {author} {\bibfnamefont {C.}~\bibnamefont
  {Volpe}}, \bibinfo {author} {\bibfnamefont {D.}~\bibnamefont
  {V\"a\"an\"anen}}, \ and\ \bibinfo {author} {\bibfnamefont {C.}~\bibnamefont
  {Espinoza}},\ }\href {\doibase 10.1103/PhysRevD.87.113010} {\bibfield
  {journal} {\bibinfo  {journal} {Phys. Rev. D}\ }\textbf {\bibinfo {volume}
  {87}},\ \bibinfo {pages} {113010} (\bibinfo {year} {2013})},\ \Eprint
  {http://arxiv.org/abs/1302.2374} {arXiv:1302.2374 [hep-ph]} \BibitemShut
  {NoStop}%
\bibitem [{\citenamefont {{Balantekin}}\ and\ \citenamefont
  {{Fuller}}(2013)}]{2013PrPNP..71..162B}%
  \BibitemOpen
  \bibfield  {author} {\bibinfo {author} {\bibfnamefont {A.~B.}\ \bibnamefont
  {{Balantekin}}}\ and\ \bibinfo {author} {\bibfnamefont {G.~M.}\ \bibnamefont
  {{Fuller}}},\ }\href {\doibase 10.1016/j.ppnp.2013.03.008} {\bibfield
  {journal} {\bibinfo  {journal} {Progress in Particle and Nuclear Physics}\
  }\textbf {\bibinfo {volume} {71}},\ \bibinfo {pages} {162} (\bibinfo {year}
  {2013})},\ \Eprint {http://arxiv.org/abs/1303.3874} {arXiv:1303.3874
  [nucl-th]} \BibitemShut {NoStop}%
\bibitem [{\citenamefont {{Vlasenko}}\ \emph {et~al.}(2014)\citenamefont
  {{Vlasenko}}, \citenamefont {{Fuller}},\ and\ \citenamefont
  {{Cirigliano}}}]{2014PhRvD..89j5004V}%
  \BibitemOpen
  \bibfield  {author} {\bibinfo {author} {\bibfnamefont {A.}~\bibnamefont
  {{Vlasenko}}}, \bibinfo {author} {\bibfnamefont {G.~M.}\ \bibnamefont
  {{Fuller}}}, \ and\ \bibinfo {author} {\bibfnamefont {V.}~\bibnamefont
  {{Cirigliano}}},\ }\href {\doibase 10.1103/PhysRevD.89.105004} {\bibfield
  {journal} {\bibinfo  {journal} {\prd}\ }\textbf {\bibinfo {volume} {89}},\
  \bibinfo {eid} {105004} (\bibinfo {year} {2014})},\ \Eprint
  {http://arxiv.org/abs/1309.2628} {arXiv:1309.2628 [hep-ph]} \BibitemShut
  {NoStop}%
\bibitem [{\citenamefont {{Serreau}}\ and\ \citenamefont
  {{Volpe}}(2014)}]{2014PhRvD..90l5040S}%
  \BibitemOpen
  \bibfield  {author} {\bibinfo {author} {\bibfnamefont {J.}~\bibnamefont
  {{Serreau}}}\ and\ \bibinfo {author} {\bibfnamefont {C.}~\bibnamefont
  {{Volpe}}},\ }\href {\doibase 10.1103/PhysRevD.90.125040} {\bibfield
  {journal} {\bibinfo  {journal} {\prd}\ }\textbf {\bibinfo {volume} {90}},\
  \bibinfo {eid} {125040} (\bibinfo {year} {2014})},\ \Eprint
  {http://arxiv.org/abs/1409.3591} {arXiv:1409.3591 [hep-ph]} \BibitemShut
  {NoStop}%
\bibitem [{\citenamefont {{Cirigliano}}\ \emph {et~al.}(2015)\citenamefont
  {{Cirigliano}}, \citenamefont {{Fuller}},\ and\ \citenamefont
  {{Vlasenko}}}]{2015PhLB..747...27C}%
  \BibitemOpen
  \bibfield  {author} {\bibinfo {author} {\bibfnamefont {V.}~\bibnamefont
  {{Cirigliano}}}, \bibinfo {author} {\bibfnamefont {G.~M.}\ \bibnamefont
  {{Fuller}}}, \ and\ \bibinfo {author} {\bibfnamefont {A.}~\bibnamefont
  {{Vlasenko}}},\ }\href {\doibase 10.1016/j.physletb.2015.04.066} {\bibfield
  {journal} {\bibinfo  {journal} {Physics Letters B}\ }\textbf {\bibinfo
  {volume} {747}},\ \bibinfo {pages} {27} (\bibinfo {year} {2015})},\ \Eprint
  {http://arxiv.org/abs/1406.5558} {arXiv:1406.5558 [hep-ph]} \BibitemShut
  {NoStop}%
\bibitem [{\citenamefont {{Volpe}}(2015)}]{2015IJMPE..2441009V}%
  \BibitemOpen
  \bibfield  {author} {\bibinfo {author} {\bibfnamefont {C.}~\bibnamefont
  {{Volpe}}},\ }\href {\doibase 10.1142/S0218301315410098} {\bibfield
  {journal} {\bibinfo  {journal} {International Journal of Modern Physics E}\
  }\textbf {\bibinfo {volume} {24}},\ \bibinfo {eid} {1541009} (\bibinfo {year}
  {2015})},\ \Eprint {http://arxiv.org/abs/1506.06222} {arXiv:1506.06222
  [astro-ph.SR]} \BibitemShut {NoStop}%
\bibitem [{\citenamefont {Blaschke}\ and\ \citenamefont
  {Cirigliano}(2016)}]{Blaschke:2016xxt}%
  \BibitemOpen
  \bibfield  {author} {\bibinfo {author} {\bibfnamefont {D.~N.}\ \bibnamefont
  {Blaschke}}\ and\ \bibinfo {author} {\bibfnamefont {V.}~\bibnamefont
  {Cirigliano}},\ }\href {\doibase 10.1103/PhysRevD.94.033009} {\bibfield
  {journal} {\bibinfo  {journal} {Phys. Rev. D}\ }\textbf {\bibinfo {volume}
  {94}},\ \bibinfo {pages} {033009} (\bibinfo {year} {2016})},\ \Eprint
  {http://arxiv.org/abs/1605.09383} {arXiv:1605.09383 [hep-ph]} \BibitemShut
  {NoStop}%
\bibitem [{\citenamefont {Richers}\ \emph {et~al.}(2019)\citenamefont
  {Richers}, \citenamefont {McLaughlin}, \citenamefont {Kneller},\ and\
  \citenamefont {Vlasenko}}]{Richers:2019grc}%
  \BibitemOpen
  \bibfield  {author} {\bibinfo {author} {\bibfnamefont {S.~A.}\ \bibnamefont
  {Richers}}, \bibinfo {author} {\bibfnamefont {G.~C.}\ \bibnamefont
  {McLaughlin}}, \bibinfo {author} {\bibfnamefont {J.~P.}\ \bibnamefont
  {Kneller}}, \ and\ \bibinfo {author} {\bibfnamefont {A.}~\bibnamefont
  {Vlasenko}},\ }\href {\doibase 10.1103/PhysRevD.99.123014} {\bibfield
  {journal} {\bibinfo  {journal} {Phys. Rev. D}\ }\textbf {\bibinfo {volume}
  {99}},\ \bibinfo {pages} {123014} (\bibinfo {year} {2019})},\ \Eprint
  {http://arxiv.org/abs/1903.00022} {arXiv:1903.00022 [astro-ph.HE]}
  \BibitemShut {NoStop}%
\bibitem [{\citenamefont {Froustey}\ \emph {et~al.}(2020)\citenamefont
  {Froustey}, \citenamefont {Pitrou},\ and\ \citenamefont
  {Volpe}}]{Froustey:2020mcq}%
  \BibitemOpen
  \bibfield  {author} {\bibinfo {author} {\bibfnamefont {J.}~\bibnamefont
  {Froustey}}, \bibinfo {author} {\bibfnamefont {C.}~\bibnamefont {Pitrou}}, \
  and\ \bibinfo {author} {\bibfnamefont {M.~C.}\ \bibnamefont {Volpe}},\ }\href
  {\doibase 10.1088/1475-7516/2020/12/015} {\bibfield  {journal} {\bibinfo
  {journal} {JCAP}\ }\textbf {\bibinfo {volume} {12}},\ \bibinfo {pages} {015}
  (\bibinfo {year} {2020})},\ \Eprint {http://arxiv.org/abs/2008.01074}
  {arXiv:2008.01074 [hep-ph]} \BibitemShut {NoStop}%
\bibitem [{\citenamefont {{O'Connor}}(2015)}]{2015ApJS..219...24O}%
  \BibitemOpen
  \bibfield  {author} {\bibinfo {author} {\bibfnamefont {E.}~\bibnamefont
  {{O'Connor}}},\ }\href {\doibase 10.1088/0067-0049/219/2/24} {\bibfield
  {journal} {\bibinfo  {journal} {Astrophys. J. Suppl.}\ }\textbf {\bibinfo
  {volume} {219}},\ \bibinfo {eid} {24} (\bibinfo {year} {2015})},\ \Eprint
  {http://arxiv.org/abs/1411.7058} {arXiv:1411.7058 [astro-ph.HE]} \BibitemShut
  {NoStop}%
\bibitem [{\citenamefont {Grohs}\ \emph {et~al.}(2023)\citenamefont {Grohs},
  \citenamefont {Richers}, \citenamefont {Couch}, \citenamefont {Foucart},
  \citenamefont {Kneller},\ and\ \citenamefont {McLaughlin}}]{Grohs:2022fyq}%
  \BibitemOpen
  \bibfield  {author} {\bibinfo {author} {\bibfnamefont {E.}~\bibnamefont
  {Grohs}}, \bibinfo {author} {\bibfnamefont {S.}~\bibnamefont {Richers}},
  \bibinfo {author} {\bibfnamefont {S.~M.}\ \bibnamefont {Couch}}, \bibinfo
  {author} {\bibfnamefont {F.}~\bibnamefont {Foucart}}, \bibinfo {author}
  {\bibfnamefont {J.~P.}\ \bibnamefont {Kneller}}, \ and\ \bibinfo {author}
  {\bibfnamefont {G.~C.}\ \bibnamefont {McLaughlin}},\ }\href {\doibase
  10.1016/j.physletb.2023.138210} {\bibfield  {journal} {\bibinfo  {journal}
  {Phys. Lett. B}\ }\textbf {\bibinfo {volume} {846}},\ \bibinfo {pages}
  {138210} (\bibinfo {year} {2023})},\ \Eprint
  {http://arxiv.org/abs/2207.02214} {arXiv:2207.02214 [hep-ph]} \BibitemShut
  {NoStop}%
\bibitem [{\citenamefont {{Foucart}}\ \emph {et~al.}(2016)\citenamefont
  {{Foucart}}, \citenamefont {{O'Connor}}, \citenamefont {{Roberts}},
  \citenamefont {{Kidder}}, \citenamefont {{Pfeiffer}},\ and\ \citenamefont
  {{Scheel}}}]{Foucart:2016rxm}%
  \BibitemOpen
  \bibfield  {author} {\bibinfo {author} {\bibfnamefont {F.}~\bibnamefont
  {{Foucart}}}, \bibinfo {author} {\bibfnamefont {E.}~\bibnamefont
  {{O'Connor}}}, \bibinfo {author} {\bibfnamefont {L.}~\bibnamefont
  {{Roberts}}}, \bibinfo {author} {\bibfnamefont {L.~E.}\ \bibnamefont
  {{Kidder}}}, \bibinfo {author} {\bibfnamefont {H.~P.}\ \bibnamefont
  {{Pfeiffer}}}, \ and\ \bibinfo {author} {\bibfnamefont {M.~A.}\ \bibnamefont
  {{Scheel}}},\ }\href {\doibase 10.1103/PhysRevD.94.123016} {\bibfield
  {journal} {\bibinfo  {journal} {Phys. Rev.}\ }\textbf {\bibinfo {volume}
  {D94}},\ \bibinfo {pages} {123016} (\bibinfo {year} {2016})},\ \Eprint
  {http://arxiv.org/abs/1607.07450} {arXiv:1607.07450 [astro-ph.HE]}
  \BibitemShut {NoStop}%
\bibitem [{\citenamefont {{Richers}}\ \emph
  {et~al.}(2021{\natexlab{a}})\citenamefont {{Richers}}, \citenamefont
  {{Willcox}}, \citenamefont {{Ford}},\ and\ \citenamefont
  {{Myers}}}]{2021PhRvD.103h3013R}%
  \BibitemOpen
  \bibfield  {author} {\bibinfo {author} {\bibfnamefont {S.}~\bibnamefont
  {{Richers}}}, \bibinfo {author} {\bibfnamefont {D.~E.}\ \bibnamefont
  {{Willcox}}}, \bibinfo {author} {\bibfnamefont {N.~M.}\ \bibnamefont
  {{Ford}}}, \ and\ \bibinfo {author} {\bibfnamefont {A.}~\bibnamefont
  {{Myers}}},\ }\href {\doibase 10.1103/PhysRevD.103.083013} {\bibfield
  {journal} {\bibinfo  {journal} {\prd}\ }\textbf {\bibinfo {volume} {103}},\
  \bibinfo {eid} {083013} (\bibinfo {year} {2021}{\natexlab{a}})},\ \Eprint
  {http://arxiv.org/abs/2101.02745} {arXiv:2101.02745 [astro-ph.HE]}
  \BibitemShut {NoStop}%
\bibitem [{\citenamefont {{Froustey}}\ \emph {et~al.}(2024)\citenamefont
  {{Froustey}}, \citenamefont {{Richers}}, \citenamefont {{Grohs}},
  \citenamefont {{Flynn}}, \citenamefont {{Foucart}}, \citenamefont
  {{Kneller}},\ and\ \citenamefont {{McLaughlin}}}]{2024PhRvD.109d3046F}%
  \BibitemOpen
  \bibfield  {author} {\bibinfo {author} {\bibfnamefont {J.}~\bibnamefont
  {{Froustey}}}, \bibinfo {author} {\bibfnamefont {S.}~\bibnamefont
  {{Richers}}}, \bibinfo {author} {\bibfnamefont {E.}~\bibnamefont {{Grohs}}},
  \bibinfo {author} {\bibfnamefont {S.~D.}\ \bibnamefont {{Flynn}}}, \bibinfo
  {author} {\bibfnamefont {F.}~\bibnamefont {{Foucart}}}, \bibinfo {author}
  {\bibfnamefont {J.~P.}\ \bibnamefont {{Kneller}}}, \ and\ \bibinfo {author}
  {\bibfnamefont {G.~C.}\ \bibnamefont {{McLaughlin}}},\ }\href {\doibase
  10.1103/PhysRevD.109.043046} {\bibfield  {journal} {\bibinfo  {journal}
  {\prd}\ }\textbf {\bibinfo {volume} {109}},\ \bibinfo {eid} {043046}
  (\bibinfo {year} {2024})},\ \Eprint {http://arxiv.org/abs/2311.11968}
  {arXiv:2311.11968 [astro-ph.HE]} \BibitemShut {NoStop}%
\bibitem [{\citenamefont {Froustey}\ \emph {et~al.}(2025)\citenamefont
  {Froustey}, \citenamefont {Kneller},\ and\ \citenamefont
  {McLaughlin}}]{Froustey:2024sgz}%
  \BibitemOpen
  \bibfield  {author} {\bibinfo {author} {\bibfnamefont {J.}~\bibnamefont
  {Froustey}}, \bibinfo {author} {\bibfnamefont {J.~P.}\ \bibnamefont
  {Kneller}}, \ and\ \bibinfo {author} {\bibfnamefont {G.~C.}\ \bibnamefont
  {McLaughlin}},\ }\href {\doibase 10.1103/PhysRevD.111.063022} {\bibfield
  {journal} {\bibinfo  {journal} {Phys. Rev. D}\ }\textbf {\bibinfo {volume}
  {111}},\ \bibinfo {pages} {063022} (\bibinfo {year} {2025})},\ \Eprint
  {http://arxiv.org/abs/2409.05807} {arXiv:2409.05807 [hep-ph]} \BibitemShut
  {NoStop}%
\bibitem [{\citenamefont {Kneller}\ \emph {et~al.}(2025)\citenamefont
  {Kneller}, \citenamefont {Froustey}, \citenamefont {Grohs}, \citenamefont
  {Foucart}, \citenamefont {McLaughlin},\ and\ \citenamefont
  {Richers}}]{Kneller:2024buy}%
  \BibitemOpen
  \bibfield  {author} {\bibinfo {author} {\bibfnamefont {J.~P.}\ \bibnamefont
  {Kneller}}, \bibinfo {author} {\bibfnamefont {J.}~\bibnamefont {Froustey}},
  \bibinfo {author} {\bibfnamefont {E.~B.}\ \bibnamefont {Grohs}}, \bibinfo
  {author} {\bibfnamefont {F.}~\bibnamefont {Foucart}}, \bibinfo {author}
  {\bibfnamefont {G.~C.}\ \bibnamefont {McLaughlin}}, \ and\ \bibinfo {author}
  {\bibfnamefont {S.}~\bibnamefont {Richers}},\ }\href {\doibase
  10.1103/PhysRevD.111.063046} {\bibfield  {journal} {\bibinfo  {journal}
  {Phys. Rev. D}\ }\textbf {\bibinfo {volume} {111}},\ \bibinfo {pages}
  {063046} (\bibinfo {year} {2025})},\ \Eprint
  {http://arxiv.org/abs/2410.00719} {arXiv:2410.00719 [hep-ph]} \BibitemShut
  {NoStop}%
\bibitem [{\citenamefont {{Richers}}\ \emph
  {et~al.}(2021{\natexlab{b}})\citenamefont {{Richers}}, \citenamefont
  {{Willcox}},\ and\ \citenamefont {{Ford}}}]{2021PhRvD.104j3023R}%
  \BibitemOpen
  \bibfield  {author} {\bibinfo {author} {\bibfnamefont {S.}~\bibnamefont
  {{Richers}}}, \bibinfo {author} {\bibfnamefont {D.}~\bibnamefont
  {{Willcox}}}, \ and\ \bibinfo {author} {\bibfnamefont {N.}~\bibnamefont
  {{Ford}}},\ }\href {\doibase 10.1103/PhysRevD.104.103023} {\bibfield
  {journal} {\bibinfo  {journal} {\prd}\ }\textbf {\bibinfo {volume} {104}},\
  \bibinfo {eid} {103023} (\bibinfo {year} {2021}{\natexlab{b}})},\ \Eprint
  {http://arxiv.org/abs/2109.08631} {arXiv:2109.08631 [astro-ph.HE]}
  \BibitemShut {NoStop}%
\bibitem [{\citenamefont {{Chakraborty}}\ \emph {et~al.}(2016)\citenamefont
  {{Chakraborty}}, \citenamefont {{Hansen}}, \citenamefont {{Izaguirre}},\ and\
  \citenamefont {{Raffelt}}}]{2016JCAP...03..042C}%
  \BibitemOpen
  \bibfield  {author} {\bibinfo {author} {\bibfnamefont {S.}~\bibnamefont
  {{Chakraborty}}}, \bibinfo {author} {\bibfnamefont {R.~S.}\ \bibnamefont
  {{Hansen}}}, \bibinfo {author} {\bibfnamefont {I.}~\bibnamefont
  {{Izaguirre}}}, \ and\ \bibinfo {author} {\bibfnamefont {G.~G.}\ \bibnamefont
  {{Raffelt}}},\ }\href {\doibase 10.1088/1475-7516/2016/03/042} {\bibfield
  {journal} {\bibinfo  {journal} {\jcap}\ }\textbf {\bibinfo {volume} {2016}},\
  \bibinfo {eid} {042} (\bibinfo {year} {2016})},\ \Eprint
  {http://arxiv.org/abs/1602.00698} {arXiv:1602.00698 [hep-ph]} \BibitemShut
  {NoStop}%
\bibitem [{\citenamefont {{Hansen}}\ and\ \citenamefont
  {{Hannestad}}(2014)}]{2014PhRvD..90b5009H}%
  \BibitemOpen
  \bibfield  {author} {\bibinfo {author} {\bibfnamefont {R.~S.}\ \bibnamefont
  {{Hansen}}}\ and\ \bibinfo {author} {\bibfnamefont {S.}~\bibnamefont
  {{Hannestad}}},\ }\href {\doibase 10.1103/PhysRevD.90.025009} {\bibfield
  {journal} {\bibinfo  {journal} {\prd}\ }\textbf {\bibinfo {volume} {90}},\
  \bibinfo {eid} {025009} (\bibinfo {year} {2014})}\BibitemShut {NoStop}%
\bibitem [{\citenamefont {{Urquilla}}\ and\ \citenamefont
  {{Richers}}(2024)}]{2024PhRvD.109j3040U}%
  \BibitemOpen
  \bibfield  {author} {\bibinfo {author} {\bibfnamefont {E.}~\bibnamefont
  {{Urquilla}}}\ and\ \bibinfo {author} {\bibfnamefont {S.}~\bibnamefont
  {{Richers}}},\ }\href {\doibase 10.1103/PhysRevD.109.103040} {\bibfield
  {journal} {\bibinfo  {journal} {\prd}\ }\textbf {\bibinfo {volume} {109}},\
  \bibinfo {eid} {103040} (\bibinfo {year} {2024})},\ \Eprint
  {http://arxiv.org/abs/2401.01936} {arXiv:2401.01936 [astro-ph.HE]}
  \BibitemShut {NoStop}%
\bibitem [{\citenamefont {{Xiong}}\ \emph
  {et~al.}(2023{\natexlab{b}})\citenamefont {{Xiong}}, \citenamefont {{Wu}},\
  and\ \citenamefont {{Qian}}}]{2023PhRvD.108d3007X}%
  \BibitemOpen
  \bibfield  {author} {\bibinfo {author} {\bibfnamefont {Z.}~\bibnamefont
  {{Xiong}}}, \bibinfo {author} {\bibfnamefont {M.-R.}\ \bibnamefont {{Wu}}}, \
  and\ \bibinfo {author} {\bibfnamefont {Y.-Z.}\ \bibnamefont {{Qian}}},\
  }\href {\doibase 10.1103/PhysRevD.108.043007} {\bibfield  {journal} {\bibinfo
   {journal} {\prd}\ }\textbf {\bibinfo {volume} {108}},\ \bibinfo {eid}
  {043007} (\bibinfo {year} {2023}{\natexlab{b}})},\ \Eprint
  {http://arxiv.org/abs/2303.05906} {arXiv:2303.05906 [hep-ph]} \BibitemShut
  {NoStop}%
\bibitem [{\citenamefont {{Nagakura}}\ and\ \citenamefont
  {{Zaizen}}(2022)}]{2022PhRvL.129z1101N}%
  \BibitemOpen
  \bibfield  {author} {\bibinfo {author} {\bibfnamefont {H.}~\bibnamefont
  {{Nagakura}}}\ and\ \bibinfo {author} {\bibfnamefont {M.}~\bibnamefont
  {{Zaizen}}},\ }\href {\doibase 10.1103/PhysRevLett.129.261101} {\bibfield
  {journal} {\bibinfo  {journal} {\prl}\ }\textbf {\bibinfo {volume} {129}},\
  \bibinfo {eid} {261101} (\bibinfo {year} {2022})},\ \Eprint
  {http://arxiv.org/abs/2206.04097} {arXiv:2206.04097 [astro-ph.HE]}
  \BibitemShut {NoStop}%
\bibitem [{\citenamefont {{Zaizen}}\ and\ \citenamefont
  {{Nagakura}}(2023)}]{2023PhRvD.107j3022Z}%
  \BibitemOpen
  \bibfield  {author} {\bibinfo {author} {\bibfnamefont {M.}~\bibnamefont
  {{Zaizen}}}\ and\ \bibinfo {author} {\bibfnamefont {H.}~\bibnamefont
  {{Nagakura}}},\ }\href {\doibase 10.1103/PhysRevD.107.103022} {\bibfield
  {journal} {\bibinfo  {journal} {\prd}\ }\textbf {\bibinfo {volume} {107}},\
  \bibinfo {eid} {103022} (\bibinfo {year} {2023})},\ \Eprint
  {http://arxiv.org/abs/2211.09343} {arXiv:2211.09343 [astro-ph.HE]}
  \BibitemShut {NoStop}%
\bibitem [{\citenamefont {{Xiong}}\ \emph
  {et~al.}(2023{\natexlab{c}})\citenamefont {{Xiong}}, \citenamefont {{Wu}},
  \citenamefont {{Abbar}}, \citenamefont {{Bhattacharyya}}, \citenamefont
  {{George}},\ and\ \citenamefont {{Lin}}}]{2023PhRvD.108f3003X}%
  \BibitemOpen
  \bibfield  {author} {\bibinfo {author} {\bibfnamefont {Z.}~\bibnamefont
  {{Xiong}}}, \bibinfo {author} {\bibfnamefont {M.-R.}\ \bibnamefont {{Wu}}},
  \bibinfo {author} {\bibfnamefont {S.}~\bibnamefont {{Abbar}}}, \bibinfo
  {author} {\bibfnamefont {S.}~\bibnamefont {{Bhattacharyya}}}, \bibinfo
  {author} {\bibfnamefont {M.}~\bibnamefont {{George}}}, \ and\ \bibinfo
  {author} {\bibfnamefont {C.-Y.}\ \bibnamefont {{Lin}}},\ }\href {\doibase
  10.1103/PhysRevD.108.063003} {\bibfield  {journal} {\bibinfo  {journal}
  {\prd}\ }\textbf {\bibinfo {volume} {108}},\ \bibinfo {eid} {063003}
  (\bibinfo {year} {2023}{\natexlab{c}})},\ \Eprint
  {http://arxiv.org/abs/2307.11129} {arXiv:2307.11129 [astro-ph.HE]}
  \BibitemShut {NoStop}%
\end{thebibliography}%

\end{document}